\documentclass[11pt]{article}
\usepackage[final]{pdfpages}
\usepackage{amsmath}
\usepackage{slashed}
\usepackage{amssymb}
\catcode`@11
\def\seceqaa{\@addtoreset{equation}{section}
\def\theequation{A\arabic{equation}}}
\def\seceqbb{\@addtoreset{equation}{section}
\def\theequation{B\arabic{equation}}}
\def\seceqcc{\@addtoreset{equation}{section}
\def\theequation{C\arabic{equation}}}
\def\seceqdd{\@addtoreset{equation}{section}
\def\theequation{D\arabic{equation}}}
\def\seceqee{\@addtoreset{equation}{section}
\def\theequation{E\arabic{equation}}}
\catcode`@11
\begin{document}
\begin{titlepage}
\begin{center}
{\Large\bf  Transport Coefficients of Black MQGP $M3$-Branes}
\vskip 0.1in Mansi Dhuria$\ ^{(a)}$\footnote{email: mansidph@prl.res.in}
and
 Aalok Misra$\ ^{(b)}$\footnote{e-mail: aalokfph@iitr.ac.in
}\\
(a)Theoretical Physics Division, Physical Research Laboratory, Ahmedabad - 380 009, India\\
(b)Department of Physics, Indian Institute of Technology,
Roorkee - 247 667, Uttaranchal, India

 \vskip 0.5 true in
\date{\today}
\end{center}
\thispagestyle{empty}
\begin{abstract}
The SYZ mirror, in the `delocalized limit' of \cite{becker2004},  of $N$ $D3$-branes, $M$ fractional $D3$-branes and $N_f$ flavor $D7$-branes wrapping a non-compact four cycle in the presence of a black-hole resulting in a non-K\"{a}hler resolved warped deformed conifold (NKRWDC) in \cite{metrics}, was carried out in \cite{MQGP} and resulted in black $M3$-branes.
There are two parts of the paper. In the first we show that in the `MQGP' limit discussed in \cite{MQGP}: finite $g_s$ (expected to hence be more relevant to QGP), finite $g_sM, N_f, g_s^2MN_f$ and very large $g_sN$,  and very small $\frac{g_sM^2}{N}$, (i) the uplift, if valid globally (like \cite{local_global_theta} for fractional $D3$ branes in conifolds),   asymptotes to $M5$-branes wrapping a two-cycle (homologously an (large) integer sum of two-spheres) in $AdS_5\times M_6$, (ii) assuming the deformation parameter to be larger than the resolution parameter, by estimating the five $SU(3)$ structure torsion ($\tau$) classes $W_{1,2,3,4,5}$ we verify that $\tau\in W_4\oplus W_5: \frac{2}{3}\Re e W^{\bar{3}}_5 = W^{\bar{3}}_4$ in the UV and near $\theta_i=0$, implying the NKRWDC locally preserves supersymmetry,  (iii) the local $T^3$ of \cite{MQGP} in the large-$r$ limit  satisfies the same conditions as the maximal $T^2$-invariant special Lagrangian three-cycle of $T^*S^3$ of \cite{T2slag}, partly justifying use of SYZ mirror symmetry in the `delocalized limit' of \cite{becker2004} in \cite{MQGP}. In the second part of the paper, by either integrating out the angular coordinates of the non-compact four-cycle which a $D7$-brane wraps around, using the Ouyang embedding, in the DBI action of a $D7$-brane evaluated at infinite radial boundary, or by dimensionally reducing the eleven-dimensional EH action to five ($\mathbb{R}^{1,3},r$) dimensions and at the infinite radial boundary, we then calculate in particular the $g_s\stackrel{<}{\sim}1$(part of the 'MQGP')limit, a variety of gauge and metric-perturbation-modes' two-point functions using the prescription of \cite{son+starinets}. We hence show: (i) diffusion constant $D\sim\frac{1}{T}$, (ii) the electrical conductivity $\sigma\sim T$, (iii) the charge susceptibility $\chi\sim T^2$, (iv) [using (i) - (iii)] the Einstein's relation $\frac{\sigma}{\chi}=D$, is indeed satisfied, (v) the R-charge diffusion constant $D_R\sim\frac{1}{T}$, and (vi)(using \cite{MQGP} and Kubo's formula related to shear viscosity $\eta$) the possibility of generating $\frac{\eta}{s}=\frac{1}{4\pi}$ from solutions to the vector and tensor mode metric perturbations' EOMs, separately.
The results are also valid for the limits of \cite{metrics}:  $g_s, \frac{g_sM^2}{N}, g_s^2MN_f\rightarrow0$, $g_sN, g_sM>>1$.
\end{abstract}
\end{titlepage}

\section{Introduction}

Since the formulation of AdS/CFT correspondence \cite{maldacena,witten,aharony},  it has been of profound interest to gain theoretical insight into the  physics of strongly coupled Quark Gluon Plasma (sQGP) produced in the heavy-ion collision at  RHIC by using holographic techniques. Within this context, the holographic spectral functions play an important role because of their ability to provide a framework to compute certain transport properties of QGP such as the conductivity and viscosity of the plasma which otherwise are not computationally tractable. The calculation scheme generally involves Kubo's formulae, which involve the long distance and low frequency limit of two-point Green's functions. The methodology to calculate two-point Green's functions by applying gauge-gravity duality has been developed in the series of papers \cite{son+starinets,PSS,son_ssound} by considering $D3$-brane background. However, this approach is useful  for transport coefficients that are universal within models with gravity duals. One of the biggest upshots of the same is the viscosity-over-entropy ratio which takes on the universal value $\frac{1}{4\pi}$, known as KSS bound \cite{kouvtun_2003,kouvtun_2005, buchel}.  Though the bound was initially obtained by assuming zero  charge density,  the same was shown to hold to satisfy for wide class of theories including non-zero chemical potential, finite spatial momentum and other different backgrounds (see \cite{RchargeB,Mas,Karch+bannon,RMyersetal}). In addition to this, it is interesting to analyse the photon production in the context of non-perturbative theories as the same reveals one of the most fascinating signatures of a QGP. The study of photoemission in strongly coupled theories by using AdS/CFT correspondence was initiated in \cite{conductivity_kouvtunetal} in the presence of vanishing chemical potential, and then continued in  \cite{Mateos,Parnachev} in the presence of non-zero chemical potential by considering $D3/D7$ and $D4/D6$  backgrounds where baryonic $U(1)$ symmetry  is exhibited from $U(N_f)= SU(N_f)\times U(1)$ global symmetry present on  the world volume of stack of coincident flavoured branes.  In addition to  the finite chemical potential, spectral functions were obtained in \cite{JMasetal2,RMyersetal,Patinoetal} by considering  finite spatial momentum, finite electric field and anisotropic plasma  respectively. The validity of the results of different transport coefficients obtained in the context of different gravitational backgrounds is manifested on the basis of certain universal bounds. For example, it is known that the results of some transport coefficients  should satisfy the famous Einstein relation according to which  one can express the diffusion constant as the ratio of conductivity and susceptibility \cite{JMasetal2}. Based on \cite{son+starinets,PSS}, the diffusion coefficient can be extracted from poles in the retarded Green's functions corresponding to density of conserved charges.  Also, it  was suggested in \cite{conductivity_kouvtun} that similar to  viscosity-to-entropy ratio, there can exist a  universal bound on the ratio of conductivity-to-viscosity in theories saturated by models with gravity duals. Further,  based on the Minkowski space prescription of Son and Starinets,  two-point functions were obtained in \cite{herzog,herzog2} for conserved $R$-symmetry current and different  components of the stress-energy tensor for  $M2$- and $M5$-brane in $ADS_{4} \times S^{7}$ and $ ADS_{7}\times S^{4}$ backgrounds respectively, using which a variety of transport coefficients including diffusion coefficient, shear viscosity  as well as the second speed of sound, were calculated.

The knowledge of UV-completion of gauge theories is important to handle  issues related to finiteness of the solution at short distances as well as to capture certain aspects of large-$N$ thermal QCD. One of the earliest successful attempts to explain the RG flow  in the dual background was made by Klebanov and Strassler in \cite{Klebanov+Strassler} by embedding $D5$-branes wrapped over $S^2$ in a conifold background which was further extended to Ouyang-Klebanov-Strassler (OKS) background  in \cite{ouyang} in the presence of fundamental quarks (by including $N_f$ $D7$-branes wrapped over non-compact four-cycle(s)). This was followed by the modified OKS-BH background \cite{metrics} in the presence of  a black-hole (BH). It  would be very interesting to determine the transport coefficients by considering non-conformal modified OKS-BH background, {\it specially in the finite-string coupling limit}, which motivates an uplift to M-theory in the `delocalized limit' of \cite{becker2004}, and calculation of the transport coefficients in the same.

In this paper,  {\it after explicitly showing that the M-theory uplift in the delocalized limit of \cite{becker2004}, of the modified OKS-BH background of \cite{metrics} to black $M3$-branes, is a solution to $D=11$ supergravity  in particular in the `MQGP limit' described in \cite{MQGP}}, we intend to study the behaviour of transport coefficients of these black $M3$-branes.\footnote{ The near-horizon geometry of these branes, near the $\theta_{1,2} = 0, \pi$ branches, preserves $\frac{1}{8}$ supersymmetry.\cite{MQGP}} {\it One should keep in mind that the MQGP limit  could be more suited to demonstrating the characteristics of sQGP \cite{finite_gYM_Natsuume} by exploiting gauge-gravity duality,  and due to the finiteness of the string coupling can meaningfully only be addressed within an M-theoretic framework. } We are not aware of previous attempts at calculation of transport coefficients of the string theory duals of large-$N$ thermal QCD-like theories
{\it at finite string coupling} - this work is hence expected to shed some light on what to expect in terms of the values of a variety of transport coefficients for sQGP.
Using the background of \cite{metrics} or its type IIA mirror in the delocalized limit of \cite{becker2004} or its M-theory uplift, we have already computed some of the transport coefficients such as viscosity-to-entropy ratio and diffusion constant in \cite{MQGP} by adopting the KSS prescription \cite{kouvtun_2003}. It would be interesting to study  important  transport coefficients such as electrical conductivity, shear viscosity, etc.
in the finite string/gauge coupling-limit (part of the MQGP limit of \cite{MQGP}) by using the  M-theory uplift of non-conformal resolved warped deformed conifold background of \cite{metrics} which asymptotically, we show, corresponds to black $M5$-branes wrapping a sum of two-spheres in  $AdS_{5}\times {\cal M}_{6}$.

An overview of the paper is as follows:  In subsection {{\bf 2.1}}, we review  the framework of our M-theory uplift in the `delocalized' limit of \cite{becker2004} of type IIB background of \cite{metrics} as constructed in \cite{MQGP} using  S(trominger) Y(au) Z(aslow) mirror symmetry (in the `delocalized' limit of \cite{becker2004}). In this section, we also highlight results of various  hydro/thermodynamical quantities obtained in the  MQGP limit provided in \cite{MQGP},   and then review the thermodynamical stability of the M-theory uplift. In subsection {\bf 2.2}, we explicitly demonstrate that the uplift described in {\bf 2.1} solves the $D=11$ supergravity equations of motion, in the large-$r$ limit and near the $\theta_{1,2}=0,\pi$-branches in the delocalized limit of \cite{becker2004}, and in the MQGP limit of \cite{MQGP}.  In subsection {{\bf 2.3}}, we obtain magnetic charge of the black $M3$-branes  due to four-form fluxes $G_{4}$ through all (non)compact four-cycles and  point out that the black $M3$-branes asymptotically can be thought of as $M5$-branes wrapped around two-cycles given homologously by integer sum of two-spheres. In subsection {{\bf 2.4}}, given that supersymmetry via the existence of a s(pecial) Lag(rangian) is necessary for implementing mirror symmetry as three T-dualities a la SYZ, we show that in particular the MQGP limit of (\ref{limits_Dasguptaetal+MQGP}) and assuming that the deformation parameter is larger than the resolution parameter,  the local $T^3$ of \cite{MQGP}, in the large-$r$ limit, is the maximal $T^2$-invariant sLag of a deformed conifold as defined in \cite{T2slag}. In section {{\bf 3}}, we turn toward the  discussion of various transport coefficients in the MQGP limit. A variety of transport coefficients can be  obtained by considering fluctuations of $U(1)_{N_f}$ gauge field corresponding to $N_f$ $D7$-branes wrapping non-compact four-cycle of (resolved) warped deformed conifold as discussed in \cite{MQGP}, the fluctuations around $U(1)_R$ gauge field, and the stress energy tensor.  In this spirit, in section {{\bf 3.1}, we first  consider gauge field fluctuations around non-zero temporal component of gauge field background (non-zero chemical potential as worked out in \cite{MQGP}) in the presence of $N_f$ $D7$-branes wrapping non-compact four-cycle in the resolved warped deformed conifold background  and work out the equations of motion (EOM) of the same.  Further, we move on to the computation of  two-point retarded Green's functions by using the prescription first set out  in \cite{son+starinets}  using which one can calculate the electrical conductivity by utilising Kubo's formula. In section {{\bf 3.2},  we work out the charge susceptibility due to non-zero baryon density which turns out to be such that the ratio of the electrical conductivity and charge susceptibility satisfies the Einstein relation. Following the same strategy as described in {{\bf 3.2}}, we  compute the correlation function corresponding to $U(1)_R$ current in the context of  11-dimensional M-theory background in section {{\bf 3.3}}, the pole of which simply reads off ``diffusion coefficient". In section {{\bf 3.4}}, we  determine correlation functions of stress energy tensor modes. After classifying different stress-energy tensor modes, we compute two-point correlation functions of vector and tensor modes of black $M3$-brane. Using Kubo's formula, we therefore calculate the shear viscosity and show that is possible to obtain the shear viscosity-to-entropy ratio ($\eta/s$)  to be $1/4\pi$. We finally summarize our results, also giving a flow-chart of the flow of logic and (sub-)section-wise summary of results in the process in Fig. 1, and remark on the possible extensions of the current work in  section {{\bf 4}}. There are four appendices.  Appendix {\bf A} includes simplified expressions of possible non-zero $G_{\mu\nu\lambda\rho}$'s  obtained. In appendix {\bf B} we explicitly evaluate the five $SU(3)$ structure torsion classes for the resolved warped deformed conifold background of \cite{metrics}. Appendix {\bf C} gives the explicit embedding of the maximal $T^2$-invariant special Lagrangian three-cycle of a deformed conifold. In appendix {\bf D}, we collect various intermediate steps relevant to evaluation of R-charge correlators of {\bf 3.2}.

\section{The  Black $M3$-Branes of \cite{MQGP}}

This section has two sub-sections. In {\bf 2.1}, we will review the uplift of type IIB background of \cite{metrics} to M-theory in the delocalized limit of \cite{becker2004}, as carried out in \cite{MQGP} using SYZ mirror symmetry (in the delocalized limit of \cite{becker2004}), as well as its hydrodynamical and thermodynamical properties. In {\bf 2.2}, we discuss the charge(s) of the black $M3$-branes of {\bf 2.1} and show that asymptotically the 11D space-time is a warped product of $AdS_5$ and a six-fold with $M5$-branes wrapping a two-cycle that is homologous to an integer sum of two-spheres.

\subsection{The Uplift in the Delocalized Limit of \cite{becker2004}}

 To include fundamental quarks at finite temperature relevant to, e.g., the study of the deconfined phase of strongly coupled QCD i.e ``Quark Gluon Plasma",  a black hole and $N_f$ flavor $D7$-branes `wrapping' a (non-compact) four-cycle were inserted apart from $N\ D3$-branes placed at the tip of six-dimensional conifold  and $M\ D5$-branes are wrapping a two cycle $S^2$,  in the context of type IIB string theory in \cite{metrics} which causes, both, resolution as well as deformation of the two- and three-cycles of the conifold respectively at $r=0$ (apart from warping), resulting in a warped resolved deformed conifold. In that background, back-reaction due to presence of black hole as well as $D7$-branes is included in 10-D warp factor $h(r,\theta_1,\theta_2)$. The metric from \cite{metrics}, is given as under:
\begin{eqnarray*}
\label{metric}
ds^2 = {1\over \sqrt{h}}
\left(-g_1 dt^2+dx_1^2+dx_2^2+dx_3^2\right)+\sqrt{h}\Big[g_2^{-1}dr^2+r^2 d{\cal M}_5^2\Big];
\end{eqnarray*}
 $g_i$'s demonstrate the presence of a black hole and is given as follows:
$ g_{1,2}(r,\theta_1,\theta_2)= 1-\frac{r_h^4}{r^4} + {\cal O}\left(\frac{g^2_sM}{N}\right)$,
where $r_h$ is the horizon, and the ($\theta_1, \theta_2$) dependence come from the
${\cal O}\left(\frac{g_sM^2}{N}\right)$ corrections.  In (\ref{metric}),
\begin{eqnarray}
\label{D=5}
& & d{\cal M}_5^2 =  h_1 (d\psi + {\cos}~\theta_1~d\phi_1 + {\cos}~\theta_2~d\phi_2)^2 +
h_2 (d\theta_1^2 + {\sin}^2 \theta_1 ~d\phi_1^2)  + h_4 (h_3 d\theta_2^2 + {\sin}^2 \theta_2 ~d\phi_2^2) \nonumber\\
& & + h_5\left[{\cos}~\psi \left(d\theta_1 d\theta_2 -
{\sin}~\theta_1 {\sin}~\theta_2 d\phi_1 d\phi_2\right) +  ~{\sin}~\psi \left({\sin}~\theta_1~d\theta_2 d\phi_1 +
{\sin}~\theta_2~d\theta_1 d\phi_2\right)\right],
\end{eqnarray}
$r>>a, h_5\sim\frac{({\rm deformation\ parameter})^2}{r^3}<<1\forall r >>({\rm deformation\ parameter})^{\frac{2}{3}}$. Due to presence of Black-hole, $h_i$ appearing in the internal metric (\ref{D=5}) as well as $M, N_f$ are not constant, and up to linear order depend on $g_s, M, N_f$ as given below:
\begin{eqnarray*}
& & \hskip -0.45in h_1 = {1\over 9} + {\cal O}(\frac{g_sM^2}{N}),\  h_2 = {1\over 6} + {\cal O}(\frac{g_sM^2}{N}),\ h_4 = h_2 + \frac{6 a^2}{r^2},\ h_3 = 1 + {\cal O}(\frac{g_sM^2}{N}),\nonumber\\
&& \hskip -0.45in M_{\rm eff}/N_{f}^{\rm eff} = M/N_f + \sum_{m\ge n} (a/b)_{mn} (g_sN_f)^m (g_sM)^n,\nonumber\\
& & \hskip -0.45in L=\left(4\pi g_s N\right)^{\frac{1}{4}}.
\end{eqnarray*}
The warp factor that includes the back-reaction due to fluxes as well as black-hole, for finite, but not large $r$, is given as:
\begin{eqnarray}
\label{h_small_r}
&& \hskip -0.45in h = \frac{L^4}{r^4}\Bigg[1+\frac{3g_sM_{\rm eff}^2}{2\pi N}{\rm log}r\left\{1+\frac{3g_sN^{\rm eff}_f}{2\pi}\left({\rm
log}r+\frac{1}{2}\right)+\frac{g_sN^{\rm eff}_f}{4\pi}{\rm log}\left({\rm sin}\frac{\theta_1}{2}
{\rm sin}\frac{\theta_2}{2}\right)\right\}\Bigg].\nonumber\\
\end{eqnarray}
For large $r$ \cite{metrics},
\begin{eqnarray}
\label{h_large_r}
h={L^4}\Biggl[\frac{1}{r^{4-\epsilon_1}} + \frac{1}{r^{4-2\epsilon_2}}
- \frac{2}{r^{4-\epsilon_2}} + \frac{1}{r^{4-r^{\frac{\epsilon_2^2}{2}}}}\Biggr]\equiv\sum_{\alpha=1}^4\frac{L_{(\alpha)}^4}{r^4_{(\alpha)}},
\end{eqnarray}
where $\epsilon_1\equiv \frac{3g_sM^2}{2\pi N} + \frac{g_s^2M^2N_f}{8\pi^2N} + \frac{3g_s^2M^2N_f}{8\pi N}ln\left(sin\frac{\theta_1}{2} sin\frac{\theta_2}{2}\right)$,
$\epsilon_2\equiv\frac{g_sM}{\pi}\sqrt{\frac{2N_f}{N}}$, $r_{(\alpha)}\equiv r^{1-\epsilon_{(\alpha)}}, \epsilon_{(1)}=\frac{\epsilon_1}{2}, \epsilon_{(2)}\equiv\epsilon_{(3)}=\frac{\epsilon_2}{2}; L_{(1)}=L_{(2)}=L_{(4)}
= L^4, L_{(3)}=-2L^4$. It is conjectured that as $r\rightarrow\infty, \alpha\in[1,\infty)$.

 In \cite{MQGP}, we considered the following two limits:
\begin{eqnarray}
\label{limits_Dasguptaetal+MQGP}
& (i)   &
{\rm weak} \ (g_s)\ {\rm coupling-large\ t'Hooft\ coupling\ limit}:\nonumber\\
& &  g_s<<1, g_sN_f<<1, \frac{g_sM^2}{N}<<1, g_sM>>1, g_sN>>1\nonumber\\
& & {\rm effected\ by}: g_s\sim\epsilon^{d}, M\sim\epsilon^{-\frac{3d}{2}}, N\sim\epsilon^{-19d}, \epsilon\leq{\cal O}(10^{-2}), d>0\nonumber\\
& (i i) &{\rm MQGP\ limit}: \frac{g_sM^2}{N}<<1, g_sN>>1, {\rm finite}\
 g_s, M\ \nonumber\\
& &{\rm effected\ by}:  g_s\sim\epsilon^d, M\sim\epsilon^{-\frac{3d}{2}}, N\sim\epsilon^{-39d}, \epsilon\lesssim 1, d>0.
\end{eqnarray}
Apart from other calculational simplifications, one of the advantages of  working with either of the limits in (\ref{limits_Dasguptaetal+MQGP}) is that the ten dimensional warp factor $h$ (\ref{h_large_r}) for large r approaches the expression (\ref{h_small_r}) for finite/small $r$. The reason for the unusual scalings appearing in (\ref{limits_Dasguptaetal+MQGP}) is, as explained towards the end of this section, that the same ensures that the finite part of the $D=11$ supergravity action - the Gibbons-Hawking-York boundary term (like \cite{schmude}) - is independent of the angular ($\theta_{1,2}$) cut-off/regulator, and yields an entropy density ($s\sim r_h^3$) from a thermodynamical calculation matching the same calculated from the horizon area.

{ The three-form fluxes, including the `asymmetry factors', are given by \cite{metrics}:
\begin{eqnarray}
\label{three-form fluxes}
& &  {\widetilde F}_3  =  2M { A_1} \left(1 + {3g_sN_f\over 2\pi}~{\rm log}~r\right) ~e_\psi \wedge
\frac{1}{2}\left({\rm sin}~\theta_1~ d\theta_1 \wedge d\phi_1-{ B_1}~{\rm sin}~\theta_2~ d\theta_2 \wedge
d\phi_2\right)\nonumber\\
&&  -{3g_s MN_f\over 4\pi} { A_2}~{dr\over r}\wedge e_\psi \wedge \left({\rm cot}~{\theta_2 \over 2}~{\rm sin}~\theta_2 ~d\phi_2
- { B_2}~ {\rm cot}~{\theta_1 \over 2}~{\rm sin}~\theta_1 ~d\phi_1\right)\nonumber \\
&&  -{3g_s MN_f\over 8\pi}{ A_3} ~{\rm sin}~\theta_1 ~{\rm sin}~\theta_2 \left({\rm cot}~{\theta_2 \over 2}~d\theta_1 +
{ B_3}~ {\rm cot}~{\theta_1 \over 2}~d\theta_2\right)\wedge d\phi_1 \wedge d\phi_2\label{brend}, \nonumber\\
& &  H_3 =  {6g_s { A_4} M}\Bigr(1+\frac{9g_s N_f}{4\pi}~{\rm log}~r+\frac{g_s N_f}{2\pi}
~{\rm log}~{\rm sin}\frac{\theta_1}{2}~
{\rm sin}\frac{\theta_2}{2}\Bigr)\frac{dr}{r}\nonumber \\
&&  \wedge \frac{1}{2}\Bigr({\rm sin}~\theta_1~ d\theta_1 \wedge d\phi_1
- { B_4}~{\rm sin}~\theta_2~ d\theta_2 \wedge d\phi_2\Bigr)
+ \frac{3g^2_s M N_f}{8\pi} { A_5} \Bigr(\frac{dr}{r}\wedge e_\psi -\frac{1}{2}de_\psi \Bigr)\nonumber  \\
&&  \wedge \Bigr({\rm cot}~\frac{\theta_2}{2}~d\theta_2
-{ B_5}~{\rm cot}~\frac{\theta_1}{2} ~d\theta_1\Bigr).
\end{eqnarray}}
The asymmetry factors ${ A_i}, { B_i}$ encode information of the black hole and of the background and are given in \cite{metrics}.  The ${\cal O}(a^2/r^2)$ corrections included in asymmetry factors correspond to modified Ouyang background  in the presence of black hole. The values for the axion $C_0$ and the five form $F_5$ are given by \cite{Ouyang2004}:
\begin{eqnarray}
\label{axfive}
&&C_0 ~ = ~ {N_f \over 4\pi} (\psi - \phi_1 - \phi_2) [{\rm since}\ \int_{S^1}dC_0=N_f],\nonumber\\
&& F_5 ~ = ~ {1\over g_s} \left[ d^4 x \wedge d h^{-1} + \ast(d^4 x \wedge dh^{-1})\right].
\end{eqnarray}
It can be shown that the deviation from supersymmetry, one way of measuring which is the deviation from the condition of $G_3\equiv\tilde{F}_3 - i e^{-\phi}H_3$ being imaginary self dual: $\left|iG_3 - *_6G_3\right|^2$, is proportional to the square of the resolution parameter $a$, which (assuming a negligible bare resolution parameter) in turn is related to the horizon radius $r_h$ via: $a^2 = \frac{g_sM^2}{N}r_h^2 + \frac{g_sM^2}{N}(g_sN_f)r_h^4$  \cite{nonextremel_dasgupta}. We hence see that despite the presence of a black hole, in either limits of (\ref{limits_Dasguptaetal+MQGP}), SUSY is approximately preserved.

 For embedding parameter $\mu<<1$, $D7$-branes monodromy arguments for Ouyang embedding: $x=\left ( 9 a^2 r^4 + r^6 \right ) ^{1/4} e^{\imath/2(\psi-\phi_1-\phi_2)}\,\sin\frac{\theta_1}{2}\,\sin\frac{\theta_2}{2}=\mu^2$ (relabeling $\rho$ as $r$) ``$j(\tau(x))=\frac{4.(24 f)^3}{27 g^2 + f^3}"\sim\frac{1}{x-\mu^2}$ or $\tau\sim\frac{1}{2\pi\imath} ln x$, implying \cite{metrics}:
\begin{eqnarray*}
& & e^{-\Phi} = {1\over g_s} -\frac{N_f}{8\pi} ~{\rm log} \left(r^6 + 9a^2 r^4\right) -
\frac{N_f}{2\pi} {\rm log} \left({\rm sin}~{\theta_1\over 2} ~ {\rm sin}~{\theta_2\over 2}\right).
\end{eqnarray*}

 Working around: $r\approx\langle r\rangle, \theta_{1,2}\approx\langle\theta_{1,2}\rangle,\psi\approx\langle\psi\rangle$ 
 define the T-duality coordinates, $(\phi_1,\phi_2,\psi)\rightarrow(x,y,z)$ as:
\begin{equation}
\label{xyz defs}
x = \sqrt{h_2}h^{\frac{1}{4}}sin\langle\theta_1\rangle\langle r\rangle \phi_1,\ y = \sqrt{h_4}h^{\frac{1}{4}}sin\langle\theta_2\rangle\langle r\rangle \phi_2,\ z=\sqrt{h_1}\langle r\rangle h^{\frac{1}{4}}\psi.
\end{equation}

{ Interestingly, around $\psi=\langle\psi\rangle$ (the `delocalised' limit of \cite{becker2004}, i.e., instead of a dependence on a local coordinate, an isometry arising by freezing a coordinate and redefining it away as also summarised below), under the coordinate transformation \cite{becker2004}:
\begin{equation}
\label{transformation_psi}
\left(\begin{array}{c} sin\theta_2 d\phi_2 \\ d\theta_2\end{array} \right)\rightarrow \left(\begin{array}{cc} cos\langle\psi\rangle & sin\langle\psi\rangle \\
- sin\langle\psi\rangle & cos\langle\psi\rangle
\end{array}\right)\left(\begin{array}{c}
sin\theta_2 d\phi_2\\
d\theta_2
\end{array}
\right),
\end{equation}
the $h_5$ term becomes $h_5\left[d\theta_1 d\theta_2 - sin\theta_1 sin\theta_2 d\phi_1d\phi_2\right]$. Following closely \cite{becker2004}, and utilizing (\ref{transformation_psi}) and (\ref{xyz defs}) one sees that:
\begin{eqnarray}
\label{theta2-trans}
& & \theta_2\rightarrow -\frac{\sqrt{6}\sin\langle\psi\rangle}{\left(4\pi g_sN\right)^{\frac{1}{4}}} y + \cos\langle\psi\rangle\theta_2\nonumber\\
& & \hskip -0.3in {\rm implying}\ \cos\theta_2d\phi_2\rightarrow \cot\left[-\frac{\sqrt{6}\sin\langle\psi\rangle}{\left(4\pi g_sN\right)^{\frac{1}{4}}} y + \cos\langle\psi\rangle\theta_2\right]\left\{\frac{\sqrt{6}\cos\langle\psi\rangle dy}{\left(4\pi g_sN\right)^{\frac{1}{4}}} + \sin\langle\psi\rangle d\theta_2\right\}.
\end{eqnarray}
Defining ${\bar\theta}_2\equiv -\frac{\sqrt{6}\sin\langle\psi\rangle}{\left(4\pi g_sN\right)^{\frac{1}{4}}} y + \cos\langle\psi\rangle\theta_2 = - \sin\langle\theta_2\rangle \sin\langle\psi\rangle \phi_2 + \cos\langle\psi\rangle \theta_2\stackrel{\theta_2\sim0,\pi}{\longrightarrow}\cos\langle\psi\rangle\theta_2$, $e_\psi\rightarrow e_\psi + \cot{\bar\theta}_2\cos\langle\psi\rangle\sin\langle\theta_2\rangle d\phi_2 + \cot{\bar\theta}_2\sin\langle\psi\rangle d\theta_2$,
one hence sees that under $\psi\rightarrow\psi - \cos\langle{\bar\theta}_2\rangle\phi_2 + \cos\langle\theta_2\rangle\phi_2 - \tan\langle\psi\rangle ln\sin{\bar\theta}_2$, $e_\psi\rightarrow e_\psi$.
In the delocalized limit of \cite{becker2004}, one thus introduces an isometry along $\psi$ in addition to the isometries along $\phi_{1,2}$. This clearly is not valid globally - the deformed conifold does not possess a third global isometry. }

 To enable use of SYZ-mirror duality via three T dualities, one needs to ensure that the abovementioned local $T^3$ is a special Lagrangian (sLag) three-cycle. This will be explicitly shown in section {\bf 3}. For implementing mirror symmetry via SYZ prescription, one also needs to ensure a large base (implying large complex structures of the aforementioned two two-tori) of the $T^3(x,y,z)$ fibration.  This is effected via \cite{Chenetal}:
\begin{eqnarray}
\label{SYZ-large base}
& & d\psi\rightarrow d\psi + f_1(\theta_1)\cos\theta_1 d\theta_1 + f_2(\theta_2)\cos\theta_2 d\theta_2,\nonumber\\
& & d\phi_{1,2}\rightarrow d\phi_{1,2} - f_{1,2}(\theta_{1,2})d\theta_{1,2},
\end{eqnarray}
for appropriately chosen large values of $f_{1,2}(\theta_{1,2})$. The three-form fluxes
 remain invariant. The fact that one can choose such large values of $f_{1,2}(\theta_{1,2})$, was justified in
 \cite{MQGP}.
The guiding principle was that one requires that the metric obtained after SYZ-mirror transformation applied to the resolved warped deformed conifold is like a warped resolved conifold at least locally, then $G^{IIA}_{\theta_1\theta_2}$ needs to vanish \cite{MQGP}.

We can get a one-form type IIA potential from the triple T-dual (along $x, y, z$) of the type IIB $F_{1,3,5}$. We therefore can construct the following type IIA gauge field one-form \cite{MQGP} in the delocalized limit of \cite{becker2004}:
\begin{eqnarray*}
\label{A}
\hskip -0.4in\bullet\ A^{F_3} & = & \Biggl[\tilde{\tilde{\tilde{F}}}_{xr}x dr + \tilde{\tilde{\tilde{F}}}_{x\theta_1}x d\theta_1 + \tilde{\tilde{\tilde{F}}}_{x\theta_2}x d\theta_2 + \tilde{\tilde{\tilde{F}}}_{y\theta_1}y d\theta_1 + \tilde{\tilde{\tilde{F}}}_{y\theta_2}y d\theta_2\nonumber\\
 & & \hskip -0.7in +  \tilde{\tilde{\tilde{F}}}_{z\theta_2}z d\theta_2 +  \tilde{\tilde{\tilde{F}}}_{z\theta_1}z d\theta_1 +   \tilde{\tilde{\tilde{F}}}_{zr}z dr
 + \tilde{\tilde{\tilde{F}}}_{yr}y dr\Biggr]  \left(\theta_{1,2}\rightarrow\langle\theta_{1,2}\rangle,\phi_{1,2}\rightarrow\langle \phi_{1,2}\rangle,\psi\rightarrow\right\langle\psi\rangle,r\rightarrow\langle r\rangle);\nonumber\\
 \bullet\  A^{F_1}&=&\left(\tilde{\tilde{\tilde{F_1}}}\ _{yx} y dx + \tilde{\tilde{\tilde{F_1}}}\ _{zx} z dx\right) \left(\theta_{1,2}\rightarrow\langle\theta_{1,2}\rangle,\phi_{1,2}\rightarrow\langle \phi_{1,2}\rangle,\psi\rightarrow\right\langle\psi\rangle,r\rightarrow\langle r\rangle);\nonumber\\
 \bullet\  A^{F_5}&=&\left( \tilde{\tilde{\tilde{F_5}}}\ _{r\theta_1}r d\theta_1 + \tilde{\tilde{\tilde{F_5}}}\ _{\theta_1\theta_2}\theta_1 d\theta_2 + \tilde{\tilde{\tilde{F_5}}}\ _{r\theta_2}r d\theta_2\right) \left(\theta_{1,2}\rightarrow\langle\theta_{1,2}\rangle,\phi_{1,2}\rightarrow\langle \phi_{1,2}\rangle,\psi\rightarrow\langle\psi\rangle,r\rightarrow\langle r\rangle\right),
\end{eqnarray*}
resulting in the following local uplift:
\begin{eqnarray*}
& &    ds^2_{11} =  e^{-\frac{2{\phi}^{IIA}}{3}} \Biggl[
\frac{1}{\sqrt{h\left(r,\theta_1,\theta_2\right)}}\Bigl(-g_1 dt^2+dx_1^2+dx_2^2+dx_3^2\Bigr)+ \sqrt{h\left(r,\theta_1,\theta_2\right)}\Bigl(g_2^{-1}dr^2\Bigr) \nonumber\\
& &+  ds^2_{IIA}({\theta_{1,2},\phi_{1,2},\psi})\Biggr]  + e^{\frac{4{\phi}^{IIA}}{3}}\Bigl(dx_{11} + A^{F_1}+A^{F_3}+A^{F_5}\Bigr)^2,
\end{eqnarray*}
which are black $M3$-branes.

{ In the MQGP limit:
$G^{\cal M}_{00}\sim\epsilon^{\frac{55d}{3}}r^2\left(1 - \frac{r_h^4}{r^4}\right),
G^{\cal M}_{rr}\sim\frac{\epsilon^{-\frac{59d}{3}}}{r^2
\left(1 - \frac{r_h^4}{r^4}\right)}$. One can rewrite $G_{rr}dr^2
=\xi^\prime\frac{d\omega^2}{\omega}$ where $\xi^\prime\sim
\frac{r_h\epsilon^\prime}{\epsilon^{\frac{59d}{3}}}$ and
$\xi^\prime \frac{d\omega^2}{\omega}=dv^2$ or $\omega=\frac{v^2}{4\xi^\prime}$.
One sees that near $r=r_h\sim1, G^{\cal M}_{tt}dt^2\sim\epsilon^{38d}u^2dt^2$ implying
again $T^2\sim\frac{1}{\left(\sqrt{g_sN}\right)^2}$ in conformity with
\cite{metrics}. Similar result are obtained for the limits of \cite{metrics}.
 Now,  $G_{00}^{\cal M}, G_{rr}^{\cal M}$ have no angular dependence and hence
the temperature $T=\frac{\partial_r G_{00}}{4\pi \sqrt{G_{00}G_{rr}}}$  \cite{kouvtun_2003, kouvtun_2005} of the black $M3$-brane then turns out to be given by:
\begin{eqnarray*}
\label{T}
& &\hskip-0.4in T =\frac{\sqrt{2}}{{r_h} \sqrt{\pi} \sqrt{\frac{{g_s} \left(18 {g_s}^2 {N_f} ln ^2({r_h}) {M_{\rm eff}}^2+3 {g_s} (4 \pi -{g_s} {N_f}
   (-3+ln (2))) ln ({r_h}) {M_{\rm eff}}^2+8 N \pi ^2\right)}{{r_h}^4}}} \stackrel{{\rm Both\ limits}}{\longrightarrow}\frac{r_h}{\pi L^2}. \nonumber\\
   \end{eqnarray*}
To get a numerical estimate for $r_h$,  we see that equating $T$ to $\frac{r_h}{\pi L^2}$;  in
both limits one then obtains $r_h=1+\epsilon$, where $0<\epsilon<1$}.

The amount of near-horizon supersymmetry was determined in \cite{MQGP} by solving for the killing spinor $\epsilon$ by the vanishing superysmmetric
variation of the gravitino in $D=11$ supergravity. In weak $(g_s)$-coupling-large-t'Hooft-couplings limit \cite{MQGP}
\begin{eqnarray*}
& &\hskip -0.3in \varepsilon(\theta_{1,2},\Phi_{1,2}\equiv\phi_1\pm\phi_2,\psi,x_{10})=\nonumber\\
& & \hskip -0.3in e^{\mp\beta\Phi_1\epsilon^{-\alpha_\Phi}G^{5678}(\epsilon)\Gamma_{56}} e^{\mp\beta\Phi_2\epsilon^{-\alpha_\Phi}G^{5678}(\epsilon)
\Gamma_{56}}e^{-\beta\psi\epsilon^{-\alpha_\psi}(\mp G^{5679}(\epsilon)\Gamma_{56} \mp G^{5689}(\epsilon)\Gamma_{56})}\nonumber\\
& & \times e^{x_{10}\epsilon^{\alpha_{10}}(G^{567\overline{10}}(\epsilon)\Gamma_{56} \pm G^{568\overline{10}}(\epsilon)\Gamma_{56})}\varepsilon_0,
\end{eqnarray*}
with constraints:
 \begin{equation}
\left(G^{5678}(\epsilon)  \pm G^{567\overline{10}}(\epsilon)\Gamma_{\overline{10}} \pm G^{568\overline{10}}(\epsilon)\Gamma_{\overline{10}}\right)\varepsilon_0=0,
\end{equation}
and
\begin{equation}
\Gamma_7\varepsilon_0=\pm\varepsilon_0;\ \Gamma_8\varepsilon_0=\pm\varepsilon_0,
\end{equation}
for a constant spinor $\varepsilon_0$; $g_{{ \Phi_{1,2}\Phi_{1,2}}}=\epsilon^{-\alpha_\Phi}, g_{\psi\psi}=\epsilon^{-\alpha_\psi}, g_{x_{10}x_{10}}=\epsilon^{\alpha_{10}}$ near $\theta_{1,2}=0,\pi$. The near-horizon black $M3$-branes solution
 possess 1/8 supersymmetry near $\theta_{1,2}=0,\pi$. Similar arguments also work in the MQGP limit of \cite{MQGP}.

 { Freezing the angular dependence
on $\theta_{1,2}$ (there being no dependence on $\phi_{1,2},\psi,x_{10}$ in, both,
 weak $(g_s)$-coupling-large-t'Hooft-couplings and MQGP limits), noting that $G_{00,rr,\mathbb{ R}^3}^{IIA/{\cal M}}$ are independent of the angular coordinates (additionally possible to tune the chemical potential $\mu_C$ to a small value \cite{MQGP}, using
 the result of   \cite{kouvtun_2003}:
 \begin{equation}\frac{\eta}{s}=T\frac{\sqrt{|G^{IIA/{\cal M}}|}}{\sqrt{|G^{IIA/{\cal M}}_{tt}G^{IIA/{\cal M}}_{rr}|}}\Biggr|_{r=r_h}
 \int_{r_h}^\infty dr\frac{|G^{IIA/{\cal M}}_{00}G^{IIA}_{rr}|}{G^{IIA/{\cal M}}_{\mathbb{ R}^3}\sqrt{|G^{IIA/{\cal M}}|}}=\frac{1}{4\pi}.
 \end{equation}}

{ In the notations of \cite{kouvtun_2003}  one can pull out a common $Z(r)$ in the angular-part of the metrics as:
$Z(r)K_{mn}(y)dy^i dy^j$, (which for the type IIB/IIA backgrounds, is $\sqrt{h}r^2$) in terms of which:
\begin{eqnarray*}
& & \hskip -0.4in D = \frac{\sqrt{|G^{IIB/IIA}|}Z^{IIB/IIA}(r)}{G^{IIb/IIA}_{\mathbb{ R}^3}\sqrt{|G_{00}^{IIB/IIA}G_{rr}^{IIB/IIA}|}}\Biggr|_{r=r_h}\int_{r_h}^\infty dr
\frac{|G_{00}^{IIB/IIA}G_{rr}^{IIB/IIA}|}{\sqrt{|G^{IIB/IIA}|}Z^{IIB/IIA}(r)} = \frac{1}{2\pi T}.
\end{eqnarray*}}

{Let us review the thermodynamical Stability of Type IIB Background in Ouyang Limit as discussed in \cite{MQGP}.
Assuming $\mu(\neq0)\in\mathbb{R}$ in Ouyang's embedding \cite{real mu Ouyang et al}: $r^{\frac{3}{2}}e^{\frac{i}{2}(\psi-\phi_1-\phi_2)}\sin\frac{\theta_1}{2} \sin\frac{\theta_2}{2}=\mu$, which could be satisfied for $\psi=\phi_1+\phi_2$ and $r^{\frac{3}{2}}\sin\frac{\theta_1}{2} \sin\frac{\theta_2}{2}=\mu$. Including a $U(1)$(of $U(N_f)=U(1)\times SU(N_f)$) field strength $F=\partial_r A_t dr\wedge dt$  in addition to $B_2$, in the $D7$-brane DBI action, taking the embedding parameter $\mu$ to be less than but close to 1, the chemical potential $\mu_C$ was calculated up to ${\cal O}(\mu)$ in \cite{MQGP}:
\begin{eqnarray*}
\label{eq:mu_C_improved}
& &\mu_C\approx g_s\left[\frac{\, _2F_1\left(\frac{1}{3},\frac{1}{2};\frac{4}{3};-\frac{C^2 g_s^2}{{r_h}^6}\right)}{2 {r_h}^2}\right] +\nonumber\\
&&   g_s N_f\left[ \frac{C g_s \left(\, _2F_1\left(\frac{2}{3},\frac{3}{2};\frac{5}{3};-\frac{C^2 g_s^2}{{r_h}^6}\right) a^2+8
   {r_h}^2 \, _2F_1\left(\frac{1}{3},\frac{3}{2};\frac{4}{3};-\frac{C^2 g_s^2}{r_h^6}\right) ln (\mu
   )\right)}{32 \pi  r_h^4}\right].\nonumber\\
   & &
\end{eqnarray*}}

{ One thus sees:
\begin{eqnarray*}
\label{dmuoverdT-ii}
& &\hskip -0.2in \frac{\partial\mu_C}{\partial T}\Biggr|_{N_f,a = f r_h:f<<<1; C=\frac{\pi^3\left(4\pi g_s N\right)^{\frac{3}{2}}}{g_s}}\sim - \frac{1}{8\pi}\frac{T^6(f^2 + 4 ln\mu)}{\left(1 + T^6\right)^{\frac{3}{2}}} + f^2 g_s N_f\frac{\, _2F_1\left(\frac{2}{3},\frac{3}{2};\frac{5}{3}; - \frac{1}{T^6}\right)}{16\pi T^3}<0.
\end{eqnarray*}
So,  it is clear that  $\frac{\partial\mu_C}{\partial T}\Biggr|_{N_f} = - \frac{\partial S}{\partial N_f}\Biggr|_T < 0$.
Apart from $C_v>0$, thermodynamic stability requires: $\frac{\partial\mu_C}{\partial N_f}\left.\right|_{T}>0$, which
for $C>0, \mu = \lim_{\epsilon\rightarrow0^+}1 - \epsilon$ is satisfied!.}

Let us, towards the end of this section, review the arguments on demonstration of thermodynamical stability from $D=11$ point of view by demonstrating the positivity of the specific heat, as shown in \cite{MQGP}.  Keeping in mind that as $ r_h\sim l_s$,
 higher order $\alpha^{\prime}$ corrections become important, the action:
\begin{eqnarray*}
  \label{eq:GravitationalAction}
 & & \hskip -0.48in {\cal S}_{E} =  \frac{1}{16\pi}\int_{\cal{M}} d^{11}\!x \sqrt{G^{\cal M}} R^{\cal M}
   + \frac{1}{8\pi}\int_{\partial M} d^{10}\!x K^{\cal M} \sqrt{\hat{h}}-\frac{1}{4}\int_{M} \Bigl( |G_4|^2\nonumber\\
   & & \hskip -0.48in-{C_3\wedge G_4\wedge G_4}\bigr) +\frac{T_2}{{2\pi}^4. 3^2.2^{13}}\int_{\cal{M}} d^{11}\!x \sqrt{G^{\cal M}}(J-\frac{1}{2}E_8)+T_2 \int C_3 \wedge X_8 - {\cal S}^{\rm ct},
\end{eqnarray*}
where $T_2\equiv M2$-brane tension,  and:
\begin{eqnarray*}
\label{J}
& & J = 3.2^8 \Bigl(R^{mijn}R_{pijq} R_{m}^{\ rsp}R^{q}_{\ rsn}+\frac{1}{2}R^{mnij}R_{pqij} R_{m}^{\ rsp}R^{q}_{\ rsn}\Bigr),
\nonumber\\
& & E_8= \epsilon^{abc m_1 n_1...m_4 n_4} \epsilon_{abc m^{\prime}_1 n^{\prime}_1...m^{\prime}_4 n^{\prime}_4}R^{m^{\prime}_1 n^{\prime}_1}_{\ \ \ \ \ \ m_1 n_1}...R^{m^{\prime}_4 n^{\prime}_4}_{\ \ \ \ \ \ m_4 n_4},
\nonumber\\
& &  X_8 =\frac{1}{192 \cdot (2 \pi^2)^4}\Bigl[tr(R^4)- (tr{R^2})^2\Bigr],
 \end{eqnarray*}
for Euclideanised space-time where $\cal{M}$ is a volume of spacetime defined by $r <r_\Lambda$, where the counter-term ${\cal S}^{\rm ct}$ is added such that the Euclidean action ${\cal S}_{E}$ is finite   \cite{S_ct-Perry et al,Mann+Mcnees}. The reason why there is no $\frac{1}{\kappa_{11}^2}$ in the action is because of the tacit understanding that as one goes from the local $T^3$ coordinates $(x,y,z)$ to global coordinates $(\phi_1,\phi_2,\psi)$ via equation (\ref{xyz defs}), the $(g_sN)^{\frac{3}{4}}$ cancels off $\kappa_{11}^2$. This also helps in determining how the cut-off $r_\Lambda$ scales with $\epsilon$ that appears in the scalings of $g_s, M$ and $N$. Given that $\tilde{F}_3\wedge H_3$ is localised, i.e., receives  the dominant contributions near $\theta_{1,2}=0$,  the total D-brane charge  in the very large-$N(>>M,N_f)$ limit, can be estimated by \cite{ouyang}:
$\frac{1}{2 \kappa^{2}_{10} T_3}\int{\tilde F_3}\wedge H_3 \sim N$.
\label{F_3xH_3}
In MQGP limit \cite{MQGP} defined by considering $g_s, g_s M, {g_s}^2 M N_f\equiv{\rm finite}, g_s N>>1, \frac{g_s M^2}{N}<<1.$
\begin{eqnarray*}
&& \int{\tilde F_3}\wedge H_3\sim\int  \left\{\frac{3^4 (g_s M)^2 (g_s N_f) N_f {A_1} { A_4}(B_1+ B_4)}{8 \pi^2}\frac{(ln  r )^2 }{r}{\rm sin}~\theta_2~{\rm sin}~\theta_1
- \right. \nonumber\\
&& \left.{\hskip -0.4in} \frac{ 3^2 (g_s M)^2 (g_s N_f) N_f {A_3} { A_5}}{ 2^6 \pi^2} \frac{1}{r} \sin{\theta_1} \sin{\theta_2} (\cot^2{\frac{\theta_2}{2}}+ B_3 B_5 \cot^2{\frac{\theta_1}{2}})  \right\} d\theta_1 d\theta_2 dr d\phi_1d\phi_2 d\psi .
\end{eqnarray*}
Integrating the same,
\begin{eqnarray*}
&&  \int{\tilde F_3}\wedge H_3 \
  \sim 3^3. \ 2^{4}\pi (g_s M)^2 (g_s N_f) N_f(ln  {r_{\Lambda}})^3.
  \end{eqnarray*}
We have $
\frac{1}{ \kappa^{2}_{11}}\sim\frac{ 1}{ 3^3.\ 2^{5}\pi\frac{(g_s M)^2}{N} (g_s N_f)^2  }\frac{1} {(\ln {r_{\Lambda}})^3}.
$ Assuming $\frac{ 1}{ 3^3.\ 2^{5}\pi\frac{(g_s M)^2}{N} (g_s N_f)^2}\frac{1} {(ln  {r_{\Lambda}})^3} \sim (g_s N)^{-\frac{3}{4}}$, we get
$ {ln  r_{\Lambda}} \sim \left\{\frac{ N^{\frac{7}{4}}g^{-\frac{13}{4}}_s}{ 3^3.\ 2^{5}\pi M^2 N^{2}_{f}}\right\}^{\frac{1}{3}}.$

The action, apart from being divergent (as $r\rightarrow\infty$) also possesses pole-singularities near $\theta_{1,2}=0,\pi$. We regulate the second divergence by taking a
small $\theta_{1,2}$-cutoff $\epsilon_\theta$, $\theta_{1,2}\in[\epsilon_\theta,\pi-\epsilon_\theta]$, and demanding $\epsilon_\theta
\sim\epsilon^\gamma$, for an appropriate $\gamma$.  We then explicitly checked that the finite part
of the action turns out to be independent of this cut-off $\epsilon/\epsilon_\theta$. It was shown in  \cite{MQGP} that in \cite{metrics}'s and MQGP limit, $S^{finite}_{EH + GHY + |G_4|^2 + {\cal O}(R^4)}\sim-r_h^3$ and the counter terms are given by:\\
$\int\left(\epsilon^{-\kappa_{\rm EH-surface}^{(i)}}\sqrt{h^{\cal M}} R^{\cal M}, \epsilon^{-\kappa^{(i)}_{\rm cosmo}}\sqrt{h}, \epsilon^{-\kappa^{(i)}_{\rm flux}}\sqrt{h}|G_4|^2\right)\biggr|_{r=r_{\Lambda}}$ $\left(\frac{a_{EH}}{\epsilon^{\frac{5}{3}} } + a_{\rm GHY-boundary} - \epsilon^9 a_{G_4}\right.  $ + $\left. a_{R^4}\epsilon^{22} \right)$, \\and \\
$\int\left(\epsilon^{-\kappa_{\rm EH-surface}^{(ii)}}\sqrt{h^{\cal M}} R^{\cal M}, \epsilon^{-\kappa^{(ii)}_{\rm cosmo}}\sqrt{h}, \epsilon^{-\kappa^{(ii)}_{\rm flux}}\sqrt{h}|G_4|^2\right)\biggr|_{r=r_{\Lambda}}$ $\left(\frac{a_{EH}}{\epsilon^5} + a_{\rm GHY-boundary} - \epsilon^{19}a_{G_4}\right. $
$\left. + a_{R^4}\epsilon^{42}  \right)$, \\
respectively in the two limits - in both limits the counter terms could be given an ALD-gravity-counter-terms \cite{Mann+Mcnees} interpretation.
It was also shown in  \cite{MQGP} that the entropy (density $s$) is positive and one can approximate the same
as $s\sim  r_h^3$ - result which is what one also obtains (as shown in \cite{MQGP}) - by calculation of the horizon area.  Using the same, therefore, $C\sim\frac{T(r_h)}{\frac{\partial T}{\partial r_h}}\frac{\partial \left(  r_h^3\right)}{\partial r_h} >0$ - implying a stable uplift!

\subsection{Satisfying D=11 SUGRA EOMs Locally in the MQGP Limit}

In this subsection we explicitly verify that the uplift to M-theory in the delocalized limit of \cite{becker2004} in the MQGP limit of \cite{MQGP} (involving $g_s\stackrel{<}{\sim}1$) constitutes a bona-fide solution of $D=11$ supergravity EOMs in the presence of (wrapped) $M5$-brane sources. The same for a single $M5$-brane source are given as under \cite{Bremer-th}:
\begin{equation}
\label{FE}
R^{{\cal M}}_{MN}-\frac{1}{2}G^{{\cal M}}_{MN}R =\frac{1}{12}\left(G_{MPQR}G_N^{PQR}-\frac{1}{8} G^{{\cal M}}_{MN} G_{PQRS}G^{PQRS}\right)+ \kappa^2_{11} T^{M5}_{MN},
\end{equation}
and in the absence of Dirac 6-brane:
\begin{equation}
\label{G4EOM}
d*_{11}G_4 + G_4\wedge G_4 = -2 \kappa_{11}^2T_5 (H_3 - A_3)\wedge *_{11}J_6,
\end{equation}
where $M5$-brane current $J_6\sim\frac{dx^0\wedge dx^1\wedge dx^2\wedge dx^3\wedge d\theta_1\wedge d\phi_1}{\sqrt{-G^{\cal M}}}$; the Bianchi identity is:
\begin{equation}
\label{BIANCHI}
dG_4 = 2\kappa_{11}^2T_5 *_{11}J_6.
\end{equation}

Taking trace of (\ref{FE}),
\begin{equation}
-\frac{9}{2} R=-\frac{1}{32}G_{PQRS}G^{PQRS} +\kappa_{11}^2 T^{Q}_{Q}
\end{equation}
Incorporating the above in equation (\ref{FE}), one gets
\begin{equation}
R^{{\cal M}}_{MN}=\frac{1}{12}\left (G_{MPQR}G_N^{PQR}-\frac{1}{12} G^{{\cal M}}_{MN} G_{PQRS}G^{PQRS} \right)+\kappa_{11}^2 \left (T_{MN}-\frac{1}{9}G^{{\cal M}}_{MN} T^{Q}_{Q}\right)
\end{equation}
and  the space-time energy momentum tensor $T_{MN}$ for a single  M5-brane  wrapped around $S^2(\theta_1,\phi_1)$ is given by
\begin{equation}
\label{TMN}
T^{MN}(x)=\int_{{\cal M}_{6} }d^{6}\xi \sqrt{{\rm -det}~{G^{M5}_{\mu\nu}}} G^{(M5)\mu\nu}\partial_{\mu}X^{M} \partial_{\nu}X^N \frac{\delta^{11}(x-X(\xi))}{\sqrt{-{\rm det} \ G^{{\cal M}}_{MN}}}
\end{equation}
where $X=0,1,...,11$ and $\mu,\nu=0,1,2,3,\theta_1,\phi_1$.

Now, in the MQGP limit, the most dominant contribution is governed by $R_{\theta_1\theta_1}$. So:
\begin{equation}
R^{{\cal M}}_{\theta_1\theta_1} =\frac{1}{12}\left (G_{\theta_1PQR}G_{\theta_1}^{PQR}-\frac{1}{8} G^{{\cal M}}_{\theta_1\theta_1} G_{PQRS}G^{PQRS}\right)+\kappa_{11}^2 \left (T_{\theta_1\theta_1}-\frac{1}{9}G^{{\cal M}}_{\theta_1 \theta_1} T^{Q}_{Q}\right)
\end{equation}
Using equation (\ref{TMN}), one sees that
\begin{eqnarray}
&& T^{Q}_{Q}\sim  \frac{\sqrt{{\rm -det}~{G^{M5}_{\mu\nu}}}}{\sqrt{-{\rm det} \ G^{{\cal M}}_{MN}}}
 \end{eqnarray}
 Similarly, in both limits,
 \begin{eqnarray}
 T_{\theta_1\theta_1}= G^{{\cal M}}_{ \theta_1 M_1}  G^{{\cal M}}_{ \theta_1 M_2} T^{M_1 M_2}\sim  G^{\cal M}_{\theta_1\theta_1} \frac{\sqrt{{\rm -det}~{G^{M5}_{\mu\nu}}}}{\sqrt{-det G^{{\cal M}}_{MN}}}
 \end{eqnarray}
 So, we get
 \begin{equation}
 T_{\theta_1\theta_1}-\frac{1}{9}G^{{\cal M}}_{\theta_1 \theta_1} T^{Q}_{Q} \approx \frac{8}{9}  \times G^{{\cal M}}_{\theta_1 \theta_1}   \frac{\sqrt{{\rm -det}~{G^{M5}_{\mu\nu}}}}{\sqrt{-{\rm det} \ G^{{\cal M}}_{MN}}}
\end{equation}
Now,  we have
 \begin{eqnarray}
&& {\rm det} \ G^{{\cal M}}_{MN} \sim \frac{2 \pi  r^6 {f_2}({\theta_2})^2  {{g_s} N} \cos ^2({\theta_1}) \cot ^6({\theta_1}) \sin ^2({\theta_2}) \cos
   ^2({\theta_2})  }{1594323 3^{2/3} {g_s}^{16/3}},\nonumber\\
   && {\rm det}~{G^{M5}_{\mu\nu}}\sim \frac{9 \left(1-\frac{r^{4}_{h}}{r^4}\right)  r^8 \sin ^4({\theta_1}) (\cos (2 {\theta_2})-5)}{64 \pi  {g_s}^{15/3} N \left(\sin ^2({\theta_1})
   \cos ^2({\theta_2})+\cos ^2({\theta_1}) \sin ^2({\theta_2})\right)}\nonumber\\
   &&  G^{{\cal M}}_{\theta_1 \theta_1}\sim \frac{\sqrt{\pi } N}{\sqrt[3]{3} \sqrt[3]{\frac{1}{{g_s}}} \sqrt{{g_s} N}}
   \end{eqnarray}
   Utilizing above, we have
    \begin{eqnarray}
    \label{EM}
&& T_{\theta_1\theta_1}-\frac{1}{9}G^{{\cal M}}_{\theta_1 \theta_1} T^{Q}_{Q} \sim \frac{8}{9}   \times G^{{\cal M}}_{\theta_1 \theta_1}   \frac{\sqrt{{\rm -det}~{G^{M5}_{\mu\nu}}}}{\sqrt{-{\rm det} \ G^{{\cal M}}_{MN}}} \nonumber\\
 && \sim -\frac{   \left(\sqrt{1-\frac{r^{4}_{h}}{r^4}}\right)     r \tan ^4( {\theta_1}) \sqrt{\frac{5 - \cos (2{\theta_2})}{2 - \cos[2(\theta_1-\theta_2)] - \cos[2(\theta_1 + \theta_2)]}} \csc ( {\theta_2}) \sec( {\theta_2})\sin(\theta_1)}{\sqrt{g_s N}  {f_2}( {\theta_2})}\nonumber\\
\end{eqnarray}
 We will estimate (\ref{EM}) at an $r_\Lambda\sim e^{\left(\frac{N^{7/4}}{g_s^{13/4}M^2N_f^2}\right)^{\frac{1}{3}}}({\rm see}\ {\bf 2.2})>>$ resolution/deformation parameter and also larger than $r_h=1+\epsilon(\epsilon\rightarrow0^+)$, near $\theta_{1,2}=0$ (effected as
 $\theta_{1,2}\sim \alpha_{\theta}\epsilon^{\frac{15}{6}}),$
for $\epsilon\sim0.83, \alpha_{\theta}<<1$ yields $r_\Lambda\sim 5$ in the MQGP limit. In this limit, from (\ref{EM}), in the same spirit as \cite{Greene_et_al-wrapped_M2} one estimates:
\begin{equation}
\label{EM-final}
T_{\theta_1\theta_1}-\left.\frac{1}{9}G^{{\cal M}}_{\theta_1 \theta_1} T^{Q}_{Q}\right|_{\sim\alpha_{\theta}\epsilon^{\frac{3}{2}},\alpha_{\theta}<<1} \sim N_{wM_5}{\cal O}(10^{-2})\alpha_{\theta}^4 <<1,
\end{equation}
where $N_{wM_5}$ is the number of wrapped $M5$-branes approximately given by $N$ which in the MQGP limit and $\epsilon=0.83$ is around $10^3; \alpha_{\theta}<<10^{-\frac{3}{4}}$.
In \cite{MQGP}, it was shown that in the MQGP limit, the EH action: $\int\sqrt{-G^{\cal M}}R$ yields a divergent contribution: $\frac{r_\Lambda^4}{\alpha_\theta^5\epsilon^5}$, which can be canceled by a boundary counter term: $\epsilon^{-29/6}\left(\frac{a(\alpha_{1,2,3})}{\alpha_\theta^5 \epsilon^5}\right)\int \sqrt{-h}R$ evaluated at $r=r_\Lambda$, wherein $g_s \sim \alpha_1 \epsilon, M\sim\alpha_2\epsilon^{3/2},N\sim\alpha_3\epsilon^{-39}$. So, at $r=r_\Lambda,
\int \sqrt{-G^{\cal M}}R - \epsilon^{-29/6}\left(\frac{a(\alpha_{1,2,3})}{\alpha_\theta^5 \epsilon^5}\right)\int \sqrt{-h}R$ will not yield any finite contribution to $R_{MN}$. Further,
\begin{eqnarray}
& & F_{MPQR}G_N^{PQR}-\frac{1}{12} G^{{\cal M}}_{MN} G_{PQRS}G^{PQRS}\nonumber\\
 & & \sim \left. \frac{\sin ^9({\theta_1}) \cos ({\theta_1}) \cot ^3({\theta_2})}{\left(\sin ^2({\theta_1}) \cos ^2({\theta_2})+\cos
   ^2({\theta_1}) \sin ^2({\theta_2})\right)^2}\right|_{\theta_{1,2}\sim\alpha_{\theta}\epsilon^{\frac{3}{2}},\alpha_{\theta}<<1}\sim\alpha_\theta^2\epsilon^{\frac{15}{3}}<<1.
   \end{eqnarray}

   Further:
   \begin{eqnarray}
   \label{Bianchi-a}
& &    *_{11}J_6\sim\epsilon^{x^0x^1x^2x^3\theta_1\phi_1}_{\ \ \ \ \ \ \ \ \ \ \ \ \ \ r\theta_2\phi_2\psi x_{10}}dr\wedge d\theta_2\wedge d\phi_2\wedge d\psi\wedge dx_{10}\nonumber\\
& & \sim -\prod_{\mu\in\mathbb{R}^{1,3}}G^{\mu\mu}\prod_{m\in S^2(\theta_1,\phi_1)}G^{mm} dr\wedge d\theta_2\wedge d\phi_2\wedge d\psi\wedge dx_{10}
\nonumber\\
& & \sim \frac{g_s^5N\sin^6\theta_1}{r^8}dr\wedge d\theta_2\wedge d\phi_2\wedge d\psi\wedge dx_{10}\stackrel{r=r_\Lambda, MQGP,\epsilon=0.83,\theta_{1,2}\rightarrow0}{\longrightarrow}<<1,
\end{eqnarray}
From \cite{MQGP}, $G_4\wedge G_4=0$ and
\begin{eqnarray}
\label{A_3-H_3}
& & A_3=B_2\wedge dx_{10},\nonumber\\
 & & H_3 = H_{\theta\theta_2\phi_1}d\theta_1\wedge d\theta_2\wedge d\phi_1 + H_{\theta_1\theta_2\phi_2}d\theta_1\wedge d\theta_2\wedge d\phi_2 + H_{r\theta_1\theta_2}dr\wedge d\theta_1 d\theta_2\nonumber\\
  & & + H_{\theta_1\phi_1\psi}d\theta_1d\phi_1d\psi + H_{\theta_1\theta_2\phi_2}d\theta_1\wedge d\theta_2\wedge d\phi_2 + H_{\theta_1\theta_2\phi_1}d\theta_1\wedge d\theta_2\wedge d\phi_1 \nonumber\\
  & & + H_{\theta_1\phi_1\psi}d\theta_1\wedge d\phi_1\wedge d\psi.
  \end{eqnarray}
So from (\ref{A_3-H_3}) and (\ref{Bianchi-a}), $(H_3 - A_3)\wedge *_{11}J_6=0$.Using the same in (\ref{G4EOM}), $G_{MNPQ}$ satisfies $\partial_M\left(\sqrt{-G}G^{MNPQ}\right)=0$, which  in the MQGP limit is given by:
   \begin{eqnarray}
   \label{G4-aii}
   & & \partial_{\theta_1}\left(\sqrt{-G^{\cal M}} G^{\theta_1NPQ}\right)
   +  \partial_{\theta_2}\left(\sqrt{-G^{\cal M}} G^{\theta_2NPQ}\right)
   +  \partial_{\phi_1}\left(\sqrt{-G^{\cal M}} G^{\phi_1NPQ}\right) = 0.\nonumber\\
   & &
   \end{eqnarray}
   Using the non-zero expressions for the various four-form flux components in the MQGP limit requires to consider:

\begin{itemize}
\item
   $N=\theta_2,P=\phi_1, Q=x_{10}$ and one hence obtains:
   \begin{eqnarray}
   \label{G4-b}
   & & \partial_{\theta_1}\left(\sqrt{-G^{\cal M}}G^{\theta_1\theta_1}G^{\theta_2\theta_2}G^{\phi_1\phi_2}G^{x_{10}x_{10}}G_{\theta_1\theta_2
   \phi_1x_{10}}\right)\nonumber\\
   & &\sim \frac{1}{g_s^{\frac{29}{12}}N^{\frac{7}{4}}}\left[\frac{r {\theta_2} \tan ({\theta_1}) \sin ({\theta_2}) \tan ({\theta_2}) \left(\left(3 \tan ^2({\theta_1})+1\right) \sin ^2({\theta_2})+2 \tan^4({\theta_1})\right)}{\left(\sin ^2({\theta_1})+\sin ^2({\theta_2})\right)^2}\right]\nonumber\\
   & & \hskip 0.9in  \downarrow r=r_\Lambda, MQGP\ {\rm limit}, \theta_{1,2}\sim\alpha_{\theta}\epsilon^{\frac{15}{6}},\alpha_{\theta}<<1, \epsilon=0.83\nonumber\\
   & & \sim \alpha_{\theta}^210^{-6}<<1.
   \end{eqnarray}

 \item
 $N=\phi_1, P=\psi, Q=\phi_1$;
 \begin{eqnarray}
 \label{G4-c}
& & \partial_{\theta_1}\left(\sqrt{-G}G^{\theta_1\phi_2\psi\phi_1}\right) +
\partial_{\theta_2}\left(\sqrt{-G}G^{\theta_2\phi_2\psi\phi_1}\right)\nonumber\\
& & \hskip 0.9in  \downarrow r=r_\Lambda, MQGP\ {\rm limit}, \theta_{1,2}\sim\alpha_{\theta}\epsilon^{\frac{15}{6}},\alpha_{\theta}<<1, \epsilon=0.83\nonumber\\
& & \sim N_f\sin\phi_2\left(\frac{r_\Lambda\sin^5\theta_1}{g_s^{\frac{3}{4}}N^{\frac{3}{4}}}
+  \frac{r_\Lambda\sin^{11}\theta_1}{g_s^{\frac{3}{4}}N^{\frac{3}{4}}}\right)\sim\alpha_{\theta}^510^{-4}N_f\sin\phi_2<<1
 \end{eqnarray}

 \item
 $N=\theta_2, P=\phi_1, Q=\phi_2$:
 \begin{eqnarray}
 \label{G4-d}
& &  \partial_{\theta_1}\left(\sqrt{-G}G^{\theta_1\theta_2\phi_1\phi_2}\right)\sim N_f\sin\phi_2 \frac{r_\Lambda^2\sin^{10}\theta_2}
{g_s^{\frac{41}{12}}N^{\frac{3}{4}}}\sim\alpha_{\theta}^{10}10^{-3}N_f\sin\phi_2<<1. \end{eqnarray}

 \end{itemize}

Utilizing the expressions for the non-zero $G_{MNPQ}$ flux components and (\ref{Bianchi-a})(which tells us that $*_{11}J_6$'s only non-zero component is $\left(*_{11}J_6\right)_{r\theta_2\phi_2\psi x_{10}}$),
at $\theta_{1,2}\sim\alpha_{\theta}\epsilon^{\frac{15}{6}}, \alpha_{\theta}<<1$:
\begin{eqnarray}
\label{Bianchi-b}
\partial_{[r}G_{\theta_2\phi_2\psi x_{10}]}=0\approx \left(*_{11}J_6\right)_{r\theta_2\phi_2\psi x_{10}}.
\end{eqnarray}
Also:
\begin{eqnarray}
\label{Bianchi-c}
& & \partial_{[r]}G_{\theta_1\phi_2\psi\phi_1]}=0;   \partial_{[r}G_{\theta_2\phi_2\psi x_{10}]}=0;\nonumber\\
& & \left.\partial_{[r]}G_{\theta_1\theta_2\phi_1\phi_2]}\right|_{r=r_\Lambda,\theta_{1,2}\rightarrow0,\ {\rm MQGP\ limit}}\sim \frac{g_s^{\frac{7}{4}}M^2}{N^{\frac{1}{4}}r_\Lambda}\cos\phi_2 +
\frac{g_s^{\frac{7}{4}}MN^{\frac{1}{4}}}{r_\Lambda}\cos\phi_2\nonumber\\
& & \downarrow g_s\sim\alpha_{g_s}\epsilon,\ M\sim\alpha_M\epsilon^{-\frac{3}{2}},\ N\sim\alpha_N\epsilon^{-{39}},\ \epsilon=0.83: \alpha_{M,N}\sim\frac{1}{{\cal O}(1)} \sim10^{-2}<<1;\nonumber\\
& & \partial_{[\psi}G_{\theta_1\theta_2\phi_1\phi_2]}\nonumber\\
& & \downarrow \theta_1\sim\alpha_{\theta}\epsilon^{\frac{15}{6}}, \theta_2\sim{\cal O}(1)\alpha_{\theta}\epsilon^{\frac{15}{6}}: \alpha_{\theta}<<1; \rm MQGP\ limit: \epsilon=0.83\nonumber\\
& & \frac{N_f\left(g_sN\right)^{\frac{3}{4}}}{({\cal O}(1))^4}\sin\phi_2\sim 10^{-2}\sin\phi_2\ {\rm for}\ {\cal O}(1)\sim7.
\end{eqnarray}
We therefore conclude that the D=11 SUGRA EOMs are nearly satisfied near $r=r_\Lambda$ and the $\theta_1=\theta_2=0$ branch.

\subsection{Black $M3$-Branes as Wrapped $M5$-branes around Two-Cycle}
Let us turn our attention toward figuring out the charge of the black $M3$-brane. Utilizing the non-zero expressions of the components of the four-form flux $G_4$ as given in (\ref{G_4-Dasgupta_et_al-limit}), and:
(i)assuming the expressions obtained in the `delocalized' limit of \cite{becker2004} and hence in conformity with the non-locality of T duality transformations, are valid globally ($\langle\theta_{1,2}\rangle$ was shown in \cite{local_global_theta} to be replaced by $\theta_{1,2}$ for $D5$ branes wrapping two-cycles in conifolds), (ii) using the principal values of the integrals over $\theta_{1,2}$, i.e., assuming:
$\int_0^\pi {\cal F}(\theta_i)d\theta_i = \lim_{\epsilon_{\theta_i}\rightarrow0}\int_{\epsilon_{\theta_i}}^{\pi - \epsilon_{\theta_i}}{\cal F}(\theta_i)d\theta_i=\lim_{\epsilon_{\theta_i}\rightarrow0}\int_{\epsilon_{\theta_i}}^{\pi - \epsilon_{\theta_i}}{\cal F}(\pi - \theta_i)d\theta_i$, (iii) assuming the $f_i(\theta_i)$ introduced in (\ref{SYZ-large base}) to make the base of the local $T^3$-fibered resolved warped deformed conifold to be large, will globally be $\cot\theta_i$, and (iv) believing that the distinction between results with respect to $\phi_1$ and $\phi_2$  arising due to the asymmetric treatment of the same while constructing $A^{IIA}$ from triple T-duals of $F^{IIB}_{1,3,5}$, being artificial and hence ignoring the same, using (\ref{G_4-Dasgupta_et_al-limit}) one sees that in the limit of \cite{metrics}:
\begin{eqnarray*}
& & \left.\int_{C_4^{(1)}(\theta_1,\theta_2\phi_1,\phi_2)}G_4\right|_{\langle\psi\rangle,\langle r\rangle,\langle x_{10}\rangle}\sim \epsilon^{-14}N_f\sin\frac{\langle\psi\rangle}{2};
\ \left.\int_{C_4^{(2)}(\theta_1,\theta_2,\phi_{1/2},x_{10})}G_4\right|_{\langle\psi\rangle,\langle\phi_{2/1}\rangle,\langle r\rangle}
\sim\epsilon^{-9};\nonumber\\
& & \left.\int_{C_4^{(3)}(r,\theta_1,\theta_2,x_{10})}G_4\right|_{\langle\psi\rangle,\langle \phi_{1,2}\rangle}=0;\ \left.\int_{C_4^{(4)}(r,\theta_1,\phi_{1/2},x_{10})}G_4\right|_{\langle\theta_2\rangle,\langle\psi\rangle,\langle\phi_{2/1}\rangle}=0;
\nonumber\\
& & \left.\int_{C_4^{(5)}(r,\theta_2,\phi_{1/2},x_{10})}G_4\right|_{\langle\psi\rangle,\langle\theta_1\rangle,\langle \phi_{2/1}\rangle}=0;    \left.\int_{C_4^{(6)}(r,\theta_2,\psi,x_{10})}G_4\right|_{\langle\theta_1\rangle,\langle\phi_{1,2}\rangle}=0\nonumber\\
& &  {\rm Strictly}\
G_{r\theta_2\psi x_{10}}\sim{\cal O}\left(\frac{g_s^2MN_f}{N}\right){\cal F}_1(\theta_{1,2}) + {\cal O}(g_s^2MN_f){\cal F}_2(\theta_{1,2}).
\nonumber\\
& & {\rm We\ first\ take \ {\rm weak} \ (g_s) \ {\rm coupling-large\ t'Hooft\ coupling\ limit}}\ {\rm in\ G_{r\theta_2\psi x_{10}}\ itself\ to \  annul} \nonumber\\
&&{\rm this\ flux}. \ {\rm Also,}\ G_{r\theta_2\psi x_{10}}\sim\frac{N_f}{r}\ {\rm in\ the \ {\rm weak}\ (g_s) \ {\rm coupling-large\ t'Hooft\ coupling\ limit}}\nonumber \\
&& {\rm and\ hence\ negligible\ for\ large\ r};\nonumber\\
& & {\rm Similarly}\  \left.\int_{C_4^{(7)}(r,\theta_1,\psi,x_{10})}G_4\right|_{\langle\theta_2\rangle,\langle\phi_{1,2}\rangle}=0;
\nonumber\\
\end{eqnarray*}
\begin{eqnarray*}
& &  \left.\int_{C_4^{(8)}(\theta_1,\phi_{1/2},\psi,x_{10})}G_4\right|_{\langle\phi_{2/1}\rangle,\langle r\rangle,\langle\theta_2\rangle}=   \left.\int_{C_4^{(9)}(\theta_2,\phi_{1/2},\phi_{1/2},x_{10})}G_4\right|_{\langle\phi_{2/1}\rangle,\langle\psi\rangle,\langle\phi_{2/1}\rangle,\langle r\rangle}\sim\epsilon^{-\frac{59}{6}};\nonumber\\
& & \left.\int_{C_4^{(10)}(\theta_1,\theta_2,\psi,x_{10})}G_4\right|_{\langle\phi_{1,2}\rangle,\langle r\rangle}=0\nonumber\\
\end{eqnarray*}
\begin{eqnarray}
\label{charge-Dasgupta_et_al_limit}
& &  {\rm Strictly}\
G_{\theta_1\theta_2\psi x_{11}}\sim{\cal O}
\left[(g_s^2MN_f)({g_sN_f})\left(\frac{g_sM^2}{N}\right)\right]{\cal F}_3(\theta_{1,2}),
\nonumber\\
& & {\rm  Again \ we\ first\ take\ {\rm weak}\ (g_s)\ {\rm coupling-large\ t'Hooft\ coupling\ limit}} \ {\rm in\ G_{\theta_1\theta_2\psi x_{10}}\ itself}\nonumber\\
&& {\rm  to\ annul\ this\ flux};\nonumber\\
& & \left.\int_{C_4^{(11)}(r,\theta_2,\phi_1,\phi_2)}G_4\right|_{\langle\theta_1\rangle,\langle\psi\rangle,\langle x_{10}\rangle}=0;\ \left.\int_{C_4^{(12)}(r,\theta_2,\phi_1,\psi)}G_4\right|_{\langle\theta_1\rangle,\langle\phi_2\rangle,\langle x_{10}\rangle}=0;\nonumber\\
& & \left.\int_{C_4^{(13)}(r,\theta_1,\phi_{1/2},\psi)}G_4\right|_{\langle\theta_2\rangle,\langle\phi_{2/1}\rangle,\langle x_{10}\rangle}=0;\ \left.\int_{C_4^{(14)}(\theta_1,\theta_2,\phi_{1/2},\psi)}G_4\right|_{\langle\phi_{2/1}\rangle,\langle r\rangle,\langle x_{10}\rangle}=0\nonumber\\
& & \left.\int_{C_4^{(15)}(\theta_{1/2},\phi_1,\phi_2,\psi)}G_4\right|_{\langle\theta_{2/1}\rangle,\langle x_{10}\rangle,\langle r\rangle}=0,
\end{eqnarray}
where we calculate flux of $G_4$ through various four cycles $C_4^{I}$. We have dropped contribution of $G_{r\theta_1\theta_2\psi}, G_{r\theta_2\phi_1\phi_2}$ to $G_{r m n p }dr\wedge dx^m\wedge dx^n\wedge dx^p$ as it is ${\cal O}\left(\frac{1}{r}\right)$-suppressed as compared to the ones retained, which is hence, dropped, at large $r$. From (\ref{charge-Dasgupta_et_al_limit}), one sees that the most dominant contribution to all possible fluxes arises (near $\langle \psi \rangle$), in the large-$r$ limit, from
$G_{\theta_1\theta_2\phi_1x_{10}}$ and $G_{\theta_1\theta_2\phi_2x_{10}}$. Using \cite{MQGP}, the large-$r$ limit of the $D=11$ metric can be written as:
\begin{eqnarray}
\label{large_r_D=11_metric}
& & ds_{11}^2 = ds_{AdS_5}^2 + g_s M^2 (ln\ r)^2\Sigma_1(\theta_{1,2})d\theta_1^2 + \frac{g_s^{\frac{11}{6}}M^2}{\sqrt{N}} a^4 r^4 (ln\ r)^2\Sigma_2(\theta_{1,2}) d\theta_2^2 \nonumber\\
& &  + g_s^{\frac{11}{6}} r\ ln\ r\left[(g_s N)^{\frac{1}{4}}M
\Sigma_3(\theta_{1,2})d\theta_1  + g_s^2 a^2 r^2\Sigma_4(\theta_{1,2}) d\theta_2\right]dx_{10} + g_s^{\frac{4}{3}}dx_{10}^2  \nonumber\\
& &+ g_s^{\frac{4}{3}} M[ N_f r\ ln\ r \Sigma_5(\theta_{1,2})d\theta_1   + a^2 r^2\ ln\ r \tilde{\Sigma}_5(\theta_{1,2}) d\theta_2] d\phi_{1/2}  + ds_{{\cal M}_3(\phi_{1,2},\psi)}^2(\theta_{1,2}).
\end{eqnarray}
Hence, asymptotically, the $D=11$ space-time is a warped product of $AdS_5(\mathbb{ R}^{1,3}\times\mathbb{R}_{>0})$ and an ${\cal M}_6(\theta_{1,2},\phi_{1,2},\psi,x_{10})$ where ${\cal M}_6$ has the following fibration structure:
\begin{equation}
\hskip -0.4in
\label{M_6}
\begin{array}{cc}
&{\cal M}_6(\theta_{1,2},\phi_{1,2},\psi,x_{10})   \longleftarrow   S^1(x_{10}) \\
&\downarrow  \\
{\cal M}_3(\phi_1,\phi_2,\psi) \hskip -0.4in & \longrightarrow  {\cal M}_5(\theta_{1,2},\phi_{1,2},\psi)   \\
&\downarrow  \\
 & \hskip 0.9in {\cal B}(\theta_1,\theta_2)  \longleftarrow  [0,1]_{\theta_1}  \\
 & \downarrow  \\
& [0,1]_{\theta_2}
\end{array}.
\end{equation}
Analogous to the $F_3^{IIB}(\theta_{1,2})$ (with non-zero components being $F_{\psi\phi_1\theta_1}, F_{\psi\phi_2\theta_2}, F_{\phi_1\phi_2\theta_1}$ and $F_{\phi_1\phi_2\theta_2}$) in Klebanov-Strassler background corresponding to $D5$-branes wrapped around a two-cycle which homologously is given by $S^2(\theta_1,\phi_1) - S^2(\theta_2,\phi_2)$ \cite{Klebanov+Strassler}, the black $M3$-brane metric, asymptotically can be thought of as black $M5$-branes wrapping a two cycle homologously given by:
$n_1 S^2(\theta_1,x_{10}) + n_2 S^2(\theta_2,\phi_{1/2}) + m_1 S^2(\theta_1,\phi_{1/2}) + m_2 S^2(\theta_2,x_{10})$ for some large $n_{1,2},m_{1,2}\in\mathbb{Z}$.

A similar interpretation is expected to also hold in the MQGP limit of \cite{MQGP} - in the same however, the analogs of (\ref{G_4-Dasgupta_et_al-limit}) are very tedious and unmanageable to work out for arbitrary $\theta_{1,2}$. But, we have verified that in the MQGP limit,
$\left.\int_{C_4(\theta_{1,2},\phi_{1/2},x_{10})}G_4\right|_{\langle\phi_{2/1}\rangle,\langle\psi\rangle,\langle r\rangle}$ will be very large.
Warped products of $AdS_5$ and an $M_6$ corresponding to wrapped $M5$-branes have been considered in the past - see \cite{Gauntlett_et_al_warpedADS5}.

\subsection{K\"{a}hlerity from Torsion Classes and a Warped Deformed Conifold sLag in the `Delocalized' Limit of \cite{becker2004}, and Large-$r$ and MQGP Limits}

In this subsection, first, by calculating the five $SU(3)$ structure torsion($\tau$) classes $W_{1,2,3,4,5}$, we show that in the MQGP limit of \cite{MQGP} and assuming the deformation parameter to be larger than the resolution parameter, the non-K\"{a}hler resolved warped deformed conifold of \cite{metrics}, in the large-$r$ limit, reduces to a warped K\"{a}hler deformed conifold for which $\tau\in W_5$. Then,  in the large-$r$ limit, we show that the local $T^3$ of \cite{MQGP} satisfies the same constraints as the one satisfied by a maximal $T^2$-invariant S(special) L(agrangian) sub manifold of a $T^*S^3$ so that the application of mirror symmetry as three T-dualities a la SYZ, in the MQGP limit of \cite{MQGP}, could be implemented on the type IIB background of \cite{metrics}.

The $SU(3)$ structure torsion classes \cite{torsion},\cite{uplift-IWASAWA} can be defined in terms of J, $ \Omega $, dJ, $ d{\Omega}$ and
the contraction operator  $\lrcorner : {\Lambda}^k T^{\star} \otimes {\Lambda}^n
T^{\star} \rightarrow {\Lambda}^{n-k} T^{\star}$,  $J$ being given by:
$$ J  =  e^1 \wedge e^2 + e^3 \wedge e^4 + e^5 \wedge e^6, $$
and
the (3,0)-form $ \Omega $ being given by
$$ \Omega  =  ( e^1 + ie^2) \wedge (e^3 +
ie^4) \wedge (e^5 + ie^6). $$
The torsion classes are defined in the following way:
\begin{itemize}
\item
$W_1 \leftrightarrow [dJ]^{(3,0)}$, given by real numbers
$W_1=W_1^+ + W_1^-$
with $ d {\Omega}_+ \wedge J = {\Omega}_+ \wedge dJ = W_1^+ J\wedge J\wedge J$
and $ d {\Omega}_- \wedge J = {\Omega}_- \wedge dJ = W_1^- J \wedge J \wedge J$;

\item
$W_2 \leftrightarrow [d \Omega]_0^{(2,2)}$ :
$(d{\Omega}_+)^{(2,2)}=W_1^+ J \wedge J + W_2^+ \wedge J$
and $(d{\Omega}_-)^{(2,2)}=W_1^- J \wedge J + W_2^- \wedge J$;

\item
 $W_3 \leftrightarrow [dJ]_0^{(2,1)}$ is defined
as $W_3=dJ^{(2,1)} -[J \wedge W_4]^{(2,1)}$;

\item
 $W_4 \leftrightarrow J \wedge dJ$ : $W_4 =\frac{1}{2} J\lrcorner dJ$;

 \item
 $W_5 \leftrightarrow
[d \Omega]_0^{(3,1)}$: $W_5 = \frac{1}{2} {\Omega}_+\lrcorner d{\Omega}_+$
(the subscript 0 indicative of the primitivity of the respective forms).
\end{itemize}

Depending on the classes of torsion one can obtain different types of manifolds, some of which are:
\begin{enumerate}
\item
(complex) special-hermitian  manifolds with $ W_1=W_2=W_4=W_5=0$ which
means that $ \tau \in W_3 $;

\item
(complex) K\"{a}hler  manifolds with $ W_1=W_2=W_3=W_4=0$ which means
$\tau\in W_5 $;

\item
(complex) balanced Manifolds with $W_1=W_2=W_4=0$ which
means $\tau\in W_3\oplus W_5$;

\item
(complex)
Calabi-Yau manifolds with $ W_1=W_2=W_3=W_4=W_5=0$ which means $\tau =0$.

\end{enumerate}
The  resolved warped deformed conifold can be written in the form of the Papadopoulos-Tseytlin
 ansatz \cite{PT} in the string frame:
\begin{eqnarray}
\label{metricD10}
ds^2 =  h^{-1/2}  ds^2_{\mathbb{R}^{1,3}}
 + e^x ds_\mathcal{M}^2  = h^{-1/2}  dx^2_{1,3} + \sum_{i=1}^6 G_i^2\ ,
\end{eqnarray}
where \cite{Butti-et-al},\cite{Bena+Klebanov}:
\begin{eqnarray}
\label{Gforms}
G_1 &\equiv& e^{(x(\tau)+g(\tau)/2}\,e_1,
G_2 \equiv {\cal A}\,e^{(x(\tau)+g(\tau))/2}\,e_2 + {\cal B}(\tau)\,e^{(x(\tau)-g(\tau))/2}\,(\epsilon_2-a e_2)\ ,\nonumber\\
G_3 &\equiv& e^{(x(\tau)-g(\tau))/2}\,(\epsilon_1-a e_1)\ , \nonumber\\
G_4 &\equiv& {\cal B}(\tau)\,e^{(x(\tau)+g(\tau))/2}\,e_2 - {\cal A}\,e^{(x(\tau)-g(\tau))/2}\,(\epsilon_2-a e_2)\ , \nonumber\\
G_5 &\equiv& e^{x(\tau)/2}\,v^{-1/2}(\tau)d\tau\ , \nonumber\\
G_6 &\equiv& e^{x(\tau)/2}\,v^{-1/2}(\tau)(d\psi + \cos\theta_2d\phi_2 + \cos\theta_1d\phi_1),
\end{eqnarray}
wherein ${\cal A}\equiv{\cosh\tau+a(\tau)\over \sinh\tau},  {\cal B}(\tau)\equiv {e^{g(\tau)} \over \sinh\tau}$. The $e_i$s are one-forms on ${S}^2$
\begin{eqnarray}
\label{e_is}
e_1\equiv d\theta_1\ ,\qquad e_2\equiv  - \sin\theta_1 d\phi_1\ ,
\end{eqnarray}
and the $\epsilon_i$s a set of one-forms on ${S}^3$
\begin{eqnarray}
\label{eps_is}
\hskip -0.4in \epsilon_1 &\equiv&\sin\psi\sin\theta_2 d\phi_2+\cos\psi d\theta_2\ ,
\epsilon_2 \equiv   \cos\psi\sin\theta_2 d\phi_2 - \sin\psi d\theta_2\ ,
\epsilon_3 \equiv d\psi + \cos\theta_2 d\phi_2\ .
\end{eqnarray}

These one-forms  are quite convenient to work with since they allow us to write down very simple expressions for the holomorphic $(3,0)$ form
\begin{equation}
\label{Omega}
\Omega = (G_1 + i G_2)\wedge (G_3 + i G_4)\wedge (G_5 + i G_6)\ ,
\end{equation}
and the fundamental $(1,1)$ form
\begin{eqnarray}
\label{J}
J ={i\over 2} &\Big [&(G_1 + i G_2) \wedge (G_1- i G_2) + (G_3 + i G_4) \wedge (G_3- i G_4)
+ (G_5 + i G_6) \wedge (G_5- i G_6) \Big ]\ .\nonumber\\
& &
\end{eqnarray}
Substituting (\ref{Gforms}) into (\ref{metricD10}), one obtains:
\begin{eqnarray}
\label{metricD10-expanded}
& & ds^2_6 = \frac{1}{\sqrt{h}}ds^2_{\mathbb{R}^{1,3}} + \frac{e^{x(\tau)}}{v(\tau)}\left(d\tau^2 + \left[d\psi + \cos\theta_1 d\phi_1 + \cos\theta_2 d\phi_2\right]^2 \right) \nonumber\\
& & + d\theta_1^2\left(e^{g(\tau)+x(\tau)} + {a(\tau)}^2e^{-g(\tau)+x(\tau)}\right)+\nonumber\\
& &   d\phi_1^2\left(\frac{1}{2} \left({a}^2(\tau)+e^{2 g(\tau)}\right) e^{x(\tau)-g(\tau)} {\rm csch}^2(\tau) \sin ^2({\theta_1}) \left(2 {a(\tau)}^2(\tau)+4 {a} \cosh (\tau)+2 e^{2 g(\tau)}\right. \right.
\nonumber\\
& & \left. \left. +\cosh (2 \tau) +1\right)\right) + d\theta_2^2e^{x(\tau)-g(\tau)} \left(\left(a^2(\tau)+e^{2 g(\tau)}\right) \sin ^2(\psi ) {\rm csch}^2(\tau)\right.\nonumber\\
   & &\left.+2 a (\tau)\sin ^2(\psi ) \coth (\tau) \text{csch}(\tau)+\sin ^2(\psi ) \coth ^2(\tau)+\cos ^2(\psi
   )\right) + d\phi_2^2e^{x(\tau)-g(\tau)} \sin ^2(\text{theta2}) \nonumber\\
   & &\left(\left(a^2(\tau)+e^{2 g(\tau)}\right) \cos ^2(\psi ){\rm csch}^2(\tau)\right. \left.+2 a \cos ^2(\psi ) \coth (t) {\rm csch}(\tau)+\cos ^2(\psi ) \coth
   ^2(t)+\sin ^2(\psi )\right) \nonumber\\
   & & - 2 a(\tau) e^{-x(\tau) + g(\tau)}\cos\psi d\theta_1d\theta_2 + d\phi_1d\phi_2a e^{x(\tau)-g(\tau)} \cos (\psi ) {\rm csch}^2(\tau) \sin ({\theta_1}) \sin (\theta_2)\nonumber\\
   & & \left(2 a^2+4 a \cosh (t)+2 e^{2 g(\tau)}+\cosh (2 \tau)+1\right) -a(\tau) e^{x(\tau)-g(\tau)} \sin (\psi ) {\rm csch}^2(\tau) \sin ({\theta_1})\nonumber\\
   & & \left(2 a^2(\tau)+4 a (\tau)\cosh (\tau)+2 e^{2 g(\tau)}+\cosh (2 \tau)+1\right)d\theta_2d\phi_1 - 2 a(\tau) e^{-x(\tau) + g(\tau)}\sin\psi\sin\theta_2 d\theta_1d\phi_2.\nonumber\\
\end{eqnarray}
As $r\sim e^{\frac{\tau}{3}}$, the large-$r$ limit is equivalent to the large-$\tau$ limit, in which (\ref{metricD10-expanded}) approaches:
\begin{eqnarray}
\label{large_r_D=6}
& & ds_6^2 =  \frac{e^{x(\tau)}}{v(\tau)}\left(d\tau^2 + \left[d\psi + \cos\theta_1 d\phi_1 + \cos\theta_2 d\phi_2\right]^2 \right)\nonumber\\
& & + \left(e^{x(\tau)+g(\tau)} + a^2(\tau)e^{x(\tau) - g(\tau)}\right)d\theta_1^2 + e^{x(\tau) - g(\tau)}\sin^2\theta_1d\phi_1^2 + e^{x(\tau) - g(\tau)}d\theta_2^2
\nonumber\\
& & + e^{x(\tau) - g(\tau)}\sin^2\theta_2^2d\phi_2^2- 2 a e^{x(\tau) - g(\tau)}\left[\cos\psi\left(d\theta_1d\theta_2 - \sin\theta_1\sin\theta_2d\phi_1d\phi_2\right) + \right.\nonumber\\
& & \left. + \sin\psi\left(d\theta_1d\phi_2\sin\theta_2 + d\theta_2d\phi_1\sin\theta_1\right)\right].
\end{eqnarray}
In the MQGP limit of \cite{MQGP}, the metric (\ref{large_r_D=6}) matches the metric (\ref{metric}) and (\ref{D=5}) with the identifications:
\begin{eqnarray}
\label{identifics}
& & \frac{e^x(\tau)}{v(\tau)}\sim\frac{\sqrt{4\pi g_sN}}{9}\left(1 + {\cal O}(r_h^2e^{-\frac{2\tau}{3}})\right);\ v(\tau)\sim\frac{3}{2}\left[1 + {\cal O}\left(\left\{\frac{g_sM^2}{N}a_{\rm res}^2,r_h^2\right\}e^{-\frac{2\tau}{3}}\right)\right];\nonumber\\
& &  e^{x(\tau)}\sim\frac{\sqrt{4\pi g_sN}}{6}\left[1 + {\cal O}\left(\frac{g_sM^2}{N}a_{\rm res}^2e^{-\frac{2\tau}{3}}\right)\right];\nonumber\\
& & g(\tau)\sim0;\ a(\tau)\sim-2e^{-\tau}.
\end{eqnarray}
The reason for the choice of $a(\tau)$ is the following. The  $\cos\psi,\sin\psi$ terms in \cite{Minasian_Tsimpis_def_conif} are given by: $\gamma(\tau)\frac{\mu^2}{\rho^2}$, where $\rho^2\sim e^{\tau}$ and for large $\rho, \gamma(\tau)\sim\frac{\left(\sinh\tau\cosh\tau - \tau\right)^{\frac{1}{3}}}{\tanh\tau}\rightarrow\rho^{\frac{4}{3}}$.  Redefining $r\sim\rho^{\frac{2}{3}}$, these terms will appear with a coefficient $r^2\left(\frac{\mu^2}{r^3}\right)$. Comparing with (\ref{metric}),(\ref{D=5}), we see: $h_5\sim\frac{1}{r^3}\sim e^{-\tau}$.
Note, (\ref{identifics}) satisfies the identity: $e^{2g(\tau)} + a^2 (\tau)= 1$ \cite{iden}.
Now, $v(\tau)$ of \cite{Bena+Klebanov} is related to $p(\tau), x(\tau)$ of \cite{Butti-et-al} via: $e^{p(\tau)}=v(\tau)^{\frac{1}{6}}e^{-\frac{x(\tau)}{3}}$. Using the same, the five $SU(3)$ structure torsion classes were worked out in \cite{Butti-et-al} and are given in (\ref{W_1}).

For the six-fold to be complex, $W_1 = W_2 =0$. This was shown in \cite{Butti-et-al}, to be equivalent to the condition:
\begin{equation}
\label{complex}
\frac{e^{2g(\tau)} + 1 + a^2(\tau)}{2a(\tau)} = - \cosh\tau.
\end{equation}
Using (\ref{identifics}), we see that in the large-$r$ limit subject to discussion below (\ref{F_3xH_3}), (\ref{complex}) is approximately satisfied.
Using that in the large-$\tau$ limit, ${\cal A}\sim1$ and ${\cal B}(\tau)\sim e^{-\tau}$(without worrying about numerical pre factors in the various terms in the final expressions) the five torsion classes are evaluated in (\ref{W_1}).
Hence, looking at the most dominant terms in the  MQGP limit of \cite{MQGP}, from (\ref{W_1}), we see that :
\begin{eqnarray}
\label{W12345}
& & \hskip -1.2in W_1 \sim \frac{e^{-3\tau}}{\sqrt{4\pi g_s N}}<<1\ ({\rm in\ the\ UV});\nonumber\\
& &\hskip -1.2in W_2 \sim \left(4\pi g_sN\right)^{\frac{1}{4}}e^{-3\tau}\left(d\tau\wedge e_\psi + e_1\wedge e_2 + \epsilon_1\wedge e_2\right)<<1\ ({\rm in\ the\ UV});\nonumber\\
& &\hskip -1.2in W_3 \sim \left.\sqrt{4\pi g_sN}\left(32\sqrt{\frac{2}{3}}e^{-3\tau}\left(e_1\wedge \epsilon_1 + e_2\wedge \epsilon_2\right)\wedge e_\psi + 2\sqrt{\frac{2}{3}}e_2\wedge \epsilon_1\wedge d\tau e^{-\tau} + 32e_1\wedge \epsilon_2\wedge d\tau e^{-3\tau} \right)\right|_{\theta_1\sim0;\ {\rm UV}}<<1;\nonumber\\
& & \hskip -1.2in W_4  \sim - \frac{2}{3}e^{-g(\tau)}d\tau = 2 W^3_4 = 2 W^{\bar{3}}_4;\nonumber\\
& & \hskip -1.2in W_5^{(\bar 3)} \sim -\frac{1}{2}\left(d\tau - i e_\psi\right).
\end{eqnarray}
implying that in the UV and near $\theta_i=0$, $\tau\in W_4\oplus W_5$ such that $\frac{2}{3}\Re e W^{\bar{3}}_5 = W^{\bar{3}}_4 = - \frac{1}{3}d\tau $ implying supersymmetry is preserved locally \cite{Butti-et-al}.  Obviously, in the strict $\tau\rightarrow\infty$ limit, one obtains a Calabi-Yau three-fold in which $W_{1,2,3,4,5}=0$.

Switching gears, we will now show that in the large-$r$ limit, the local $T^3$ of \cite{MQGP}, satisfies the constraints of a special Lagrangian three-cycle of the deformed conifold. Using \cite{T2slag}, the following gives the embedding equation of a $T^2(\phi_1,\phi_2)$-invariant sLag $C_3(\phi_1,\phi_2,\psi)$ in the deformed conifold $T^*S^3$:
\begin{eqnarray}
\label{T2sLag-1}
& & K'(r^2) \Im m(z_1{\bar z}_2) = c_1,  K'(r^2) \Im m(z_3{\bar z}_4) = c_2,\Im m (z_1^2 + z_2^2) = c_3,
\end{eqnarray}
which using the same complex structure as that for the singular conifold and $K'(r^2)\stackrel{r>>1}{\longrightarrow}r^{-\frac{2}{3}}$, (\ref{T2sLag-1}) yields:
\begin{eqnarray}
\label{T2sLag-2}
& &  r^{\frac{7}{3}}\left(cos^2\frac{\theta_1}{2} cos^2\frac{\theta_2}{2} + sin^2\frac{\theta_1}{2} sin^2\frac{\theta_2}{2}\right) cos(\phi_1+\phi_2) = c_1,\nonumber\\
& & r^{\frac{7}{3}}\left(cos^2\frac{\theta_1}{2} sin^2\frac{\theta_2}{2} + sin^2\frac{\theta_1}{2} cos^2\frac{\theta_2}{2}\right) cos(\phi_1+\phi_2) = c_2, r^3 sin\theta_1 sin\theta_2 cos\psi = c_3.
\end{eqnarray}
The equations (\ref{T2sLag-2}) can be solved to yield (\ref{theta1-embedding_i})-(\ref{theta1-embedding_ii}).
   From (\ref{theta1-embedding_i}) and (\ref{theta1-embedding_ii}), one sees that $dr=r_{\phi_1}d\phi_1 + r_{\phi_2}d\phi_2$, $d\theta_{1,2}=\theta^{(1),(2)}_{\phi_1}d\phi_1 + \theta^{(1),(2)}_{\phi_2}d\phi_2 + \theta^{(1),(2)}_{\psi} d\psi$.
   One can show from (\ref{theta1-embedding_i}) and (\ref{theta1-embedding_ii})) that for $(\theta_1,\theta_2)=(0/\pi,0/\pi)$, $(\psi,\phi_1,\phi_2)\approx(0/2\pi/4\pi,\frac{\pi}{5},\frac{\pi}{4})$. Now, the K\"{a}hler form $J$ and nowhere vanishing holomorphic three-form $\Omega$ are respectively given by (\ref{J}) - (\ref{Omega}) \cite{Gwyn+Knauf}:
    In $\mu\equiv$ deformation parameter $<<1, r>>1$-limit near $\theta_1=\theta_2=0$ implying near $\psi=0$ (and $\phi_1=\frac{\pi}{5},\phi_2=\frac{\pi}{4}$), one sees that:
\begin{eqnarray}
& & \hskip -1.3in J\sim r^{\frac{1}{3}} dr\wedge (d\phi_1 + d\phi_2)=r^{\frac{1}{3}}(r_{\phi_1} - r_{\phi_2})d\phi_1\wedge d\phi_2,\nonumber\\
& & \hskip -1.3in \Omega\sim 2\imath r dr\wedge d\theta_1\wedge d\theta_2+  r^2 d\theta_1\wedge d\theta_2\wedge
  (d\psi+d\phi_1+d\phi_2).
\end{eqnarray}
One can show that near $\psi=0/2\pi/3\pi,\phi_1=\frac{\pi}{5},\phi_2=\frac{\pi}{4}$ corresponding to $\theta_{1,2}\approx 0,\pi$:
\begin{eqnarray}
\label{ders}
& & r_{\phi_{1,2}}\sim {\cal O}(10)r^{<}_\Lambda\ {\rm implying}\ r_{\phi_1}\sim r_{\phi_2},\nonumber\\
& & \theta^{(1),(2)}_{\phi_{1,2}}\sim - {\cal O}(1),\nonumber\\
& & \theta^{(1),(2)}_\psi=0,
\end{eqnarray}
which defining the embedding $i:C_3(\phi_1,\phi_2,\psi)\hookrightarrow T^*S^3$,  implies:
\begin{eqnarray}
\label{J+imOmega_pullback}
& & i^* J = 0,\nonumber\\
& & \Im m\left( i^*\Omega\right) = 0,
\end{eqnarray}
and
\begin{equation}
\label{reOmega_pullback}
\Re e \left(i^*\Omega\right) \sim r^2{\cal O}(1) d\phi_1\wedge d\phi_2\wedge d\psi.
\end{equation}
Now $x\sim\left(g_sN\right)^{\frac{1}{4}} sin\langle\theta\rangle\phi_1, y\sim\left(g_sN\right)^{\frac{1}{4}} sin\langle\theta\rangle\phi_2,z\sim\left(g_sN\right)^{\frac{1}{4}} \psi$. In the MQGP limit of \cite{MQGP}, we take $\theta_{1,2}\sim\epsilon^{\frac{15}{6}}, g_s\sim\epsilon, N\sim\epsilon^{-39}$. So, as $r\rightarrow r_\Lambda$, one obtains:
\begin{equation}
\label{repullbackOmega}
\Re e\left(i^*\Omega\right)\sim \frac{r_\Lambda^2}{\epsilon^{-\frac{141}{6}}}dx\wedge dy\wedge dz,
\end{equation}
implying that if $r_\Lambda\rightarrow\infty$ as $\epsilon^{-\frac{141}{6}}$, then $$\Re e\left(i^*\Omega\right)\sim{\rm volume \ form}\left(T^3(x,y,z)\right).$$
A similar argument can be given in the limits of \cite{metrics}.

We hence see that in the large-$r$ limit and the limits of (\ref{limits_Dasguptaetal+MQGP}), we were justified in \cite{MQGP} to have used mirror symmetry a la SYZ prescription.

\section{Two-Point Functions of D7-Brane Gauge Field, R-charge and Stress Energy Tensor, and Transport Coefficients}

In this section, our basic aim is to calculate two-point correlation functions and hence transport coefficients  due to gauge field present in type IIB background and $U(1)_R$ gauge field  and stress energy tensor modes induced in the M-theory background  by  using the gauge-gravity prescription originally formulated in \cite{son+starinets}.
Let us  briefly  discuss the strategy of calculations related to evaluation of retarded two-point correlation functions to be used in further subsections  and the basic formulae of different transport coefficients  in  terms of retarded two-point Green's function.\begin{enumerate}
\item[]
  Step 1: Given that 11-dimensional background  geometrically can be represented as $AdS_{5}\times {\cal M}^{6}$ asymptotically, we  evaluate the kinetic term  corresponding to a particular field in the gravitational action by integrating out the angular direction i.e.
 \begin{equation}
 S=  \int d^6y{\cal P}(\theta_i,\phi_i,\psi, x_{11}) \int d^{4}x du ~ Q(u) (\partial_{u}\phi)^2+....
 \end{equation}
 \item[]
 Step 2: Solve the EOM corresponding to a particular field in that background by considering fluctuations  around the same. Further, the solution can be expressed in terms of boundary fields as:
\begin{equation}
 \phi(q,u)=f_{q}(u) \phi_{0}(q),
 \end{equation}
where $ \phi({\bf x},u)=\int \frac{d^{4} q}{(2\pi)^4}e^{-iwt+i{{\bf {q.x}}}}\phi(q,u)$ in the momentum space and $u=\frac{r_h}{r}$.
The solution can be evaluated by using boundary conditions $\phi(u,q)=\phi(0) $ at $u=0$ and incoming wave boundary condition according to which $\phi(u,q)\sim e^{-iwt}$ at $u=1$.
 \item[]
Step 3: Evaluate two-point Green's functions by using
\begin{equation}
G^{R}(k)= 2 \left. F(q,u)\right|^{u=1({\rm horizon)}}_{u=0(boundary)}, ~{\rm where}~F(u,q)= {\cal P}(\theta_i,\phi_i,\psi, x_{11}) Q(u)f_{-q}(u) \partial_{u}f_{q}(u).
\end{equation}

\end{enumerate}
In general, the transport coefficients in hydrodynamics are defined as response to the system after applying small perturbations. The  small perturbations in the presence of external source can be evaluated in terms of retarded two-point correlation functions. For example, the shear viscosity from dual background is calculated from the stress energy tensor involving spatial components at zero momentum i.e.
 \begin{equation}
\eta=\lim_{w\rightarrow 0}\frac{1}{2w}\int dt d{{\bf x}} e^{iwt} \theta(t) \langle\left[T_{xy/yz}(0), T_{xy/yz}(0)\right]\rangle.\end{equation}
By applying a small perturbation to the metric of curved space, and assuming homogeneity of space,
one gets the Kubo's relation to calculate shear viscosity \cite{kouvtun_2005}:
 \begin{equation}
\eta=-\lim_{w\rightarrow 0} \frac{i}{w} G^{R}(w,q),
\end{equation}
where the retarded two-point Green's function for stress-energy tensor components is defined as:
\begin{equation}
G^{R}_{\mu\rho,\nu\sigma}(w,{{\bf q}})=-i \int d^{4}x e^{-iq.x} \theta(t) \langle\left[T_{\mu\rho}(x), T_{\nu\sigma}(0)\right]\rangle.
\end{equation}
One can compute the above correlation function for different stress energy tensor modes by using the steps defined above.

Similarly, when one perturbs the system by applying an external field, the response of the system to the external source coupled to current is governed by \cite{PSS}:
\begin{equation}
G^{R}_{\mu\nu}(w,{{\bf q}})=-i \int d^{4}x e^{-i{\bf q.x}} \theta(t) \langle\left[j^{R}_{\mu}(x), j^{R}_{\nu}(0)\right]\rangle.
\end{equation}
In the low frequency and long distance limit, the $j^0$ evolves according to overdamped diffusion equation given as
\begin{equation}
\partial_{0}j^0= D \nabla^{2} j^0,
\end{equation}
with dispersion relation $w=-iD q^2$ generally known as Fick's law.  The diffusion coefficient $`D'$  can be evaluated from the pole located at $w=-iD q^2$  in the complex $w$-plane in the retarded two-point correlation function of $j^0$.

In (d+1) dimensional $SU(N)$ gauge theory, in thermal equilibrium, the differential photon emission rate per unit volume and time
at leading order in the electromagnetic coupling constant $e$ is given as \cite{Mateos}:
\begin{equation}
\frac{d\Gamma}{d^{d}q}=\left. \frac{e^2}{(2\pi)^d 2|{{\bf q}}|}n_{B}(w)\sum^{d-1}_{s=1} \epsilon^{\mu}_{(s)}({{\bf q}})\epsilon^{\nu}_{(s)}({{\bf q}})\chi_{\mu\nu}(q)\right|_{w=q}.
\end{equation}
where $q=(w, {{\bf q}})$ is the photon momentum and $\chi_{\mu\nu}$ is the spectral density given as:
\begin{equation}
\chi_{\mu\nu}(q)=-2\Im m G^{R}_{\mu\nu}(q)
\end{equation}
and
\begin{equation}
G^{R}_{\mu\nu}(w,{{\bf q}})=-i \int d^{4}x e^{-iq.x} \theta(t) \langle\left[j^{EM}_{\mu}(x), j^{EM}_{\nu}(0)\right]\rangle
\end{equation}
is the retarded correlation function of two electromagnetic currents.
The trace of spectral function will be given as:
 \begin{equation}
 \label{eq:trace}
\chi^{\mu}_{\mu}(q)=\eta^{\mu\nu}\chi_{\mu \nu}(q)=\sum^{d-1}_{s=1} \epsilon^{\mu}_{(s)}({{\bf q}})\epsilon^{\nu}_{(s)}({{\bf q}})\chi_{\mu\nu}(q)
\end{equation}
The electrical conductivity in terms of trace of spectral function is defined as:
\begin{equation}
\sigma=\left. \frac{e^2}{2(d-1)}\lim_{w\rightarrow 0}\frac{1}{w}\chi^{\mu}_{\mu}(q)\right|_{w= q}.
\end{equation}
We utilize the above expression to calculate the conductivity in the next subsection.

 \subsection{D7-Brane Gauge field fluctuations}

The gauge field in the  type IIB background  as described in section {{\bf 2}} appears due to presence of coincident $N_f$ D7-branes wrapped around non-compact four-cycle  in the resolved warped deformed conifold background.  The $U(1)$ symmetry acting on the centre of $U(N_f)=SU(N_f)\times U(1)$ group living on the world volume of   D7-branes wrapped around non-compact four-cycle induces a non-zero chemical potential.   The spectral functions in the presence of non-zero  chemical potential are obtained by computing two-point correlation functions of gauge field fluctuations about background field  \cite{kaminskietal} which includes only a non-zero temporal component of gauge field.
\subsubsection{EOMs}
Considering fluctuations of gauge field around non-zero temporal component of gauge field, we have
\begin{equation}
{\hat A_{\mu}}(u, \vec {x})= \delta^{0}_{\mu} {A_t}(u)+ {\tilde {\tilde A}}_{\mu}(\vec x,u).
\end{equation}
It is assumed that fluctuations are gauged to have non-zero components only along Minkowski coordinates.

 The action for gauge field in the presence of flavor branes is given as:
\begin{eqnarray}
{\cal I}_{D7}=\-T_{7}\int d^{8}{\sigma} e^{-\phi(r)}\sqrt{det(i^{*}g+F)},
\end{eqnarray}
 where `$g$' corresponds to  the determinant of type IIB metric as given in equation (\ref{metric}), and $i^{*}g$ gives the pull back of 10D metric onto $D7$-brane world volume \cite{MQGP}. Introduce fluctuations around ${A_t}(u)$,
$F_{\mu\nu}\rightarrow {\hat F}_{\mu \nu}= \partial_{[\mu} {\hat A}_{\nu]}$ and $ {\hat A}_{\mu}= A_t +{\tilde {\tilde A}}_{\mu} $. Defining $$G= i^{*}g+ F,$$ the DBI Action will be given by
\begin{eqnarray}
{\cal I}_{D7}=\-T_{7}\int d^{8}{\sigma} e^{-\phi(r)}\sqrt{det(G+{\tilde {\tilde F}})}.
\end{eqnarray}
Expanding above in the fluctuations quadratic in field strength, the DBI action as worked out in \cite{kaminskietal} can be written as :
\begin{eqnarray}
&&{\cal I}^{2}_{D7}=\-T_{7}\int d^{8} {\sigma} \ e^{-\phi(r)} \sqrt{ \left|det G\right|} \times \left( G^{\mu\alpha} G^{\beta\gamma} {\tilde {\tilde F}}_{\alpha\beta}{\tilde {\tilde F}}_{\gamma \mu}-\frac{1}{2}G^{\mu\alpha} G^{\beta \gamma} {\tilde {\tilde F}}_{\mu \alpha} {\tilde {\tilde F}}_{\beta \gamma}\right),
\end{eqnarray}
 and the EOM for ${\tilde A}_{\mu}$ is:
 \begin{eqnarray}
 \partial_{\nu}\left[\sqrt{\left| det G\right|}\times \left( G^{\mu\nu} G^{\sigma\gamma}-G^{\mu \sigma}G^{\nu\gamma}- G^{[\nu\sigma]}G^{\gamma\nu}\right)\partial_{[\gamma}{\tilde {\tilde A}}_{\mu]}\right] =0.
 \end{eqnarray}
 Using above, the DBI action at the boundary will be given as:
\begin{eqnarray}
\label{eq:I2D7}
&&{\hskip -0.2in} {\cal I}^{2}_{D7}=\-T_{7}\int d^{8}{\sigma} \ e^{-\phi(r)} \sqrt{ \left|det G\right|} \times \left.\left( (G^{tu})^2 {\tilde {\tilde A}}_{t}\partial_{u} {\tilde {\tilde A}}_t - G^{rr} G^{ik} {\tilde {\tilde A}}_{i}\partial_{u} {\tilde {\tilde A}}_{k}-A_{t} G^{ut} tr(G^{-1}F)\right)\right|^{u=0}_{u=1}.\nonumber\\
\end{eqnarray}
We have already obtained the expression of $F_{rt}$ in \cite{MQGP} and the same is given as
\begin{equation}
\label{F_rt}
F_{rt}=\frac{C e^{\phi(r)}}{\sqrt{ C^2 e^{2\phi(r)}+r^6}}.
\end{equation}
Considering $u=\frac{r_h}{r}$, we have
\begin{equation}
\label{F_ut}
F_{ut}=-\frac{C u^2 e^{\phi(u)}}{r_h\sqrt{ C^2 e^{2\phi(u)}+r^6}}.
\end{equation}
Looking at the above expression, we see that at $u=0$ (boundary), $F_{ut}\rightarrow 0$. The action in equation (\ref{eq:I2D7}) will get simplified and  given as:
\begin{eqnarray}
&&{\cal I}^{(2)}_{D7}=\-T_{7}\int d^{8}{\sigma} \ e^{-\phi(u)} \sqrt{ \left|det G\right|} \times \left.\left(  - G^{uu} G^{ii} {\tilde {\tilde A}}_{i}\partial_{u} {\tilde {\tilde A}}_{i} )\right)\right|^{u=0}_{u=1}.
\end{eqnarray}
where $i \in \mathbb{R}^{1,3}(t,x_1,x_2,x_3)$  and
\begin{eqnarray}
\label{eq:Gmet}
 && G^{ x_1 x_1}=G^{x_2 x_2}=G^{x_3 x_3}= g^{x_1 x_1}= g^{x_2 x_2}=g^{x_3 x_3}=\frac{u^2 L^2}{r^{2}_{h}},G^{ tt}=\frac{u^2 L^2}{g_1 ( {r^{2}_{h}- u^4 F^{2}_{ut}})} \nonumber\\
 &&G^{uu} = \frac { g_1u^2 r^{2}_{h}}{L^2 (r^{2}_{h}- F^{2}_{ut} u^4)},\sqrt{-G}= \frac{r^{3}_{h}}{u^5}\sqrt{r^{2}_{h}- u^4 F^{2}_{ut}}.
 \end{eqnarray}
 Defining gauge invariant field components
$$ E_{x_1}= w {\tilde {\tilde A}}_{x_1} + q{\tilde {\tilde A}}_t, E_{\alpha}= w {\tilde {\tilde A}}_{\alpha},\alpha=(x_2,x_3),$$ the DBI action in terms of these co-ordinates (using equation (\ref{EOMs-ii})) will be given as:
\begin{eqnarray}
&&{\cal I}^{(2)}_{D7}=\-T_{7}\int \frac{dw dq}{(2\pi)^3}  \frac{ e^{-\phi(u)}  r^{2}_{h} }{u}  \left.\left[  \frac{{E}_{x_1}\partial_{u} E_{x_1}}{\left(q^2-\frac{w^2}{g_1}\right)}-  \frac{1}{w^2}   {E}_{x_2}\partial_{u} E_{x_2} )\right]\right|^{u=0}_{u=1}.
\end{eqnarray}
Defining the longitudinal electric field as $E_{x_1}(q,u)= E_{0}(q) \frac{E_{q}(u)}{E_{q}(u=0)}$, the flux factor as defined in \cite{son+starinets} in the zero momentum limit will be given as:
\begin{eqnarray}
{\cal F}(q,u)=  -\frac{ e^{-\phi(u)}  r^{2}_{h} }{w^2 u}  \frac{E_{-q}(u)\partial_{u}E_{q}(u)}{ E_{-q}(u=0)E_{q}(u=0)},
\end{eqnarray}
and  the retarded green function for $E_{x_1}$ will be  ${\cal G}(q,u)= -2 {\cal F}(q,u)$. The retarded green function for ${\tilde {\tilde A}}_{x_1}$ is $w^2$ times above expression and for $q=0$, it gives
\begin{eqnarray}
{\cal G}_{x_1 x_!}= -2{\cal F}(q,u)= \left.\frac{ 2 e^{-\phi(u)}  r^{2}_{h} }{ u}\frac{ \partial_{u}E_{q}(u)}{ E_{q}(u)}\right|_{u=0}.
\end{eqnarray}
The spectral functions in zero momentum limit will be given as:
\begin{equation}
\label{correlator_sigma}
{\cal X}_{x_1 x_1}(w,q=0)= -2 Im {\cal G}_{x_1 x_1}(w,0) = { e^{-\phi(u)}  r^{2}_{h} }Im\left[\frac{1}{u} \frac{ \partial_{u}E_{q}(u)}{ E_{q}(u)}\right]_{u=0}.
\end{equation}
 Since we are interested to obtain two-point Green's function in the zero-momentum limit ($q=0$),  the transverse and longitudinal components of electric field will be simply given as $E_{x_1}=w {\tilde{\tilde A}}_{x_1}, E_{x_2}= w {\tilde{\tilde A}}_{x_2}$. In these set of co-ordinates, the above equations take the form as given below:
\begin{eqnarray}
&& E^{\prime\prime}_{x_1}+\frac{\partial_{u}\left(\sqrt{-G} G^{uu} G^{x_1 x_1}\right)}{\sqrt{-G} G^{uu} G^{x_1 x_1}}E^{\prime}_{x_1}-\frac{G^{tt}}{G^{uu}} w^2E_{x_1}=0,\nonumber\\
&& E^{\prime\prime}_{x_2}+\frac{\partial_{u}\left(\sqrt{-G} G^{uu} G^{x_2 x_2}\right)}{\sqrt{-G} G^{uu} G^{x_2 x_2}}E^{\prime}_{x_2}-\frac{G^{tt}}{G^{uu}} w^2E_{x_2}=0,
\end{eqnarray}
where the prime denotes derivative w.r.t $u$. The co-ordinates in Minkowski directions are chosen such that the momentum four-vector exhibits only one spatial component i.e $ (-w, q, 0, 0)$. Writing $A_\mu(\vec x,u)=\int \frac{d^4q}{(2\pi)^4}e^{-iw t+iq x_1}\tilde{A}_\mu(q,u)$,
incorporating the values of various metric components,
   setting $E(u)= E_{x_1}(u)=E_{x_2}(u)$, and $w_3=\frac{w}{\pi T}$, and substituting the value of $F_{ut}$ and $g_1$, we have:
  \begin{eqnarray}
  \label{E_eom}
  E^{\prime\prime}(u) + \left(-\frac{4u^3}{1-u^4} - \frac{1}{u} + \sqrt{\frac{3C^2u^5}{g_s^2 r_h^6 + C^2 u^6}}\right)E^\prime(u)
 + \frac{w_3^2}{(1-u^4)^2} E(u)=0.
 \end{eqnarray}
From (\ref{E_eom}), one sees that $u=1$ is a regular singular point and the roots of the indicial equation about the same are given by: $\pm\frac{iw_3 }{4 }$; choosing `incoming wave' solution, solutions to (\ref{E_eom}) sought will be of the form:
 \begin{equation}
 \label{E_sol_i}
 E(u) = \left(1-u\right)^{-\frac{iw_3}{4}}{\cal E}(u)\equiv \left(1-u\right)^{-\frac{iw_3}{4}}{\cal E}(u),
 \end{equation}
 where ${\cal E}(u)$ is analytic in $u$. As one is interested in solving for ${\cal E}(u)$, analytic in $u$, near $u=0$, (\ref{E_eom}) will be approximated by:
 \begin{equation}
 \label{E_eom-ii}
  (u-1)^2 E^{\prime\prime}(u) - \frac{(u-1)^2 E^\prime(u)}{u}
 + \frac{w_3^2}{(1+u^2)^2(1+u)^2} E(u)\approx0.
 \end{equation}
One converts (\ref{E_eom-ii}) into a differential equation in ${\cal E}$. Performing a perturbation theory in powers of $w_3$, one looks for a solution of the form:
\begin{equation}
\label{E_sol_ii}
{\cal E}(u) = {\cal E}^{(0)}(u) + w_3 {\cal E}^{(1)}(u) + w_3^2 {\cal E}^{(2)}(u) + ...
\end{equation}
Near $u=0$:
the solutions are given as under:
\begin{eqnarray}
\label{E_sol_iii}
& & {\cal E}^{(0)}(u\sim0) = \frac{C_1^{(0)}}{2}u^2 + C_2^{(0)};\nonumber\\
& & {\cal E}^{(1)}(u\sim0) = \frac{C_1^{(0)}u^2}{2} + C^{(0)}_2 + \frac{i\beta}{4}\left(\frac{C_1^{(1)}u^3}{2} - \frac{C_2^{(1)}u^2}{2}\right).
\end{eqnarray}
To get non-zero conductivity, we will required $C_1^{(0)}\in\mathbb{C}$.

 \subsubsection{Electrical Conductivity, Charge Susceptibility and Einstein Relation}

 The conductivity, using (\ref{correlator_sigma}),(\ref{E_sol_iii}) and $T=\frac{r_h}{\pi\sqrt{4\pi g_sN}}$ \cite{MQGP},  will be given as:
\begin{equation}
\sigma= \lim_{w\rightarrow 0} \frac{{\cal X}_{x_1 x_1 }(w,q=0)}{w}=\frac{r_h^2C_1^{(0)}}{g_s\pi T}=\frac{\pi({4 \pi g_s N})T}{g_s}.
\end{equation}

Another physically relevant quantity  is the charge susceptibility, which is thermodynamically defined as response of the charge  density to the change in chemical potential \cite{chi}.  \begin{eqnarray}
\label{chi-a}
& & \chi=\left.\frac{\partial n_q}{\partial \mu}\right|_{T},
\end{eqnarray}
 where $n_q = \frac{\delta S_{DBI}}{\delta F_{rt}}$, and the chemical potential is defined as $
\mu=\int_{r_h}^{r_B} { F_{rt}} dr$.
Using the same,  the charge susceptibility will be given as:
\begin{eqnarray}
\chi=\left(\int_{r_h}^{r_B} \frac{d F_{rt}}{dn_q}\right)^{-1}.
\end{eqnarray}

From \cite{MQGP},
\begin{equation}
\label{n_q}
n_q\sim \frac{r^3F_{rt}\left(\frac{1}{g_s} - \frac{N_f ln\mu}{2\pi}\right)}{\sqrt{1 - F_{rt}^2}},\ \mu\equiv{\rm embedding\ parameter},
\end{equation}
which using (\ref{F_rt}) implies:
\begin{eqnarray}
\label{chi_ii}
& & \hskip -0.5in \frac{1}{\chi}\sim\frac{1}{\left(\frac{1}{g_s} - \frac{N_f ln\mu}{2\pi}\right)}\int_{r_h}^\infty \frac{dr}{r^3}\left(1 - F_{rt}^2\right)^{\frac{3}{2}}\nonumber\\
& & \hskip -0.5in = \frac{1}{6
   {r_h}^2\frac{({g_s} {N_f} \log (\mu )-1)}{C
   {g_s}} \left(C^2 {g_s}^2+{g_s}^2 {N_f}^2 {r_h}^6 \log
   ^2(\mu )-2 {g_s} {N_f} {r_h}^6 \log (\mu )+{r_h}^6\right)}\nonumber\\
   & & \hskip -0.5in\times\Biggl[\left(C^2 {g_s}^2-2 {g_s}^2 {N_f}^2 {r_h}^6 \log ^2(\mu )+4
   {g_s} {N_f} {r_h}^6 \log (\mu )-2 {r_h}^6\right) \,
   _2F_1\left(\frac{1}{3},\frac{1}{2};\frac{4}{3};-\frac{C^2 {g_s}^2}{{r_h}^6
   ({g_s} {N_f} \log (\mu )-1)^2}\right)\nonumber\\
   & & \hskip -0.5in+5 {r_h}^6 ({g_s} {N_f}
   \log (\mu )-1)^2 \, _2F_1\left(-\frac{1}{2},\frac{1}{3};\frac{4}{3};-\frac{C^2
   {g_s}^2}{{r_h}^6 ({g_s} {N_f} \log (\mu )-1)^2}\right)\Biggr].
\end{eqnarray}
For $\mu=1-\epsilon,\epsilon\rightarrow0^+, C<<1$ \cite{MQGP}, in the MQGP limit \cite{MQGP} (and also the weak $(g_s)$ coupling-Strong t'Hooft coupling limit \cite{metrics}), one sees: $\chi\sim \frac{r_h^2}{g_s}\sim \pi({4 \pi N})T^2$.

The diffusion coefficient corresponding to non-zero charge density appear by demanding the longitudinal component of electric field strength to be zero i.e $E_{x_2}=0$. As given in \cite{chi}, the condition provides with the expression of diffusion coefficient to be expressed as:
\begin{eqnarray}
\label{D}
D= \left.e^{-\phi}\sqrt{G G^{00}G^{uu}}G^{ii}\right|_{u=1}\int^{u=0}_{u=1}\frac{du}{e^{-\phi}\sqrt{G}G^{00}G^{uu}}.
\end{eqnarray}
Subtituting  the values  for above metric components as given in equation (\ref{eq:Gmet}), after integration, we get $D=\frac{L^2}{r_h}+{\cal O}(C^2 g_s^2/r_h^8) \sim \frac{1}{T}$. Using equations  (\ref{chi-a}), (\ref{chi_ii}) and (\ref{D}), we get $\frac{\sigma}{\chi}\sim\frac{1}{T}\sim D$, hence verifying the Einstein relation.

 \subsection{R-Charge Correlators}

The $U(1)_R$-charges are defined in the bulk gravitational background  dual to the  rank of isometry group corresponding to the spherical directions transverse to the AdS space. Given that the 11-dimensional M-theory background  corresponds to  black $M3$-branes which asymptotically can be expressed as $M5$-branes wrapped around two-cycles defined homologously as integer sum of two-spheres (as described in section {{\bf 2}}),  there will be a rotational (R)- symmetry group  dual to the isometry group $U(1)\times U(1)$  corresponding to $\phi_{1/2}$ and $\psi$ in the directions transverse to $M5$-branes wrapped around $S^{2}({\theta_1,\phi_{(2/1)}})+ S^{2}({\theta_2,x_{10}})$. To determine the diffusion coefficient due to the ${R}$-charge, one needs to evaluate the two-point correlation function of ${\tilde A}_{\mu}$ which basically will be a metric perturbation of the
form $h_{M\mu}$ where M is a spherical direction and $\mu$ is an asymptotically AdS direction.  As a first step, the ${\tilde A}_\mu$ EOM is:
\begin{equation}
\label{gauge-EOM}
\partial_\beta\left[g^{\mu\nu}g^{\alpha\beta}\sqrt{g}{\tilde F}_{\nu\alpha}\right]=0.
\end{equation}

By defining u= $\frac{r_h}{r}$ so that $g_1=g_2= 1-u^4$, the black $M3$-brane metric of \cite{MQGP} reduces to the form as given below:
\begin{eqnarray}
\label{ds5squared}
ds^2 = -\frac{g_1 g^{-\frac{2}{3}}_{s} r^{2}_{h} }{u^2 L^2}dt^2 +  \frac{g^{-\frac{2}{3}}_{s} r^{2}_{h} }{u^2 L^2}\left(d{x}^{2}_{1} +d{x}^{2}_{2} +d{x}^{2}_{3}  \right)+   \frac{g^{-\frac{2}{3}}_{s}L^2 }{u^2g_1}du^2  + d{\cal M}^2_{6},
\end{eqnarray}
where $d{\cal M}^2_{6}=d{\cal M}^2_{5} +  g^{\frac{4}{3}}_{s} \left(d^{2}_{x_{11}} + A^{F_1} + A^{F_2} +A^{F_5}\right)^2$,
and in  weak $(g_s)$ coupling-strong t'Hooft coupling/ MQGP limit, the non-zero component include
\begin{eqnarray}
&& d{\cal M}^2_{5}=  G^{\cal M}_{\theta_1 \theta_1} d{\theta^{2}_1}+ G^{\cal M}_{\theta_1\theta_2} d{\theta_1} d{\theta_2} + G^{\cal M}_{\phi_1 \theta_1}d\phi_1 d\theta_1 + G^{\cal M}_{\phi_2 \theta_1} d\phi_2 d\theta_1+  G^{\cal M}_{z \theta_1} d\psi d\theta_1+\nonumber\\
&&+ G^{\cal M}_{10 \theta_1} dx_{10} d\theta_1+ G^{\cal M}_{\theta_2 \theta_2}d{\theta^{2}_2}+   G^{\cal M}_{\phi_1 \theta_2}d\phi_1 d\theta_2, G^{\cal M}_{\phi_2 \theta_2}d\phi_2 d\theta_2+  G^{\cal M}_{\psi\theta_2}dyz d\theta_2 +G^{\cal M}_{10 \theta_2} dx_{10} d\theta_2\nonumber\\
&&+  G^{\cal M}_{\phi_1\psi_1} d\phi_1^2 + G^{\cal M}_{\phi_1 \phi_2} d\phi_1 d\phi_2+ G^{\cal M}_{\phi_1 \psi} d\phi_1 d\psi +  G^{\cal M}_{yy} d\phi_2^2 + G^{\cal M}_{\phi_2 \psi} d\phi_2 d\psi + G^{\cal M}_{\psi\psi } d\psi^2+  G^{\cal M}_{10\ 10} dx^{2}_{10}.\nonumber\\
\end{eqnarray}
The simplified expressions of aforementioned metric components are given in \cite{MQGP}.

Writing ${\tilde A}_\mu(x_1)=\int \frac{d^4q}{(2\pi)^4}e^{-i\omega t+iq x_1}\tilde{A}_\mu(q,u)$ and working in the $A_u=0$ gauge, by setting $\mu=u,x_1,t,\alpha(\in {\mathbb R}^3)$ in (\ref{gauge-EOM}) one ends up with the following equations:
\begin{eqnarray}
\label{EOMs-ii}
& & w_3 \tilde{A}^{\prime}_t + g_1   q_3\tilde{A}^{\prime}_{x_1} = 0,\nonumber\\
& & \tilde{A}_{x_1}^{\prime\prime} -\frac{1}{u}\tilde{A}_{x_1}^\prime +\frac{g^{\prime}_1}{g_1} \tilde{A}_{x_1}^\prime -\frac{1}{g^{2}_1}\left( w_3 q_3 \tilde{A}_t + w^{2}_{3} \tilde{A}_{x_1}\right)= 0,\nonumber\\
& & \tilde{A}_t^{\prime\prime} - \frac{1}{u}\tilde{A}_t^\prime -\frac{1}{g_1}\left( w_3 q_3 \tilde{A}_t + q^{2}_{3} \tilde{A}_{x_1}\right) = 0,\nonumber\\
& & \tilde{A}_\alpha^{\prime\prime} - \frac{1}{u}\tilde{A}_\alpha^\prime +\frac{g^{\prime}_1}{g_1} \tilde{A}_{\alpha}^\prime+\frac{1}{g^{2}_1}\left( w^{2}_3 - g_1 q^{2}_{3} \right)\tilde{A}_{\alpha}= 0.
\end{eqnarray}
where $\alpha=(x_2,x_3)$. The simplest to tackle is the last equation as it is decoupled from the previous three equations. One notices that the horizon $u=1$ is a regular singular point and the exponents of the indicial equation about the same are given by: $\pm\frac{i w_3}{4}$; one chooses the incoming-wave solution and hence write $\tilde{A}_\alpha(u)=(1 - u)^{-\frac{i w_3}{4}}\tilde{\cal A}_\alpha(u)$. The last equation in (\ref{EOMs-ii}) then is rewritten as:
\begin{eqnarray}
\label{cal-A_alpha-eom-i}
& & (1 - u)^2\tilde{\cal A}_\alpha^{\prime\prime}(u) + \left(\frac{i w_3}{2}(1 - u) - (1 - u)\left[\frac{4u^3}{(u + 1)(u^2 + 1)}
- \frac{(u - 1)}{u}\right]\right)\tilde{\cal A}_\alpha^\prime(u) \nonumber\\
& & + \left(\frac{i w_3}{4}\left[\frac{i w_3}{4} + 1\right] + \frac{\left[w_3^2 + (u^4 - 1)q_3^2\right]}{(u + 1)^2(u^2 + 1)^2}\right)\tilde{\cal A}_\alpha(u)
=0.
\end{eqnarray}
We make the following double perturbative ansatz for the solution to $\tilde{\cal A}_\alpha(u)$'s EOM (\ref{cal-A_alpha-eom-i}):
\begin{equation}
\label{cal-A_alpha_soln-i}
\tilde{\cal A}_\alpha(u) = \tilde{\cal A}_\alpha^{(0,0)}(u) + q_3 \tilde{\cal A}_\alpha^{(1,0)}(u) + w_3 \tilde{\cal A}_\alpha^{(0,1)}(u) + q_3 w_3 \tilde{\cal A}_\alpha(u)^{(1,1)} + q_3^2\tilde{\cal A}_\alpha^{(2,0)}(u)+...
\end{equation}
Plugging (\ref{cal-A_alpha_soln-i}) into (\ref{cal-A_alpha-eom-i}) yields as
solutions, near $u=0$ up to ${\cal O}(u)$, (\ref{cal-A_alpha_soln_iii}).
In the $w\rightarrow0,q\rightarrow0$-limit, one can take $\tilde{A}_\alpha(0)\sim c_2$.
Also,
\begin{eqnarray}
\label{cal-A_alpha_soln_iii}
& & (1 - u)^{-\frac{i w_3}{4}}\tilde{\cal A}_\alpha^{(0,0)\ \prime}(u) = - c_1u + {\cal O}(u^2);\nonumber\\
& & (1 - u)^{-\frac{i w_3}{4}}\tilde{\cal A}_\alpha^{(0,1)\ \prime}(u) = -\frac{i c_2}{4}+\frac{1}{16} \left(5 \pi  c_1+{w3} c_2-16 c_3\right) u+O\left(u^2\right)\nonumber\\
& & (1 - u)^{- \frac{i w-3}{4}}\tilde{\cal A}_\alpha^{(2,0)}(u) = \left( - c_3 + c_2 ln u\right) u + {\cal O}(u^2).
\end{eqnarray}
The kinetic terms relevant to the evaluation of two-point correlators of ${\tilde A}_{u,t,\alpha}$ in the MQGP limit are:
\begin{equation}
\label{kinetic-vector}
S= g_s^{-\frac{4}{3}} r_h^2 L^2\frac{1}{{\kappa^{2}_{11}}\epsilon^{\frac{15}{2}}} \int d^4x du \frac{1}{u}
\left[ (1 - u^4) \left({\tilde A}^\prime_{\alpha}\right)^2  + (1 - u^4) \left({\tilde A}^\prime_{x}\right)^2- \left({\tilde A}^\prime_{t}\right)^2\right]
\end{equation}
Hence, the retarded Green's function $G^R_{\alpha\alpha}$ will be given as:
\begin{eqnarray}
\label{G-alpha_alpha}
& &   G^{R\ {\rm finite}}_{\alpha\alpha}\sim \left.\frac{1}{u}\frac{\tilde{\cal A}_{\alpha,-q}(u)}{\tilde{\cal A}_{\alpha,-q}(0)}\partial_u
\left(\frac{\tilde{\cal A}_{\alpha,q}(u)}{\tilde{\cal A}_{\alpha,q}(0)}\right)\right|_{u=0}\nonumber\\
& &  \sim  T\Biggl[i w + \nonumber\\
&& {\hskip -0.3in} \frac{q^2}{\pi T}
\left(\frac{\frac{1}{24} \left[-3 \left\{8 c_2 c_3+c_1 \left[8 c_4+c_1 (2 {Li}_3(-i)+2 {Li}_3(i)+3 \zeta (3))\right]\right\}+i \pi ^3 c_1^2-\pi ^2
   c_2 c_1-6 i \pi  c_3 c_1\right]}{-\frac{1}{64}  \pi ^2 c_1^2+\frac{1}{8} \pi  \left(- i c_2 + 2  c_3\right) c_1 - i c_4 c_1-\frac{ c_2^2}{4}
+ ic_2 c_3}\right)\Biggr].\nonumber\\
& &
\end{eqnarray}
On comparison with $G^R_{\alpha\alpha}\sim i w + 2 D_R q^2$ \cite{herzog}, one sees that there is a three-parameter family (which in the $c_2>>c_{1,3,4}$-limit are $c_{1,3,4}$) of solutions to the $R$-charge fluctuation $\tilde{A}_\alpha$  which would generate $D_R^{\alpha}\sim\frac{1}{\pi T}$\footnote{In $c_2>>c_{1,3,4}$-limit, $D_R^{\alpha}=\frac{24 c_3 + \pi^2 c_1}{12\pi \tilde{A}_\alpha^0}$.}.

We now go to the $\tilde{A}_{x_1,t}$ EOMs and observe that one can decouple $\tilde{A}_{x_1}$ and $\tilde{A}_t$, and obtain, e.g., the following third order differential equation for $\tilde{A}_{x_1}$:
\begin{eqnarray}
\label{Ax-decoupled-eom_i}
&& (1 - u^4)^2\tilde{A}_{x_1}^{\prime\prime\prime}(u) - (1 - u^4)\left\{(1 - u^4) + 12 u^3\right\}\tilde{A}_{x_1}^{\prime\prime}(u) - \left[4u^2(1 - u^4) - 16 u^6 - q_3^2(1 - u^4) \right.\nonumber\\
&& \left. + w_3^2\right]\tilde{A}_{x_1}^\prime = 0.
\end{eqnarray}
One notes that $u=1$ is a regular singular point of the second order differential equation (\ref{Ax-decoupled-eom_i}) in $\tilde{A}_{x_1}^\prime$. The exponents of the indicial equation are $- 1 \pm \frac{i w_3}{4}$; choosing the incoming-boundary-condition solution, we write
$\tilde{A}_{x_1}^\prime = (1 - u)^{- 1 - \frac{i w_3}{4}}\tilde{\cal A}_{x_1}^\prime$ and assume a double perturbative series for solution to the second order differential equation satisfied by $\tilde{\cal A}_{x_1}^\prime$ of the form:
\begin{equation}
\label{sol_calA_x-i}
\tilde{\cal A}_{x_1}^\prime(u) = \tilde{\cal A}_{x_1}^{(0,0)\ \prime}(u) + w_3 \tilde{\cal A}_{x_1}^{(0,1)\ \prime}(u) + q_3^2 \tilde{\cal A}_{x_1}^{(2,0)\ \prime}(u) + {\cal O}(w_3^2,w_3q_3^2).
\end{equation}

The equations that one obtains at various orders of (\ref{sol_calA_x-i}), if to be solved exactly, are intractable. We will be content with their solutions near $u=0$. They are given in (\ref{calAx_solutions_near_u_0}) and are of the form:
\begin{eqnarray*}
\label{calAx_solutions_near_u_0-ii}
& &  \tilde{\cal A}_{x_1}^{(0,0)\ \prime}   \equiv a + b u + c u^2 + {\cal O}(u^2).
\end{eqnarray*}
 The solutions of $\tilde{\cal A}_{x_1}^{(2,0)\ \prime}$ and $\tilde{\cal A}_{x_1}^{(0,1)\ \prime}$ require more work. Going even up to ${\cal O}(u^2)$ for the same is intractable. Given that we would be needing for the purpose of evaluation of the retarded Green's function $G^R_{x_1 x_1}$ solutions only up to ${\cal O}(u\rightarrow0)$, we give below solutions up to ${\cal O}(u)$.

One can show that $\tilde{\cal A}_{x_1}^{(0,1)\ \prime}(u)$, up to ${\cal O}(u)$, will be given by the expansion of (\ref{Ax'01_soln-i})  up to ${\cal O}(u)$.
 After some MATHEMATICAlgebra, one can show:
\begin{equation}
\label{Ax'01_up_to_u}
\tilde{\cal A}_{x_1}^{01}(u) \approx (7.4 - 1.8 i)c_1 + (25.5 - 45.4 i)c_2 + \biggl[( - 4.8 + 5 i) c_1 - (55.7 - 108.3 i)c_2\biggr] u + {\cal O}(u^2)
\end{equation}

One can similarly show that $\tilde{\cal A}_{x_1}^{(2,0)\ \prime}(u)$, up to ${\cal O}(u)$, will be given by the expansion, up to ${\cal O}(u)$, of (\ref{Ax'20_soln-i}). Again after some MATHEMATICAlgebra, one hence obtains:
\begin{equation}
\label{calAx'02up_to_u}
\tilde{\cal A}_{x_1}^{(2,0)}(u) \approx (7.4 - 1.8 i)c_1 + (40.2 - 5.4 i)c_2 + \biggl[(-4.8 + 5 i)c_1 - (12.1 - 26.7 i)c_2\biggr]u + {\cal O}(u^2).
\end{equation}
Now, writing:
\begin{equation}
\label{Ax'up_to_u}
\left(\begin{array}{c}
\tilde{\cal A}_{x_1}^{(0,0)\ \prime}(u)\\
\tilde{\cal A}_{x_1}^{(0,1)\ \prime}(u) \\
\tilde{\cal A}_{x_1}^{(2,0)\ \prime}(u)
\end{array}\right)
= \left(\begin{array}{cc}
a(c_2) & b(c_1,c_2) \\
\tilde{a}(c_1,c_2) & \tilde{B}(\tau)(c_1,c_2) \\
\tilde{\tilde{a}}(c_1,c_2) & \tilde{\tilde{B}}(\tau)(c_1,c_2)
\end{array}\right)
\left(\begin{array}{c}
1 \\ u
\end{array}\right),
\end{equation}
one obtains:
\begin{eqnarray}
\label{Ax'}
& & \tilde{A}_{x_1}^\prime = (1 - u)^{- 1 - \frac{i w_3}{4}}\biggl[a(c_2) + w_3 \tilde{a}(c_1,c_2) + q_3^2\tilde{\tilde{a}}(c_1,c_2) + \biggl(b(c_1,c_2) + w_3 \tilde{B}(\tau)(c_1,c_2) \nonumber\\
&&+ q_3^2\tilde{\tilde{B}}(\tau)(c_1,c_2)\biggr)u  + {\cal O}(u^2)\biggr],\nonumber\\
& & \tilde{A}_{x_1}^{\prime\prime}(u)  = \left(1 + \frac{i w_3}{4}\right)(1 - u)^{- 2 - \frac{i w_3}{4}}\biggl[a(c_2) + w_3 \tilde{a}(c_1,c_2) + q_3^2\tilde{\tilde{a}} + \biggl(b(c_1,c_2)  \nonumber\\
 &&+ w_3 \tilde{B}(\tau)(c_1,c_2) + q_3^2\tilde{\tilde{B}}(\tau)(c_1,c_2)\biggr)u   + {\cal O}(u^2)\biggr] + (1 - u)^{ - 1 - \frac{i w_3}{4}}\biggl[\biggl(b(c_1,c_2) + w_3 \tilde{B}(\tau)(c_1,c_2)  \nonumber\\
 &&+ q_3^2 \tilde{\tilde{B}}(\tau)(c_1,c_2)\biggr)  + {\cal O}(u)\biggr].
\end{eqnarray}
As in the differential equation:
\begin{equation}
\label{A'tA'x}
\tilde{A}_{x_1}^{\prime\prime}(u) - \left(\frac{4u^3}{1 - u^4} + \frac{1}{u}\right)\tilde{A}_{x_1}^\prime(u) - \frac{w_3q_3\tilde{A}_t + w_3^2\tilde{A}_{x_1}}{(1 - u^4)} = 0,
\end{equation}
we require $\tilde{A}_{x_1}^\prime(u)$ to be well-defined at $u=0$, one has to impose the following constraint on $c_1, c_2$:
\begin{equation}
\label{constraint_c1c2}
a(c_2) + w_3 \tilde{a}(c_1,c_2) + q_3^2\tilde{\tilde{a}}(c_1,c_2) = 0.
\end{equation}
As,
\begin{equation}
\label{as}
\left(\begin{array}{c}
a \\
\tilde{a} \\
\tilde{\tilde{a}}
\end{array}\right)
\approx
\left(\begin{array}{cc}
0 & \alpha \\
a_1 & b_1 \\
a_2 & b_2
\end{array}\right)
\left(\begin{array}{c}
c_1\\c_2\end{array}\right),
\end{equation}
one sees that:
\begin{equation}
\label{c1}
c_1 = c_2\left(\frac{ - \alpha - w_3 b_1 - q_3^2 b_2}{w_3 a_1 + q_3^2 a_2}\right).
\end{equation}
Substituting (\ref{Ax'}) into (\ref{A'tA'x}) evaluated at $u=0$, one obtains:
\begin{equation}
\label{constrat_bc_AxAt}
w_3 q_3 \tilde{A}_t^0 + w_3^2 \tilde{A}_{x_1}^0 = 0.
\end{equation}
As,
\begin{equation}
\label{bs}
\left(\begin{array}{c}
b \\ \tilde{B}(\tau) \\ \tilde{\tilde{B}}(\tau)
\end{array}\right)
\approx
\left(\begin{array}{cc}
0 & \beta \\
p_1 & q_1 \\
p_2 & q_2
\end{array}\right)
\left(\begin{array}{c}
c_1 \\ c_2
\end{array}\right),
\end{equation}
one realizes that:
\begin{equation}
\label{Ax'LO}
\tilde{A}_{x_1}^\prime(u) = (1 - u)^{- 1 - \frac{i w_3}{4}}c_2 u \frac{\left[w_3(a_1 \beta - p_1 \alpha) + q_3^2(a_2\beta - p_2\alpha)
+ {\cal O}(w_3^2,w_3q_3^2,u^2)\right]}{(a_1 w_3 + a_2 q_3^2)}.
\end{equation}
This yields:
\begin{equation}
\label{Ax}
\tilde{A}_{x_1}(u) =
(1 - u)^{ - \frac{i w_3}{4}} c_2\frac{\left[w_3(a_1 \beta - p_1 \alpha) + q_3^2(a_2\beta - p_2\alpha)
+ {\cal O}(w_3^2,w_3q_3^2,u^2)\right]}{(a_1 w_3 + a_2 q_3^2)}\frac{4 - i w_3 u}{w_3(4 i + w_3)} + c_3.
\end{equation}
Thus, the retarded Green's function $G^R_{xx}$ will be given by:
\begin{equation}
\label{D_1overT}
G^R_{x_1 x_1}\sim \lim_{u\rightarrow0}\frac{1}{u} \tilde{A}_{x_1,q}\partial_u\tilde{A}_{x_1,-q}(u).
\end{equation}
The constants of integration $c_2, c_3$ must satisfy:
\begin{equation}
\label{constraint-c_23}
c_2\frac{\left[w_3(a_1 \beta - p_1 \alpha) + q_3^2(a_2\beta - p_2\alpha)
\right]}{(a_1 w_3 + a_2 q_3^2)}\frac{4}{w_3(4 i + w_3)} + c_3 = \tilde{A}_{x_1}^0.
\end{equation}
Choose:
\begin{equation}
\label{c_2}
c_2\frac{\left[w_3(a_1 \beta - p_1 \alpha) + q_3^2(a_2\beta - p_2\alpha)
\right]}{w_3(4 i + w_3)}\sim {\cal O}(w_3),
\end{equation}
which can be fine tuned to ensure $c_3\approx \tilde{A}_{x_1}^0$.
Thus, the retarded Green's function $G^R_{x_1 x_1}$ will be given by:
\begin{equation}
\label{D_1overT}
G^R_{x_1 x_1}\sim \lim_{u\rightarrow0}\frac{1}{u} \left(\frac{\tilde{A}_{x_1,q}(u)}{\tilde{A}_{x_1,q}(u=0)}\right)
\partial_u\left(\frac{\tilde{A}_{x_1,-q}(u)}{\tilde{A}_{x_1,-q}(u=0)}\right)\sim T\left(\frac{w^2}{iw - \frac{i a_2}{\pi T a_1}q^2}\right).
\end{equation}
Now $a_1=a_2$ and if one were to consider the ${\cal O}(w_3q_3^0)$ term to be $i w_3\tilde{A}_{x_1}^{(0,1)}(u)$, then upon comparison with $G^R_{x_1 x_1}\sim \frac{w^2}{i w - D_R q^2}$ \cite{herzog}, one sees that $D_R^{x_1}=\frac{1}{\pi T}$. By requiring $D_R^{x_1} = D_R^{\alpha}$, one sees that in fact that there is a two-parameter family of solutions that would generate $D_R=\frac{1}{\pi T}$ as, in the $c_2>>c_{1,3,4}$-limit as an example, one generates a constraint:
$\frac{24 c_3 + \pi^2 c_1}{12 A_\alpha^0}=1$.

\subsection{Stress Energy Tensor Modes}

The two point function of stress-energy tensors are obtained by considering small perturbations of the five-dimensional  metric, $g_{\mu\nu}= g_{\mu\nu(0)}+h_{\mu\nu}$.
Up to first order in the metric perturbation (dropping $G_4$ flux contributions) the Einstein equation  as given in \cite{herzog}:
\begin{equation}
\label{eq:einsteineq}
{\cal R}^{(1)}_{\mu\nu}=\frac{2}{d-2} \Lambda h_{\mu\nu},
\end{equation}
where $d$ is dimension of AdS space.
 To linear order in $h_{\mu\nu}$, Ricci scalar  will be given by \cite{Miedema}:
\begin{eqnarray}
\label{eq:R1munu}
{\hskip -0.4in}{\cal R}^{(1)}_{\mu\nu}= \Gamma^{\alpha}_{(1)\mu\nu,\alpha}-\Gamma^{\alpha}_{(1)\mu\alpha,\nu}+ \Gamma^{\beta}_{(0)\mu\nu}\Gamma^{\alpha}_{(1)\alpha\beta}+ \Gamma^{\beta}_{(1)\mu\nu}\Gamma^{\alpha}_{(0)\alpha\beta}-\Gamma^{\beta}_{(0)\mu\alpha}\Gamma^{\alpha}_{(1)\nu\beta}-\Gamma^{\beta}_{(1)\mu\alpha}\Gamma^{\alpha}_{(0)\nu\beta},
\end{eqnarray}
where
\begin{equation}
\label{eq:gamma1munu}
\Gamma^{(1)\alpha}_{\mu\nu}=-g^{\alpha \beta}_{(0)}h_{\beta\gamma}\Gamma^{\gamma}_{\mu\nu}+\frac{1}{2}g^{\alpha\beta}_{(0)}\left(h_{\beta\mu,\nu}+h_{\beta\nu,\mu}-h_{\mu\nu,\beta}\right).
\end{equation}
The five-dimensional metric in M-theory background in the limits of (\ref{limits_Dasguptaetal+MQGP}), is as follows:
\begin{eqnarray}
\label{ds5squared1}
ds^2 = -\frac{g_1 g^{-\frac{2}{3}}_{s} r^{2}_{h} }{u^2 L^2}dt^2 +  \frac{g^{-\frac{2}{3}}_{s} r^{2}_{h} }{u^2 L^2}\left(d{x}^{2}_{1}  +d{x}^2_{2}  +d{x}^{2}_{3} \right)+   \frac{g^{-\frac{2}{3}}_{s}L^2 }{u^2g_1}du^2.
\end{eqnarray}

We assume the perturbation of metric of $M3$-branes to be dependent on $x_1$ and $t$ only i.e after Fourier decomposing the same, we have $h_{\mu\nu}({\vec x},t)=\int \frac{d^{4}q}{(2\pi)^4 }e^{-iwt+iqx_1} h_{\mu\nu}(q,w)$ and choose the gauge where $h_{\mu u}=0$.  In case of $M3$-branes, there will be rotation group $SO(2)$ acting on the directions transverse to $u,t$, and $x_1$.  Based on the the spin of different metric perturbations under this group , the same  can be classified into groups as follows:
\begin{enumerate}
\item[(i)]
 vector modes: $h_{x_1 x_2}, h_{t x_2}\neq 0$ or $h_{x_1 x_3},  h_{t x_3}\neq 0$, with all other $h_{\mu \nu}=0$.
\item[(ii)]
Scalar modes: $h_{x_1 x_1}=h_{x_2 x_2}=h_{x_3 x_3}=h_{tt}\neq 0$, $h_{x_1 t}\neq 0$, with all other $h_{\mu \nu}=0$.
\item[(iii)]
Tensor modes: $h_{x_2 x_3}\neq 0$, with all other $h_{\mu \nu}=0$.
\end{enumerate}
 We are interested to calculate shear viscosity in the context of $M3$-brane by obtaining correlator functions corresponding to vector and tensor modes.

\subsubsection{Metric Vector Mode Fluctuations}

The vector mode fluctuations will be  given by considering non-zero $h_{t x_2}$ and $h_{x_1 x_2}$ components with all other $h_{\mu\nu}=0$ \cite{herzog}. Since the aforementioned metric is conformally flat near $u=0$, one can make a Fourier decomposition at large $r$ such that:
 \begin{eqnarray}
 && h^{x_2}_{t}= e^{-iwt+iqx_1} H_{t}(u),\nonumber\\
 && h^{x_2}_{x_1}=e^{-iwt+iqx_1} H_{x_1}(u).
 \end{eqnarray}
 Using the aforementioned in  equation (\ref{eq:R1munu}) and (\ref{eq:gamma1munu}), we get the following  linearized Einstein equation for $H_t$ and $H_{x_1}$.
 \begin{eqnarray}
 \label{eq:eomvec}
 && w_3 H^{\prime}_{t}+g_1 q_3 H^{\prime}_{x_1}=0,\nonumber\\
 &&H^{\prime\prime}_{x_1}-\frac{(4-\frac{5}{2} g_1)}{ug_1}H^{\prime}_{x_1} +\frac{1}{g^{2}_{1}}\left(w_3 q_3 H_t +w^{2}_{3} H_{x_1}\right)+\left( \frac{8}{u^2 g_1} - {\cal O}(1)\frac{g^{\frac{2}{3}}_{s} L^2 \Lambda}{u^2 g_1}\right)H_{x_1}=0,\nonumber\\
 && H^{\prime\prime}_{t}-\frac{3}{u}H^{\prime}_{t} -\frac{1}{g_1}\left(w_3 q_3 H_{x_1} +q^{2}_{3} H_t\right)+ \left(\frac{8}{u^2 g_1} -{\cal O}(1)\frac{g^{\frac{2}{3}}_{s} L^2 \Lambda}{u^2 g_1} \right)H_t=0.
 \end{eqnarray}
where $\Lambda$ is the cosmological constant arising from $\left|G_4\right|^2$ and higher order corrections (${\cal O}(R^4)$). It is shown in \cite{MQGP} that the higher order corrections are very subdominant as compared to flux term in both limits.

 The dominant flux  term as calculated in \cite{MQGP} is given by
\begin{equation}
\label{eq:G42}
\frac{\left|G^{2}_{4}\right|}{\sqrt{G^M}} \sim H^{2}_{\theta_1 \theta_2 \phi_2} G^{\cal M}\ ^{\theta_1\theta_1}G^{\cal M}\ ^{\theta_2 \theta_2}G^{\cal M}\ ^{\phi_2 \phi_2}G^{\cal M}\ ^{10\ 10},
 \end{equation}
and the simplified components are given as:
 \begin{eqnarray}
 && G^{\cal M}\ ^{\theta_1 \theta_1} \sim \frac{\sqrt[3]{3} \sqrt[3]{\frac{1}{ {g_s}}} \sqrt{ {g_s} N}}{\sqrt{\pi } N}\nonumber\\
&& G^{\cal M}\ ^{\theta_2\theta_2} \sim \frac{27 \sqrt[3]{3} \sqrt{ {g_s} N} \tan ^2( {\theta_1}) \csc ^2( {\theta_2})}{2 \sqrt{\pi } \sqrt[3]{ {g_s}} N  {f_2}( {\theta_2})^2}\nonumber\\
  &&
 G^{\cal M}\ ^{\phi_2 \phi_2}\sim\frac{6912 \sqrt[3]{3} \sin ^7( {\theta_1}) \cos ( {\theta_1}) \cot ^3( {\theta_2}) \csc ^4( {\theta_2})}{\sqrt{\pi }
   \left(\frac{1}{ {g_s}}\right)^{2/3} \sqrt{ {g_s} N} (\cos (2  {\theta_1})-5)^3} \nonumber\\
    && G^{\cal M}\ ^{10\ 10} \sim  3 \sqrt[3]{3} \left(\frac{1}{{g_s}}\right)^{4/3}\nonumber\\
   && H_{\theta_1\theta_2\phi_2} \sim \frac{\sqrt{\pi }  {f_2}( {\theta_2}) \sqrt{ {g_s} N} \sin ( {\theta_1}) \cos ( {\theta_1}) \sin ^3( {\theta_2}) \sin (2  {\theta_2}) \cos
   ( {\theta_2})}{ {3} \left(\sin ^2( {\theta_1}) \cos ^2( {\theta_2})+\cos ^2( {\theta_1}) \sin ^2( {\theta_2})\right)^2}.
   \end{eqnarray}
  Incorporating the aforementioned expressions in  (\ref{eq:G42}) and assuming that integrand receives dominant contribution along $\theta_{1,2}= 0,\pi$, we introduce a cut-off 
  $\theta_{1,2} \sim \epsilon^{\frac{15}{6}} $ in the MQGP limit of \cite{MQGP}.  Utilising this and integrating over rest of the 11-dimensional components,   \begin{eqnarray}
   && \Lambda= \int du d{\cal M}_{6}  \left|\frac{ G^{2}_{4} }{\sqrt{G^M}}\right|^2
   \sim c_1 \frac{2{g_s}^{\frac{2}{3}}\epsilon^{\frac{15}{3}}}{3 L^2},
   \end{eqnarray}
   where $c_1$ is constant.
   Analogous to a partial cancelation in \cite{herzog} of the coefficient of $H_x$ originating from $R_{\mu\nu}^{(1)}$ and the cosmological constant contribution from the $\Lambda h_{\mu\nu}$ term in (\ref{eq:einsteineq}), assuming that $ \frac{2 c_1\epsilon^{\frac{5}{3}}}{3}=8$ in weak $(g_s)$ coupling-string t'Hooft limit and $ \frac{2 \tilde{c}_1\epsilon^{\frac{15}{3}}}{3}=8$ in MQGP limit after incorporating value of $\Lambda$ in (\ref{eq:eomvec}),  we assume that the $\frac{1}{u^2g_1}$-term will be canceled out with term appearing in the coefficient of $H_{x_1,t}$ in $H_{x_1}(x_1,t)$'s EOM (\ref{eq:eomvec}). Hence the linearized set of EOM will be given as
    \begin{eqnarray}
 \label{eq:eomvec1}
 && w_3 H^{\prime}_{t}+g_1 q_3 H^{\prime}_{x_1}=0,\nonumber\\
 &&H^{\prime\prime}_{x_1}-\frac{(4-\frac{5}{2}g_1)}{ug_1}H^{\prime}_{x_1} +\frac{1}{g^{2}_{1}}\left(w_3 q_3 H_t +w^{2}_{3} H_{x_1}\right) =0,\nonumber\\
 && H^{\prime\prime}_{t}-\frac{3}{u}H^{\prime}_{t} -\frac{1}{g_1}\left(w_3 q_3 H_{x_1} +q^{2}_{3} H_t\right) =0.
 \end{eqnarray}
  To obtain two-point function at the asymptotic boundary, one needs to determine the kinetic term for vector modes $H_{x_1}$ and $H_{t}$. The same can be calculated using the  Einstein-Hilbert action up to quadratic order in $h_{\mu\nu}$ given as:
 \begin{eqnarray}
 &&{\hskip -0.3in}  S= \frac{1}{2 K^{2}_{11}}\int \sqrt{g} g^{\mu\nu}\left(\Gamma^{(1)\alpha}_{\mu\beta}\Gamma^{(1)\beta}_{\mu\nu}-\Gamma^{(1)\alpha}_{\mu\nu}\Gamma^{(1)\beta}_{\alpha\beta}\right)\nonumber\\
 &&
+  \frac{1}{2 \kappa_{11}^2}\int \sqrt{g}g^{\mu\nu}\left(\Gamma^{(2) \alpha}_{\mu\beta}\Gamma^{\beta}_{\nu\alpha}+\Gamma^{\alpha}_{\mu\beta}\Gamma^{(2)\beta}_{\nu\alpha}-\Gamma^{(2)\alpha}_{\mu\nu}\Gamma^{\beta}_{\alpha\beta}-\Gamma^{\alpha}_{\mu\nu}\Gamma^{(2)\beta}_{\alpha\beta}\right)\nonumber\\
 && + \frac{1}{16 \kappa_{11}^2}\int \sqrt{g}\left[8 h^{\mu}_{\sigma}h^{\sigma\nu}-4h h^{\mu\nu}+\left(h^2-2h_{\sigma\tau}h^{\sigma\tau}\right)g^{\mu\nu}\right]\left(\Gamma^{\alpha}_{\mu\beta}\Gamma^{\beta}_{\nu\alpha}-\Gamma^{\alpha}_{\mu\nu}\Gamma^{\beta}_{\alpha\beta}\right) \nonumber\\
 && +\frac{1}{4 \kappa_{11}^2}\int \sqrt{g}\left(g^{\mu\nu}-2h^{\mu\nu}\right)\left(\Gamma^{(1)\alpha}_{\mu\beta}\Gamma^{\beta}_{\nu\alpha}+\Gamma^{\alpha}_{\mu\beta}\Gamma^{(1)\beta}_{\nu\alpha}-\Gamma^{(1)\alpha}_{\mu\nu}\Gamma^{\beta}_{\alpha\beta}-\Gamma^{\alpha}_{\mu\nu}\Gamma^{(1)\beta}_{\alpha\beta}\right).
 \end{eqnarray}
 where
 \begin{eqnarray}
&& \Gamma^{(1)\alpha}_{\mu\nu}=\frac{1}{2}g^{\alpha\beta}\left(\nabla_{\nu}h_{\beta\mu}+\nabla_{\mu}h_{\beta\nu}-\nabla_{\beta}h_{\mu\nu}\right)\nonumber\\
&& \Gamma^{(2)\alpha}_{\mu\nu}= -g^{\alpha\beta}g^{\gamma\delta}h_{\beta\delta}\left(\nabla_{\mu}h_{\gamma\nu}+\nabla_{\nu}h_{\mu\gamma}-\nabla_{\gamma}h_{\mu\nu}\right).
\end{eqnarray}
The only  first two terms in the action will be relevant to get the kinetic term for vector modes. Solving the same, we get:
 \begin{eqnarray}
 S=\frac{1}{8 \kappa_{11}^2} \int d^{11}x {\sqrt{G^{\cal M}}}G^{{\cal M}uu} G^{\cal M}_{x_2 x_2}\left[-G^{{\cal M}x_1 x_1} \left(H^{\prime}_{x_1}\right)^2-G^{{\cal M}tt} \left(H^{\prime}_{t}\right)^2+...\right].
 \end{eqnarray}
 The very simplified form of 11-dimensional metric in $ \theta_i\rightarrow0 $ limit will be given by
  \begin{equation}
 \sqrt{G^{\cal M}}\sim \frac{g^{-\frac{8}{3}}_s L^2 r^{4}_{h}}{u^5 }\cot^3{\theta_2} \sin{\theta_2}f_2(\theta_2).
 \end{equation}
Using the fact that integrand possess maximum contribution along $\theta=0,\pi$, we assume that result of integration along $\theta_{1,2}$ will be given by sum of the contribution of integrand at $\theta_{1,2}=0,\pi$. In \cite{MQGP},  we have introduced a cut-off 
$\theta_{1,2}\sim \alpha_{\theta}\epsilon^{\frac{15}{6}}$ in `MQGP' limit where
$\epsilon\lesssim 1,\alpha_{\theta}<<1$ in the MQGP limit of \cite{MQGP}. Using the same and equation (\ref{ds5squared1}), the simplified action, in the MQGP limit, will be given as:
  \begin{equation}
 S\sim\epsilon^{-\frac{15}{2}}\frac{r^{4}_{h}}{K^{2}_{11}g^{2}_{s}}\int du \ d^{4}x\frac{1}{u^3}\left[\left( H^{\prime}_{t}\right)^2- g_1 \left(H^{\prime}_{x_1}\right)^2\right].
 \end{equation}
According to the Kubo's formula as mentioned in the beginning of  section {\bf 4},  shear viscosity is defined as
$
 \eta= -
 \lim_{w\rightarrow0}\left[\frac{1}{w} \Im m G_{x_1 x_2,x_1 x_2}\right].
$
In the $q_3\rightarrow0$ limit, the $H_t$ and $H_{x_1}$ decouple and the EOM for `$H_{x_1}$' becomes:
\begin{equation}
\label{Hx_eom-q3=0_i}
H_{x_1}^{\prime\prime}(u) - \frac{[4 - \frac{5}{2}(1 -u^4)]}{u(1-u^4)}H_{x_1}^\prime(u) + \left[\frac{\omega_3^2}{(1-u^4)^2}
+ \frac{\alpha}{u^2(1 - u^4)}\right]H_{x_1}(u)=0,
\end{equation}
where $u=1$ is thus seen to be a regular singular point with exponents of the corresponding indicial equation given by: $\pm\frac{iw_3}{4}$. Choosing the `incoming boundary condition' exponent, we will look for solutions of the form
$H_{x_1}(u)=(1-u)^{-\frac{iw_3}{4}}{\cal H}_{x_1}(u)$, ${\cal H}_{x_1}(u)$ being analytic in $u$. Assuming a perturbative ansatz for ${\cal H}_{x_1}(u): {\cal H}_{x_1}(u)= {\cal H}_{x_1}^{(0)}(u) + w_3 {\cal H}_{x_1}^{(1)}(u) + w_3^2{\cal H}_{x_1}^{(2)} + {\cal O}(w_3^3)$, we obtain:
\begin{eqnarray}
\label{Hx00alpha-i}
& {\hskip -0.2in} {\cal O}(w_3^0): & (u-1)^2 {\cal H}_{x_1}^{(0)\ \prime\prime}(u) + \frac{[4 - \frac{5}{2}(1 - u^4)](u-1)}{u(1+u)(1+u^2)}{\cal H}_{x_1}^{(0)\ \prime}(u) + \frac{\alpha(u - 1)}{u^2(u+1)(u^2+1)}{\cal H}_{x_1}^{(0)}(u)=0,\nonumber\\
& &
\end{eqnarray}
where  $\alpha\equiv 8 - {\cal O}(1)g_s^{-\frac{2}{3}}L^2\Lambda$. The solution to (\ref{Hx00alpha-i}) for arbitrary $\alpha$ is given as:
\begin{eqnarray}
\label{Hx00alpha-ii}
& & \hskip -0.2in {\cal H}_{x_1}^{(0)}(u,\alpha) = (-1)^{\frac{1}{16} \left(5-\sqrt{16 \alpha +25}\right)} u^{\frac{5}{4}-\frac{1}{4} \sqrt{16 \alpha +25}}\times  \nonumber\\
 & & \hskip -0.2in \left(\tilde{c}_1 \ _2F_1\left[\frac{1}{16}\left(5 - \sqrt{25 + 16 \alpha}\right), \frac{1}{16}\left(11 - \sqrt{25 + 16 \alpha}\right) ,1-\frac{1}{8} \sqrt{16 \alpha
   +25};u^4\right] \right.\nonumber\\
   &  & \hskip -0.2in  \left.(-1)^{\frac{1}{8} \sqrt{16 \alpha +25}} \tilde{c}_2 u^{\frac{1}{2} \sqrt{16 \alpha +25}}  \ _2F_1 \left[\frac{1}{16}\left(5 + \sqrt{25 + 16 \alpha}\right), \frac{1}{16}\left(11 + \sqrt{25 + 16 \alpha}\right),\frac{1}{8} \left(\sqrt{16 \alpha
   +25}+8\right);u^4\right]\right).\nonumber\\
& &
\end{eqnarray}
One hence notices that $H_{x_1}^{(0)}(u=0,\alpha)\neq0$ for $\alpha=0$ for which we will henceforth write:
\begin{equation}
\label{Hx00}
{\cal H}_{x_1}^{(0)}(u) = -\frac{2}{5} c_1 u^{5/2}+\frac{1}{4} i \pi  c_1+c_2 + {\cal O}(u^{\frac{13}{2}}).
\end{equation}
In terms of $\tilde{c}_{1,2}$ ${\cal H}_{x_1}^{(0)}(u)=\tilde{c}_1 + \tilde{c}_2e^{\frac{5i\pi}{8}} u^{\frac{5}{2}}\ _2F_1\left(\frac{5}{8},1,\frac{13}{8};u^4\right)$.

Assuming $\alpha=0$ henceforth, ${\cal H}^{(1)}_{x_1}(u)$ will be determined by the following differential equation:
\begin{eqnarray}
\label{Hx01_i}
& & (u - 1)^2{\cal H}^{(1) \prime\prime}_{x_1}(u) - \frac{i}{2}(u - 1){\cal H}^{(0)\ \prime}_{x_1}(u) + \frac{i}{4}{\cal H}^{(0)}_{x_1}(u) -\frac{i}{4}\frac{[4 - \frac{5}{2}(1 - u^4)]}{u(u + 1) (u^2 + 1)}{\cal H}^{(0)}_{x_1}(u) \nonumber\\
& & + \frac{i}{4}\frac{[4 - \frac{5}{2}(1 - u^4)](u - 1)}{u(u + 1) (u^2 + 1)}{\cal H}^{(1) \ \prime}_{x_1}(u) = 0.
\end{eqnarray}
Near $u=0$, (\ref{Hx01_i}) is solved to yield:
\begin{eqnarray}
\label{Hx01_ii}
{\cal H}_{x_1}^{(1)}(u) = c_4 + \frac{1}{16} \left(-4 c_2-i \pi  c_1\right) i u+\frac{2}{5} i e^{3/2} c_3 u^{5/2} + u^{7/2} \left(-\frac{3 c_1 i}{70}+\frac{6}{7} i e^{3/2} c_3\right) + {\cal O}(u^{\frac{9}{2}}).\nonumber\\
\end{eqnarray}
Imposing boundary condition on $H_{x_1}(u)$ at $u=0$, assuming $c_{1,2}>>c_4$, yields: $H_{x_1 0}\approx \frac{i \pi c_1}{4} + c_2$. Hence, setting $\kappa_{11}^2\sim{\cal O}(1)(g_sN)^{\frac{3}{4}}$, and defining $\cot^3\langle\theta_2\rangle \sin\langle\theta_2\rangle f_2(\langle\theta_2\rangle)|_{\theta_2\sim\epsilon^{\frac{3}{2}}}\sim{\cal O}(1)\epsilon^{-\beta} \equiv \epsilon^{-\frac{15}{2}}$ for the  the MQGP limit \cite{MQGP}:
\begin{eqnarray}
\label{eta}
& &{\hskip -0.4in}  \lim_{w\rightarrow0}\left[\lim_{q\rightarrow0}\frac{1}{w}\Im m G^R_{x_1 x_2,x_1 x_2}(w,q)\right]_{u=0}\nonumber\\
&&{\hskip -0.4in} \sim {\cal O}(1)T^3\frac{(4\pi)^2N^{\frac{5}{4}}}{{\cal O}(1) g_s^{\frac{3}{4}}}\epsilon^{-\beta^{(i)/(ii)}}\lim_{w\rightarrow0}\frac{1}{u^3}\Im m\left[\left(\frac{H_{x_1,q}(u)}{\frac{i \pi c_1}{4} + c_2}\right)\left(\frac{H_{x_1,-q}^\prime(u)}{\frac{i \pi c_1}{4} + c_2}\right)\right]_{u=0}\nonumber\\
& & {\hskip -0.4in} \sim \frac{(4\pi)^2{\cal O}(1)N^{\frac{5}{4}}}{{\cal O}(1)\pi g_s^{\frac{3}{4}}}\epsilon^{-\beta^{(i)/(ii)}}\Im m\left(\frac{
-\frac{1}{64} i \left(\pi  c_1-4 i c_2\right)^2}{\left[\frac{i \pi c_1}{4} + c_2\right]^2}\right)T^3=\frac{4\pi N^{\frac{5}{4}}{\cal O}(1)\epsilon^{-\beta^{(i)/(ii)}}}{{\cal O}(1) g_s^{\frac{3}{4}}}T^3.
\end{eqnarray}
Using $s = {\cal O}(1)r_h^3 = {\cal O}(1)(4\pi)^{\frac{3}{2}}(g_sN)^{\frac{3}{2}} T^3$ \cite{MQGP}, we obtain $\frac{\eta}{s}=\frac{{\cal O}(1)\epsilon^{-\beta^{(i)/(ii)}}}{\sqrt{4\pi}g_s^{\frac{9}{4}}N^{\frac{1}{4}}\left({\cal O}(1)\right)^2}$, which,
writing $g_s=\alpha_{g_s}^{(i)/(ii)}\epsilon, N=\alpha_N^{(i)/(ii)}\epsilon^{-19/-39}$ in the aforementioned two limits (i) and (ii), is $\frac{{\cal O}(1)\epsilon^{\frac{5}{3}}}{({\cal O}(1))^2\left(\alpha_{g_s}^{(i)}\right)^{\frac{9}{4}}\left(\alpha_N^{(i)}\right)^{\frac{1}{4}}\sqrt{4\pi}}$ in the limit of \cite{metrics}. If one assumes that the introduction of $M$ fractional $D3$-branes and $N_f$ flavour $D7$-branes does not have a significant effect on the 10D warp factor $h$, then in the limit of \cite{metrics} effected as the first limit of (\ref{limits_Dasguptaetal+MQGP}), one can show that $\epsilon$ can not be taken to be much smaller than around 0.01. One can choose appropriate $\alpha_{g_s,N}^{(i)}\sim\frac{1}{{\cal O}(1)}$ such that $\frac{{\cal O}(1)\epsilon^{\frac{5}{3}}}{({\cal O}(1))^2\left(\alpha_{g_s}^{(i)}\right)^{\frac{9}{4}}\left(\alpha_N^{(i)}\right)^{\frac{1}{4}}}=\frac{1}{\sqrt{4\pi}}$, implying one can generate $\frac{\eta}{s}=\frac{1}{4\pi}$.

In the more important MQGP limit, one obtains: $\frac{\eta}{s}=\frac{{\cal O}(1)\epsilon^{\frac{15}{3}}}{({\cal O}(1))^2\left(\alpha_{g_s}^{(ii)}\right)^{\frac{9}{4}}\left(\alpha_N^{(ii)}\right)^{\frac{1}{4}}}$. Now, in the MQGP limit, $\epsilon$ is less than but close to unity, hence yet again we can choose $\alpha_{g_s,N}^{(ii)}\sim\frac{1}{{\cal O}(1)}$ such that $\frac{{\cal O}(1)\epsilon^{\frac{15}{3}}}{({\cal O}(1))^2\left(\alpha_{g_s}^{(ii)}\right)^{\frac{9}{4}}\left(\alpha_N^{(ii)}\right)^{\frac{1}{4}}}=\frac{1}{\sqrt{4\pi}}$ hence implying $\frac{\eta}{s}=\frac{1}{4\pi}$.

\subsubsection{Metric Tensor Mode Fluctuations}
To obtain the correlations function corresponding to  tensor mode, we consider a fluctuation of $M3$-brane metric of the form $h_{x_2 x_3}\neq 0$ with all other $h_{\mu\nu}=0$ \cite{herzog}.  By Fourier decomposing the same,
 \begin{eqnarray}
 h^{x_3}_{x_2}(u,\vec{x}) = e^{ - i w t + i q x_1}\phi(u).
 \end{eqnarray}
Using equations (\ref{eq:einsteineq}) and (\ref {eq:R1munu}), the linearized Einstein EOM for $\phi(u)$ by  will be given as:
\begin{equation}
\label{phi}
\phi^{\prime\prime}(u) - \frac{(3 + u^2)}{u(1 - u^4)} \phi^\prime(u) + \frac{1}{(1 - u^4)}\biggl[w_3^2 - (1 - u^4) q_3^2\biggr]\phi(u) = 0.
\end{equation}
The horizon $u=1$ is a regular singular point and the roots of the indicial equation around this are $\pm\frac{i w_3}{4}$; choosing the incoming-wave boundary condition, $\phi(u) = (1 - u)^{- \frac{i w_3}{4}}\Phi(u)$
and writing a double perturbative ansatz:
\begin{equation}
\label{Phi-ii}
\Phi(u) = \Phi^{(0,0)}(u) + w_3 \Phi^{(0,1)}(u) + q_3^2 \Phi^{(2,0)}(u) + ...,
\end{equation}
the solutions  near $u=0$, up to ${\cal O}(u^4)$ are given below:
\begin{eqnarray}
\label{Phi-iv}
& & \Phi^{(0,0)}(u) = c_1+c_2-\frac{c_1 \tanh ^{-1}\left(\frac{1}{\sqrt{2}}\right)}{\sqrt{2}} -\frac{1}{4} c_1 u^4
+ {\cal O}(u^5);\nonumber\\
& & \Phi^{(0,1)}(u) = \left(c_4-\frac{2 e^3 c_3}{9}\right)+\frac{1}{8} i u \left(-2 c_1-2 c_2+\sqrt{2} c_1 \tanh ^{-1}\left(\frac{1}{\sqrt{2}}\right)\right)-\frac{1}{4} \left(e^3
   c_3\right) u^4 + {\cal O}\left(u^5\right);\nonumber\\
   & & \Phi^{(2,0)}(u) = \frac{1}{8} u^2 \left[c_1 \left(\sqrt{2} \tanh ^{-1}\left(\frac{1}{\sqrt{2}}\right)-2\right)-2 c_2\right]+u^3 \left[c_1+c_2-\frac{c_1 \tanh
   ^{-1}\left(\frac{1}{\sqrt{2}}\right)}{\sqrt{2}}\right]\nonumber\\
   & & +\frac{1}{64} u^4 \left[-108 \left\{c_1 \left(\sqrt{2} \tanh
   ^{-1}\left(\frac{1}{\sqrt{2}}\right)-2\right)-2 c_2\right\} \log (u)+342 c_2-16 e^3 c_3\right.\nonumber\\
   & & \left.-171 c_1 \left(\sqrt{2} \tanh
   ^{-1}\left(\frac{1}{\sqrt{2}}\right)-2\right)\right]  + {\cal O}\left(u^5\right).
\end{eqnarray}
Writing:
\begin{equation}
\label{phi_solution}
\phi(u) = ( 1 - u)^{- \frac{i w_3}{4}}\left[a_1 + f_1 u^4 + w_3\left(a_2 + b_2 u + f_2 u^4\right) + q_3^2\left(c_3 u^2 + d_3 u^3 + g_3 u^4 ln u\right)\right],
\end{equation}
the boundary condition yields: $\phi_0\approx a_1 + w_3 a_2\stackrel{c_{1,2}>>c_{3,4}}{\longrightarrow}a_1$. The kinetic term for $\phi$ is given by:
\begin{equation}
\label{kinetic-phi}
cot^3\langle\theta_2\rangle f_2(\langle\theta_2\rangle) sin\langle\theta_2\rangle\left(\frac{r_h^4}{g_s^2\kappa_{11}^2 }\right)\int du d^4x \frac{g_1(u) (\phi^\prime)^2}{u^3}.
\end{equation}
 Hence, using (\ref{kinetic-phi}) and Kubo's formula:
\begin{eqnarray}
\label{eta-ii}
&  \eta=&\lim_{w\rightarrow0}\frac{1}{w}\left(\lim_{q\rightarrow0}\Im m \left(G^R_{x_2 x_3,x_2 x_3}\right)\right)\nonumber\\
&& \sim\frac{(4\pi)^2{\cal O}(1)N^{\frac{5}{4}}}{{\cal O}(1) g_s^{\frac{3}{4}}}\epsilon^{-\beta^{(i)/(ii)}}\left.\lim_{w\rightarrow0}\frac{1}{w}\left(\lim_{q\rightarrow0}\frac{1}{u^3}\Im m \left(\phi_q(u)\phi^\prime_{-q}(u)\right)\right)\right|_{u=0}\nonumber\\
& & \sim T^3\frac{(4\pi)^2{\cal O}(1)N^{\frac{5}{4}}}{{\cal O}(1)\pi g_s^{\frac{3}{4}}}\epsilon^{-\beta^{(i)/(ii)}}\frac{4 \Im m (f_2)}{\pi a_1},
\end{eqnarray}
which utilizing $s = {\cal O}(1)r_h^3 = {\cal O}(1)(4\pi)^{\frac{3}{2}}(g_sN)^{\frac{3}{2}} T^3$ \cite{MQGP}, yields $\frac{\eta}{s}=-\frac{2{\cal O}(1)\epsilon^{-\beta^{(i)/(ii)}}}{\sqrt{\pi}g_s^{\frac{9}{4}}N^{\frac{1}{4}}\left({\cal O}(1)\right)^2}\frac{e^3 \Im m (c_3)}{\pi \phi_0}$. Therefore for $\Im m(c_3): -\frac{e^3{\cal O}(1)\Im m (c_3)\epsilon^{\frac{5}{3},\frac{15}{3}}}{({\cal O}(1))^2\left(\alpha_{g_s}^{(i),(ii)}\right)^{\frac{9}{4}}\left(\alpha_N^{(i),(ii)}\right)^{\frac{1}{4}}}=\frac{1}{4\sqrt{\pi}}$, one obtains: $\frac{\eta}{s}=\frac{1}{4\pi}$.

\section{Summary and Outlook}

Given the strong-coupling nature of QGP it is believed that the same will be better described in the limit of finite gauge coupling (or string coupling from the string theory dual perspective) \cite{finite_gYM_Natsuume}. With this as the basic motivation, the  black $M3$-branes in \cite{MQGP}  were obtained as a solution to the  M-theory uplift of resolved warped deformed conifold in the `delocalized' limit of \cite{becker2004} (in conformity with the non-locality of T duality transformations),  constructed by using  modified `OKS-BH' background  \cite{metrics} given in the context of Type IIB string theory involving $N$ D3-branes placed at the tip, $N_f$ $D7$-branes wrapped around a four-cycle in and $M$ $D5$-branes wrapping an $S^2$ inside a resolved warped deformed conifold, in particular in the MQGP limit discussed in \cite{MQGP}: {\it finite} $g_s$,
finite $g_sM, N_f, g_s^2MN_f$, very large $g_sN$,  and very small $\frac{g_sM^2}{N}$. Given the finite string coupling, such a limit could have been meaningfully discussed only in M theory, which is what we did in \cite{MQGP}.  The thermodynamical stability of the M-theory uplift in the same limit was demonstrated in \cite{MQGP} by showing positivity of specific heat. Also, it was shown in \cite{MQGP} that the black $M3$-branes' near-horizon geometry near the $\theta_{1,2} = 0$, branches, preserved $\frac{1}{8}$ supersymmetry.  By using the KSS prescription \cite{kouvtun_2003}, we had calculated in \cite{MQGP} the diffusion coefficient   to be $\frac{1}{T}$ in both type IIB and type IIA backgrounds, and the $\eta/s$ turned out to be $\frac{1}{4\pi}$ in the type IIB, Type IIA at finite string coupling (as part of the MQGP limit).

The flow-chart of our calculations and results are summarized in Fig.1.
After having explicitly shown that the uplift obtained in \cite{MQGP} in the `delocalized' limit of \cite{becker2004}, is a solution to the $D=11$ SUGRA EOMs in the MQGP limit, we have looked at the following two aspects at finite string coupling (and hence from an M-theory perspective) as part of the MQGP limit of \cite{MQGP}:
 \begin{itemize}
 \item
         {\bf Geometry of black $M3$-branes in the MQGP limit of \cite{MQGP}}: By evaluating the flux(/charge) of(/corresponding to) $G_{4}$   by integrating  over all  (non)compact four cycles,  the  black $M3$-branes,  asymptotically, were shown to be black $M5$-branes wrapping a two-cycle homologously given by (large)integer sum of two-spheres in $AdS_5\times M_6$. As shown in \cite{nonextremel_dasgupta}, the supersymmetry breaking measured by violation of the ISD condition of the flux $G_3$, is proportional to the square of the resolution parameter which in turn (turning off a bare resolution parameter, or assuming it be extremely small) goes like ${\cal O}\left(\frac{g_sM^2}{N}r_h^2\right)$; in the MQGP limit of \cite{MQGP}, we hence disregard the same. By comparing the non-K\"{a}hler resolved warped deformed conifold (NKRWDC) metric with the one of \cite{PT,Butti-et-al,Bena+Klebanov} and hence evaluating the five $SU(3)$ torsion($\tau$) classes $W_{1,2,3,4,5}$, we show that in the MQGP limit of \cite{MQGP}, for extremely large radial coordinates, the NKRWDC  (i)locally, preserves supersymmetry as $\tau\in W_4\oplus W_5: \frac{2}{3}\Re e W^{\bar{3}}_5 = W^{\bar{3}}_4$ (similar to Klebanov-Strassler solution \cite{Butti-et-al}),   and (ii) is asymptotically a Calabi-Yau as $W_{1,2,3,4,5}=0$ (as expected). Further,  to permit use of SYZ symmetry,  in addition to the large base of the $T^3$ used for triple T dualities in \cite{MQGP}, one requires this three-cycle to be  special Lagrangian. We  explicitly prove that the local three-torus $T^{3}$ of \cite{MQGP} satisfies the constraints satisfied by  the maximal $T^{2}$-invariant special Lagrangian submanifold of a deformed conifold of \cite{T2slag},  in the MQGP limit of \cite{MQGP}.

         \item
         {\bf Transport Coefficients of Black $M3$-Branes in MQGP Limit of \cite{MQGP}}:
         Exploiting the aforementioned asymptotic $AdS_5\times M_6$  background and  based on the prescription of \cite{son+starinets} , we have evaluated  at finite string coupling (as part of  the MQGP limit), different transport coefficients   by calculating fluctuations of  metric as well  as gauge field corresponding to $D7$-brane living on the world volume of non-compact four cycle in the (warped) deformed conifold and R-symmetry group present in the M-theory.  However, to do the same, we need to extract out the five dimensional AdS metric by integrating out all $r$-independent angular directions.  Going ahead, to calculate the two-point correlator/spectral  functions, we  evaluate the EOMs for $U(1)_{N_f}$-gauge field, $U(1)_{R}$ gauge field as well as vector and tensor modes and then evaluate the solutions by   double perturbative ansatze  up to ${\cal O}(w_3, q^{2}_3)$. The electrical conductivity, diffusion coefficient and charge susceptibility  obtained due to  $U(1)_{N_f}$ gauge field satisfy Einstein's relation, which is a reasonable check of our results. Similarly, we show that one can calculate  shear viscosity  from vector and tensor modes by using Kubo's formula  such that $\eta/s$ turns out\footnote{The entropy was calculated in \cite{MQGP} from the partition function of 11-dimensional M-theory background in the MQGP limit, or equivalently the horizon area.} to be  $\frac{1}{4\pi}$, which is expected for any theory obeying gauge-gravity correspondence.

\end{itemize}

 \begin{figure}
 \begin{center}
 \includegraphics[scale=0.8]
 {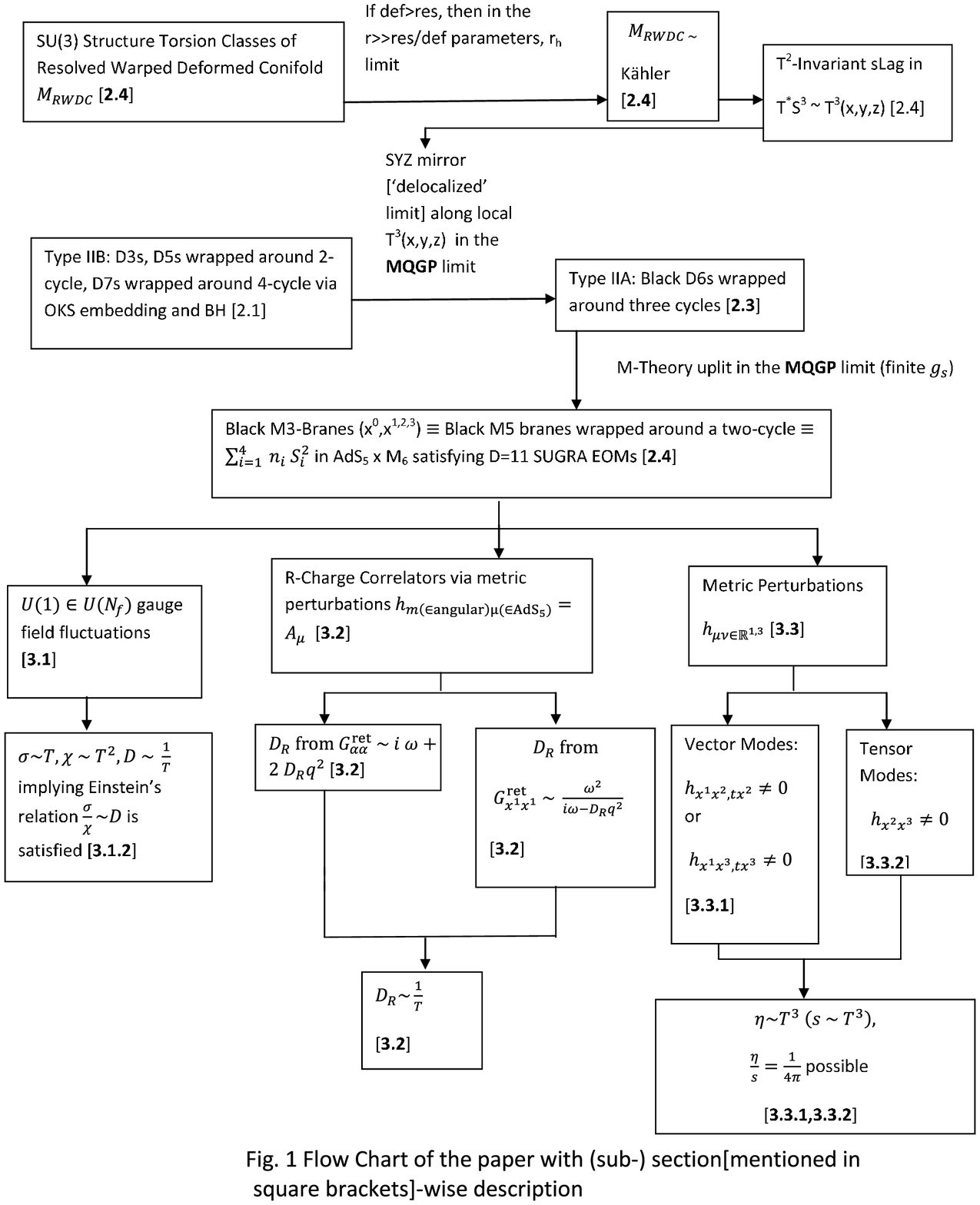}
 \end{center}
\end{figure}
Our results, we feel, are significant in the sense that the same have been obtained in the context of M-theory background \cite{MQGP} that are valid even for  {\it finite  coupling constant} $g_s$,  which in fact could be  more appealing to studying aspects of `strongly coupled Quark Gluon plasma plasma' and might bring one closer to the results obtained using experimental data at RHIC. We repeat, as we wrote in sec. {\bf 1}, we are not aware of previous attempts at evaluation of transport coefficients such as shear viscosity ($\eta$), diffusion constant ($D$), electrical conductivity ($\sigma$), charge susceptibility ($\chi$), etc. of large-$N$ thermal QCD-like theories at finite gauge coupling or equivalently finite string coupling, and hence correctly addressable only from M-theory perspective. It is for this reason that we feel that our results:  $\eta\sim T^3, D, D_R\sim1/T,  \chi\sim T^2$ (such that $D=\sigma/\chi$), etc. (apart from scalings w.r.t $N$, etc.) serve as M-theory predictions for sQGP.

 For the future, it would be interesting to extend these calculations to compute  ``second speed of sound" by working out two-point correlator functions of scalar modes of stress energy tensor.
 One can also calculate thermal conductivity corresponding to R-charge correlators and check if the ratio of thermal conductivity and viscosity satisfies Wiedemann-Franz law. Also, one should obtain the holographic spectral function by using non-abelian $SU(N_f)$ gauge field background and  produce the expected continuos meson spectra as a function of non-zero chemical potential due  to presence of black hole (BH) in the  background of \cite{MQGP}.

\section*{Acknowledgments}
One of us (MD) would like to thank CSIR, Govt. of India for a senior research fellowship when she was at IIT Roorkee. One of us (AM) would like to thank Northeastern University (specially P. Nath), Department of Mathematical Sciences at the University of Liverpool (specially T. Mohaupt), Swansea University (specially C. Nunez), Neils Bohr Institute (specially N. Obers and J. Hartnong) and the CERN theory group for a warm hospitality and partial support, where part of this work was done. AM would like to thank K. Dasgupta and P.Ouyang for useful clarifications, N. Obers and P. Kumar for interesting discussions and specially C. Nunez for enlightening conversations/discussions. AM was also partly supported by the CSIR, Govt. of India, project number CSR-656-PHY.

\appendix

\section{$G_4$ components }
\setcounter{equation}{0} \seceqaa

We now list the non-zero four-flux components $G_4$ in the limit of \cite{metrics} and disregarding the asymmetry w.r.t. $\phi_1$ and $\phi_2$ which can be eliminated by symmetrizing $A^{IIB}_1$ w.r.t. the same:
  \begin{eqnarray}
\label{G_4-Dasgupta_et_al-limit}
&&(i) \ G_{r\theta_1\phi_2\theta_2}\sim \left(\frac{81 {g_s} M {N_f} {\sin\ \phi_2} \cot \left(\frac{{\theta_1}}{2}\right) \sin ({\theta_2})
   \sqrt[4]{\frac{{g_s} N}{r^4}}}{4 \sqrt{2} \pi ^{5/4} \sqrt{{g_s} N}}+\frac{3 \sqrt{3} {N_f} {\sin\ \phi_1} r \sin ({\theta_1})
   \sqrt[4]{\frac{{g_s} N}{r^4}}}{\sqrt{2} \pi  \sqrt[4]{{g_s} N}}\right) \times\nonumber\\
   && \frac{1}{96 \pi ^{3/2}
   r^2 \sqrt{\frac{{g_s} N}{r^4}} \left(2 \cot ^2({\theta_1})+2 \cot ^2({\theta_2})+3\right)^2}\times\Biggl({g_s} \sin ({\theta_2}) \Bigl(64 \pi ^2 N \cos ({\theta_1}) \cot ({\theta_2})
   {f_1}'({\theta_1}) \nonumber\\
   &&  \bigl(2 \cot ^2({\theta_1})+2 \cot ^2({\theta_2})+3\bigr)+3 {g_s}^2 {M^{2}_{\rm eff}} {N_f}
   {f_1}({\theta_1}) \log (r) \cos ({\theta_1}) \cot \left(\frac{{\theta_1}}{2}\right) \cot ({\theta_2})\nonumber\\
   && \left(2 \cot ^2({\theta_1})+2
   \cot ^2({\theta_2})+3\right)-64 \pi ^2 N {f_1}({\theta_1}) \sin ({\theta_1}) \cot ({\theta_2}) \left(2 \cot ^2({\theta_1})+2 \cot
   ^2({\theta_2})+3\right)\nonumber\\
   &&+256 \pi ^2 N {f_1}({\theta_1}) \cot ^2({\theta_1}) \csc ({\theta_1}) \cot ({\theta_2})\Bigr)\Biggr),\nonumber\\
    && \Rightarrow\lim_{\epsilon_{\theta_2}\rightarrow0}\int_{\epsilon_{\theta_2}}^{\pi-\epsilon_{\theta_2}} G_{r\theta_1\theta_2\phi_2}d\theta_2 = 0;\nonumber\\
     &&  (ii) \ G_{r\theta_2\phi_2\phi_1}\sim \frac{1}{3} \sqrt{\pi } r^2 \sin ({\theta_1}) \sin ({\theta_2}) \sqrt{\frac{{g_s} N}{r^4}} \left(\frac{2 \sqrt[4]{\pi } {f_1}({\theta_1})
   \cos ({\theta_1}) \cot ({\theta_2}) \sqrt[4]{\frac{{g_s} N}{r^4}}}{\sqrt{3} \left(2 \cot ^2({\theta_1})+2 \cot
   ^2({\theta_2})+3\right)}\right.\nonumber\\
   &&\left.-\frac{2 \sqrt[4]{\pi } {g_s} N {f_1}({\theta_1}) \cos ({\theta_1}) \cot ({\theta_2})}{\sqrt{3} r^4
   \left(\frac{{g_s} N}{r^4}\right)^{3/4} \left(2 \cot ^2({\theta_1})+2 \cot ^2({\theta_2})+3\right)}\right) \times \nonumber\\
   &&\left(\frac{\sqrt{\frac{3}{2}}
   {N_f} {\sin\ \phi_2} \sin ^2({\theta_2}) \left(8 \cos ^3({\theta_1}) \cos ({\theta_2}) \cot ({\theta_2})-16 \cos ^4({\theta_1}) \cot
   ({\theta_2})\right)}{\pi  (\cos (2 {\theta_1})-5) \left(2 \sin ^2({\theta_1}) \cos ^2({\theta_2})+2 \cos ^2({\theta_1}) \sin
   ^2({\theta_2})\right)}\right.\nonumber\\
   &&\left.-\frac{\sqrt{\frac{2}{3}} {N_f} 2\sin\left(\frac{\psi}{2}\right)  \csc ({\theta_2}) \left(-2 \cos ({\theta_1}) \cos ({\theta_2})+\sin
   ^2({\theta_1})-\cos ^2({\theta_1})+5\right)}{\pi  (\cos (2 {\theta_1})-5)}\right),\nonumber\\
   & & \Rightarrow \int_0^{2\pi}G_{r\theta_2\phi_2\phi_1}d\phi_2\sim \frac{N_f {\cal F}(\langle\theta_1\rangle,\theta_2) }{r}\sin\left(\frac{\langle\psi\rangle}{2}\right)\ {\rm which\ we\ disregard\ in\ large}\ r\ {\rm limit};\nonumber\\
   &&(iii) \ G_{r\theta_2\psi\theta_1}\sim \frac{1}{2560 \sqrt{2} \pi ^{11/4} N^{3/4} r^2}\left({g_s}^{5/4} M {N_f} {\theta_1} \cot \left(\frac{{\theta_1}}{2}\right) \left(9 \log (r) \Bigl(768 \pi ^2 a^2 N\right.\right.\nonumber\\
   && \left.\left.+{g_s}^2 {M^{2}_{\rm eff}}
   {N_f} r \log \left(\sin \left(\frac{{\theta_1}}{2}\right) \sin \left(\frac{{\theta_2}}{2}\right)\right)+4 \pi  {g_s} {M^{2}_{\rm eff}}
   r\Bigr)\right.\right.\nonumber\\
   &&\left.\left. +2268 a^2 {g_s}^2 {M^{2}_{\rm eff}} {N_f} \log ^2(r) \log \left(\sin \left(\frac{{\theta_1}}{2}\right) \sin
   \left(\frac{{\theta_2}}{2}\right)\right)+32 \pi ^2 N r\right)\right),\nonumber\\
   & & \Rightarrow\lim_{\epsilon_{\theta_{1,2}}\rightarrow0}\int_{\epsilon_{\theta_1}}^{\pi - \epsilon_{\theta_1}}d\theta_1\int_{\epsilon_{\theta_2}}^{\pi - \epsilon_{\theta_2}}d\theta_2 G_{r\theta_2\psi\theta_1}\sim  \epsilon^{-5}\frac{N_f}{r}\ {\rm which\ we\ disregard\ in\ large}\ r\ {\rm limit};\nonumber
   \end{eqnarray}
   \begin{eqnarray}
   &&(iv) \  G_{r \theta_2\psi\phi_1}\sim-\left(\frac{81 {g_s}
   M {N_f} {\sin\ \phi_2} \cot \left(\frac{{\theta_1}}{2}\right) \sin ({\theta_2}) \sqrt[4]{\frac{{g_s} N}{r^4}}}{4 \sqrt{2} \pi ^{5/4}
   \sqrt{{g_s} N}}+\frac{3 \sqrt{3} {N_f} {\sin\ \phi_1} r \sin ({\theta_1}) \sqrt[4]{\frac{{g_s} N}{r^4}}}{\sqrt{2} \pi  \sqrt[4]{{g_s}
   N}}\right)\times\nonumber\\
   && \frac{1}{27 \left(3 \sin ^2({\theta_1}) \sin ^2({\theta_2})+2 \sin ^2({\theta_1}) \cos ^2({\theta_2})+2 \cos ^2({\theta_1}) \sin
   ^2({\theta_2})\right)^2}\times\nonumber\\
   && \left(\sqrt{\frac{2 \pi }{3}} r^2 \sin ^2({\theta_1}) \sin ^3({\theta_2}) \sqrt{\frac{{g_s} N}{r^4}} \left(4 \cos ^3({\theta_1}) \left(6
   \sqrt{6}-4 \cot ({\theta_1})\right) \cot ({\theta_2}) \csc ({\theta_2})\right.\right.\nonumber\\
   && \left.\left. -12 \sin ({\theta_1}) \cos ^2({\theta_1}) \sin (2 {\theta_2})
   \csc ^3({\theta_2})+36 \sqrt{6} \sin ^2({\theta_1}) \cos ({\theta_1}) \cot ({\theta_2}) \csc ({\theta_2})\right)\right.\Biggr),
   \nonumber\\
   & & \Rightarrow \lim_{\epsilon_{\theta_2}\rightarrow0}\int_{\epsilon_{\theta_2}}^{\pi - \epsilon_{\theta_2}}d\theta_2\int_0^{2\pi}d\phi_1 G_{r \theta_2\psi\phi_1} = 0;\nonumber\\
   && (v) \ G_{r\phi_2\psi\theta_1}\sim - \left(\frac{81 {g_s} M {N_f} {\sin\ \phi_2} \cot \left(\frac{{\theta_1}}{2}\right) \sin ({\theta_2})
   \sqrt[4]{\frac{{g_s} N}{r^4}}}{4 \sqrt{2} \pi ^{5/4} \sqrt{{g_s} N}}+\frac{3 \sqrt{3} {N_f} {\sin\ \phi_1} r \sin ({\theta_1})
   \sqrt[4]{\frac{{g_s} N}{r^4}}}{\sqrt{2} \pi  \sqrt[4]{{g_s} N}}\right)\times\nonumber\\
   && \left(\frac{\sqrt{\pi } r^2 \sin ^3({\theta_2}) \sqrt{\frac{{g_s} N}{r^4}} (4 \cos (2 ({\theta_1}-{\theta_2}))-4 \cos (2
   ({\theta_1}+{\theta_2})))}{12 \left(3 \sin ^2({\theta_1}) \sin ^2({\theta_2})+2 \sin
   ^2({\theta_1}) \cos ^2({\theta_2})+2 \cos ^2({\theta_1}) \sin ^2({\theta_2})\right)^2}\right),\nonumber\\
  & & \Rightarrow\lim_{\epsilon_{\theta_1}\rightarrow0}\int_{\epsilon_{\theta_1}}^{\pi-\epsilon_{\theta_1}}d\theta_1\int_0^{2\pi} d\phi_1
  G_{r\phi_2\psi\theta_1}=0;\nonumber\\
   && (vi) G_{\theta_1\phi_2\psi\theta_2}\sim -\frac{\pi ^{3/4} \sqrt[4]{N} r^2 {\theta_1} \sin ^3({\theta_2}) \sqrt{\frac{{g_s} N}{r^4}} (4 \cos (2 ({\theta_1}-{\theta_2}))-4 \cos (2
   ({\theta_1}+{\theta_2})))}{360 \sqrt{2} {g_s}^{3/4} \left(3 \sin ^2({\theta_1}) \sin ^2({\theta_2})+2 \sin ^2({\theta_1}) \cos
   ^2({\theta_2})+2 \cos ^2({\theta_1}) \sin ^2({\theta_2})\right)^2},\nonumber\\
   & &\lim_{\epsilon_{\theta_2}\rightarrow0}\int_{\epsilon_{\theta_2}}^{\pi-\epsilon_{\theta_2}}d\theta_1G_{\theta_1\phi_2\psi\theta_2}=0;\nonumber\\
   && (vii) \ G_{\theta_2\phi_2\psi\phi_1} \sim\left(\frac{\sqrt{\frac{3}{2}} {N_f} {\sin\ \phi_2} \sin
   ^2({\theta_2}) \left(8 \cos ^3({\theta_1}) \cos ({\theta_2}) \cot ({\theta_2})-16 \cos ^4({\theta_1}) \cot ({\theta_2})\right)}{\pi
   (\cos (2 {\theta_1})-5) \left(2 \sin ^2({\theta_1}) \cos ^2({\theta_2})+2 \cos ^2({\theta_1}) \sin
   ^2({\theta_2})\right)}\right.\nonumber\\
   && -\left. \frac{\sqrt{\frac{2}{3}} {N_f} 2\sin\left(\frac{\psi}{2}\right)  \csc ({\theta_2}) \left(-2 \cos ({\theta_1}) \cos ({\theta_2})+\sin
   ^2({\theta_1})-\cos ^2({\theta_1})+5\right)}{\pi  (\cos (2 {\theta_1})-5)}\right)\times\nonumber\\
   && {\hskip -0.2in} \left(\frac{\pi ^{3/4} r^3 \sin ^3({\theta_1}) \sin ({\theta_2}) \left(\frac{{g_s} N}{r^4}\right)^{3/4} (3 \cos (2 ({\theta_1}-{\theta_2}))+3
   \cos (2 ({\theta_1}+{\theta_2}))+2 \cos (2 {\theta_1})-14 \cos (2 {\theta_2})+6) }{12 \sqrt{3} \left(3 \sin ^2({\theta_1}) \sin
   ^2({\theta_2})+2 \sin ^2({\theta_1}) \cos ^2({\theta_2})+2 \cos ^2({\theta_1}) \sin ^2({\theta_2})\right)^2}\right),\nonumber\\
   & & \Rightarrow\int_0^{2\pi}d\phi_2\int_0^{4\pi}d\psi G_{\theta_2\phi_2\psi\phi_1}=0;\nonumber
   \end{eqnarray}
   \begin{eqnarray}
   && (viii)\ G_{\theta_1\theta_2\phi_2\phi_1}\sim \left(\frac{\sqrt{\frac{3}{2}} {N_f} {\sin\ \phi_2} \sin ^2({\theta_2}) \left(8 \cos
   ^3({\theta_1}) \cos ({\theta_2}) \cot ({\theta_2})-16 \cos ^4({\theta_1}) \cot ({\theta_2})\right)}{\pi  (\cos (2 {\theta_1})-5)
   \left(2 \sin ^2({\theta_1}) \cos ^2({\theta_2})+2 \cos ^2({\theta_1}) \sin ^2({\theta_2})\right)}\right.\nonumber\\
   && \left. -\frac{\sqrt{\frac{2}{3}} {N_f} 2\sin\left(\frac{\psi}{2}\right)
   \csc ({\theta_2}) \left(-2 \cos ({\theta_1}) \cos ({\theta_2})+\sin ^2({\theta_1})-\cos ^2({\theta_1})+5\right)}{\pi  (\cos (2
   {\theta_1})-5)}\right)\times \nonumber\\
   && \frac{1}{96 \sqrt{3} \pi ^{5/4} r \sqrt[4]{\frac{{g_s}
   N}{r^4}} \left(2 \cot ^2({\theta_1})+2 \cot ^2({\theta_2})+3\right)^2}\Biggl({g_s} \sin ({\theta_1}) \sin ({\theta_2} \Bigl(64 \pi ^2 N \cos ({\theta_1}) \cot ({\theta_2}) {f_1}'({\theta_1})\nonumber\\
   &&  \left(2 \cot ^2({\theta_1})+2 \cot
   ^2({\theta_2})+3\right)+3 {g_s}^2 {M^{2}_{\rm eff}} {N_f} {f_1}({\theta_1}) \log (r) \cos ({\theta_1}) \cot
   \left(\frac{{\theta_1}}{2}\right) \cot ({\theta_2}) \nonumber\\
   && \left(2 \cot ^2({\theta_1})+2 \cot ^2({\theta_2})+3\right)-64 \pi ^2 N
   {f_1}({\theta_1}) \sin ({\theta_1}) \cot ({\theta_2}) \left(2 \cot ^2({\theta_1})+2 \cot ^2({\theta_2})+3\right)\nonumber\\
   && +256 \pi ^2 N
   {f_1}({\theta_1}) \cot ^2({\theta_1}) \csc ({\theta_1}) \cot ({\theta_2})\Bigr)\Biggr),\nonumber\\
   & & \Rightarrow\lim_{\epsilon_{\theta_{1,2}}\rightarrow0}\int_{\epsilon_{\theta_1}}^{\pi-\epsilon_{\theta_1}}d\theta_1 \int_{\epsilon_{\theta_2}}^{\pi-\epsilon_{\theta_2}}d\theta_2\int_0^{2\pi}d\phi_2G_{\theta_1\theta_2\phi_2\phi_1}\sim  \epsilon^{-14}N_f\sin\left(\frac{\langle\psi\rangle}{2}\right); \nonumber\\
   &&(ix) \ G_{r \theta_1\theta_2 11}  \sim\frac{2 {g_s} M}{r^3} \Biggl(2 r^2 \log (r) \Bigl(\frac{2 {f_2}({\theta_2}) \cot ({\theta_1}) \left(2 \cot ^2({\theta_1})+3\right) \sin
   ^2({\theta_2}) \cos ({\theta_2})}{\left(2 \cot ^2({\theta_1})+3\right) \sin ^2({\theta_2})+2 \cos
   ^2({\theta_2})}\nonumber\\
   && -\frac{{f_1}({\theta_1}) (\cos (2 {\theta_1})-5) \cot ({\theta_1}) \csc ({\theta_1}) \cot ({\theta_2})}{2 \cot
   ^2({\theta_1})+2 \cot ^2({\theta_2})+3}\Bigr)\Biggr)\nonumber\\
   && +\frac{2 {g_s} M}{r^3}\Biggl(\sin ({\theta_2}) \Bigl(\frac{{f_1}({\theta_1}) \left(r^2+2 r^2 \log (r)\right) (\cos (2
   {\theta_1})-5) \cot ({\theta_1}) \csc ({\theta_1}) \cot ({\theta_2}) \csc ({\theta_2})}{2 \cot ^2({\theta_1})+2 \cot
   ^2({\theta_2})+3}\nonumber\\
   && +\frac{2 {f_2}({\theta_2}) \left(r^2+2 r^2 \log (r)\right) \cot ({\theta_1}) \left(-2 \cot ^2({\theta_1})-3\right) \sin
   ({\theta_2}) \cos ({\theta_2})}{\left(2 \cot ^2({\theta_1})+3\right) \sin ^2({\theta_2})+2 \cos ^2({\theta_2})}\Bigr)\Biggr),\nonumber\\
   & & \Rightarrow\lim_{\epsilon_{\theta_{1,2}}\rightarrow0}\int_{\epsilon_{\theta_1}}^{\pi-\epsilon_{\theta_1}}d\theta_1\int_{\epsilon_{\theta_2}}^{\pi-\epsilon_{\theta_2}}d\theta_2 G_{r \theta_1\theta_2 11}=0;\nonumber\\
   && (x) \ G_{r\theta_1\phi_1 11} \sim\frac{{g_s} {M^{2}_{\rm eff}} r {f_1}({\theta_1}) \sin ^4({\theta_1}) (\cos (2 {\theta_2})-5) \csc ^2({\theta_2}) \sqrt{\frac{{g_s}
   N}{r^4}}}{16 \sqrt{\pi } N \left(\sin ^2({\theta_1}) \left(2 \cot ^2({\theta_2})+3\right)+2 \cos ^2({\theta_1})\right)},\nonumber\\
   & & \Rightarrow\lim_{\epsilon_{\theta_1}\rightarrow0}\int_{\epsilon_{\theta_1}}^{\pi-\epsilon_{\theta_1}}d\theta_1G_{r\theta_1\phi_1 11}=0;\nonumber\\
   &&(xi)\ G_{r\theta_1\psi 11}\sim \frac{1}{256 \pi ^3 N r^2}\times \nonumber\\
   && \Biggl(3 {g_s}^2 M {N_f} \cot \left(\frac{{\theta_1}}{2}\right) \Bigl(-36 \log ^2(r) \left(96 \pi ^2 a^2 N-63 a^2 {g_s}^2 {M^{2}_{\rm eff}}
   {N_f} \log \left(\sin \left(\frac{{\theta_1}}{2}\right) \sin \left(\frac{{\theta_2}}{2}\right)\right)\right)\nonumber\\
   && +9 \log (r) \left(768 \pi ^2 a^2
   N+{g_s}^2 {M^{2}_{\rm eff}} {N_f} r \log \left(\sin \left(\frac{{\theta_1}}{2}\right) \sin \left(\frac{{\theta_2}}{2}\right)\right)+4 \pi
   {g_s} {M^{2}_{\rm eff}} r\right)+32 \pi ^2 N r\Bigr)\Biggr)\nonumber\\
   & & \sim \frac{1}{r}\left(\frac{g_s^2MN_f}{N}{\cal F}_3(\theta_{1,2})\right)\nonumber
   \end{eqnarray}
   \begin{eqnarray}
   && (xii) \ G_{r\theta_2\psi 11}\sim -\frac{1}{512 \pi ^3 N r^2 (\cos (2 {\theta_1})-5)}\Biggl(3 {g_s}^2 M {N_f} \csc \left(\frac{{\theta_2}}{2}\right) \Bigl(2 \cos \left({\theta_1}-\frac{3 {\theta_2}}{2}\right)\nonumber\\
   &&+2 \cos
   \left({\theta_1}-\frac{{\theta_2}}{2}\right) +\cos \left(2 {\theta_1}-\frac{{\theta_2}}{2}\right)+2 \cos
   \left({\theta_1}+\frac{{\theta_2}}{2}\right)+\cos \left(\frac{1}{2} (4 {\theta_1}+{\theta_2})\right)+2 \cos \left({\theta_1}+\frac{3
   {\theta_2}}{2}\right)\nonumber\\
   && -10 \cos \left(\frac{{\theta_2}}{2}\right)\Bigr) \Biggl(-36 \log ^2(r) \Bigl(a^2 \left(-63 {g_s}^2 {M^{2}_{\rm eff}}
   {N_f}-84 \pi  {g_s} {M^{2}_{\rm eff}}+32 \pi ^2 N\right)\nonumber\\
   && -21 a^2 {g_s}^2 {M^{2}_{\rm eff}} {N_f} \log \left(\sin
   \left(\frac{{\theta_1}}{2}\right) \sin \left(\frac{{\theta_2}}{2}\right)\right)-{g_s}^2 {M^{2}_{\rm eff}} {N_f} r\Bigr)+9 \log (r) \Bigl(256
   \pi ^2 a^2 N\nonumber\\
   && +{g_s}^2 {M^{2}_{\rm eff}} {N_f} r \log \left(\sin \left(\frac{{\theta_1}}{2}\right) \sin
   \left(\frac{{\theta_2}}{2}\right)\right)+4 \pi  {g_s} {M^{2}_{\rm eff}} r\Bigr)-2592 a^2 {g_s}^2 {M^{2}_{\rm eff}} {N_f} \log ^4(r)\nonumber\\
   && -1728 \pi
   a^2 {g_s} {M^{2}_{\rm eff}} \log ^3(r)+32 \pi ^2 N r\Biggr)\Biggr)\sim \frac{1}{r}\left(\frac{g_s^2MN_f}{N}{\cal F}_1(\theta_{1,2}) + g_s^2MN_f{\cal F}_2(\theta_{1,2})\right)\sim0;\nonumber\\
     &&(xiii)  \ G_{\theta_1 \phi_2\psi 11}\sim -\frac{\sqrt{\pi } r^2 \sin ^3({\theta_2}) \sqrt{\frac{{g_s} N}{r^4}} (4 \cos (2 ({\theta_1}-{\theta_2}))-4 \cos (2
   ({\theta_1}+{\theta_2})))}{12 \left(3 \sin ^2({\theta_1}) \sin ^2({\theta_2})+2 \sin ^2({\theta_1}) \cos ^2({\theta_2})+2 \cos
   ^2({\theta_1}) \sin ^2({\theta_2})\right)^2},\nonumber\\
   & & \Rightarrow\lim_{\epsilon_{\theta_1}\rightarrow0}\int_{\epsilon_{\theta_1}}^{\pi-\epsilon_{\theta_1}}d\theta_1G_{\theta_1 \phi_2\psi 11}=0;\nonumber\\
   &&(xiv) \ G_{\theta_2 \phi_2\psi 11}\sim \frac{1}{12 \left(3 \sin ^2({\theta_1}) \sin ^2({\theta_2})+2 \sin
   ^2({\theta_1}) \cos ^2({\theta_2})+2 \cos ^2({\theta_1}) \sin ^2({\theta_2})\right)^2}\times \nonumber\\
   && {\hskip -0.2in} \Biggl(\sqrt{\pi } r^2 \sin ^2({\theta_1}) \sin ({\theta_2}) \sqrt{\frac{{g_s} N}{r^4}} (3 \cos (2 ({\theta_1}-{\theta_2}))+3 \cos (2
   ({\theta_1}+{\theta_2}))+2 \cos (2 {\theta_1})-14 \cos (2 {\theta_2})+6)\Biggr),\nonumber\\
   & & \Rightarrow\lim_{\epsilon_{\theta_2}\rightarrow0}\left.\int_{\epsilon_{\theta_2}}^{\pi-\epsilon_{\theta_2}}d\theta_2 G_{\theta_2 \phi_2\psi 11}\right|_{\theta_1\sim\epsilon_{\theta_1}}   \sim \epsilon^{-\frac{17}{2}}; \nonumber\\
   &&(xiv) \ G_{\theta_1 \phi_1\psi 11}\sim \frac{1}{27 (\cos (2
   {\theta_1})-5)^2 \left(3 \sin ^2({\theta_1}) \sin ^2({\theta_2})+2 \sin ^2({\theta_1}) \cos ^2({\theta_2})+2 \cos ^2({\theta_1}) \sin
   ^2({\theta_2})\right)^2}\times \nonumber\\
   && \Biggl(2 \sqrt{\frac{2 \pi }{3}} r^2 \sin ^3({\theta_1}) \sin ({\theta_2}) \sqrt{\frac{{g_s} N}{r^4}} \Biggl(81 \left(6 \sqrt{6} \sin
   ^6({\theta_1}) \sin ^3({\theta_2})+4 \sqrt{6} \sin ^6({\theta_1}) \sin ({\theta_2}) \cos ^2({\theta_2})\right)\nonumber\\
   && +3 \sin ^3({\theta_1})
   \cos ^3({\theta_1}) (112 \cos (2 {\theta_2})-14 \cos (4 {\theta_2})+126) \csc ({\theta_2})+54 \sin ^4({\theta_1}) \cos ^2({\theta_1})
   \Bigl(15 \sqrt{6} \sin ^3({\theta_2})\nonumber\\
   && -8 \sqrt{6} \cos ^3({\theta_2}) \cot ({\theta_2})+2 \sqrt{6} \sin ({\theta_2}) \cos
   ^2({\theta_2})\Bigr)-8 \cos ^6({\theta_1}) \Bigl(-4 \cos ^3({\theta_2})\nonumber\\
   &&  \left(4 \cot ({\theta_1}) \cot ({\theta_2})-3 \sqrt{6} \cot
   ({\theta_2})\right)+9 \left(2 \sqrt{6} \cot ^2({\theta_1})+5 \sqrt{6}\right) \sin ^3({\theta_2})\nonumber\\
   && +6 \left(-2 \sqrt{6} \cot ^2({\theta_1})-4
   \cot ({\theta_1})+\sqrt{6}\right) \sin ({\theta_2}) \cos ^2({\theta_2})\Bigr)+32 \sin ({\theta_1}) \cos ^5({\theta_1}) \Bigl(16 \cos
   ^3({\theta_2}) \cot ({\theta_2})\nonumber\\
   && +24 \sin ({\theta_2}) \cos ^2({\theta_2})\Bigr)-12 \sin ^2({\theta_1}) \cos ^4({\theta_1}) \Bigl(-9
   \sqrt{6} \sin ^3({\theta_2})+44 \sqrt{6} \cos ^3({\theta_2}) \cot ({\theta_2})\nonumber\\
   && +30 \sqrt{6} \sin ({\theta_2}) \cos ^2({\theta_2})\Bigr)+9
   \sin ^5({\theta_1}) \cos ({\theta_1}) (16 \cos (2 {\theta_2})-2 \cos (4 {\theta_2})+18) \csc ({\theta_2})\Biggr)\Biggr),\nonumber\\
   & & \Rightarrow\lim_{\epsilon_{\theta_1}\rightarrow0}\left.\int_{\epsilon_{\theta_1}}^{\pi-\epsilon_{\theta_1}}d\theta_1 G_{\theta_2 \phi_2\psi 11}\right|_{\theta_2\sim\epsilon_{\theta_1}}  \sim \epsilon^{-8};  \nonumber
   \end{eqnarray}
   \begin{eqnarray}
   && (xv) \ G_{\theta_2 \phi_1\psi 11}\sim -\frac{1}{27 \left(3 \sin
   ^2({\theta_1}) \sin ^2({\theta_2})+2 \sin ^2({\theta_1}) \cos ^2({\theta_2})+2 \cos ^2({\theta_1}) \sin ^2({\theta_2})\right)^2}\times \nonumber\\
   && \Biggl(\sqrt{\frac{2 \pi }{3}} r^2 \sin ^2({\theta_1}) \sin ^3({\theta_2}) \sqrt{\frac{{g_s} N}{r^4}} \Bigl(4 \cos ^3({\theta_1}) \left(6
   \sqrt{6}-4 \cot ({\theta_1})\right) \cot ({\theta_2}) \csc ({\theta_2})\nonumber\\
   && -12 \sin ({\theta_1}) \cos ^2({\theta_1}) \sin (2 {\theta_2})
   \csc ^3({\theta_2})+36 \sqrt{6} \sin ^2({\theta_1}) \cos ({\theta_1}) \cot ({\theta_2}) \csc ({\theta_2})\Bigr)\Biggr),\nonumber\\
   & & \Rightarrow\lim_{\epsilon_{\theta_2}\rightarrow0}\int_{\epsilon_{\theta_2}}^{\pi-\epsilon_{\theta_2}}d\theta_2
   G_{\theta_2 \phi_1\psi 11} = 0; \nonumber\\
   &&(xvi)\  G_{\theta_2\theta_1 \psi 11}\sim -\frac{9 {g_s}^4 M {M^{2}_{\rm eff}} {N_f}^2 \log ^2(r) \cot \left(\frac{{\theta_1}}{2}\right) \cot \left(\frac{{\theta_2}}{2}\right)}{512 \pi ^3
   N}\nonumber\\
   && \sim(g_s^2MN_f)(g_sN_f)\left(\frac{g_sM^2}{N}\right){\cal F}_4(\theta_{1,2})\sim0;\nonumber\\
   && (xvi)\ G_{\theta_1\theta_2\phi_2 11}\sim \frac{1}{96 \pi ^{3/2} r^2 \sqrt{\frac{{g_s} N}{r^4}}
   \left(2 \cot ^2({\theta_1})+2 \cot ^2({\theta_2})+3\right)^2}\Biggl({g_s} \sin ({\theta_2})\nonumber\\
   && \Biggl(64 \pi ^2 N \cos ({\theta_1}) \cot ({\theta_2}) {f_1}'({\theta_1}) \left(2 \cot ^2({\theta_1})+2
   \cot ^2({\theta_2})+3\right)+3 {g_s}^2 {M^{2}_{\rm eff}} {N_f} {f_1}({\theta_1}) \log (r) \cos ({\theta_1}) \nonumber\\
   && \cot
   \left(\frac{{\theta_1}}{2}\right) \cot ({\theta_2}) \left(2 \cot ^2({\theta_1})+2 \cot ^2({\theta_2})+3\right)-64 \pi ^2 N
   {f_1}({\theta_1}) \sin ({\theta_1}) \cot ({\theta_2}) \nonumber\\
   &&\left(2 \cot ^2({\theta_1})+2 \cot ^2({\theta_2})+3\right) +256 \pi ^2 N
   {f_1}({\theta_1}) \cot ^2({\theta_1}) \csc ({\theta_1}) \cot ({\theta_2})\Biggr)\Biggr),\nonumber\\
& & \Rightarrow\lim_{\epsilon_{\theta_2}\rightarrow0}\int_{\epsilon_{\theta_2}}^{\pi-\epsilon_{\theta_2}}d\theta_2
G_{\theta_1\theta_2\phi_2 11} = 0.
    \end{eqnarray}

\section{$SU(3)$ torsion classes of Resolved Warped Deformed Conifolds}
\setcounter{equation}{0}\seceqbb

Based on the results of \cite{Butti-et-al} and \cite{Bena+Klebanov}, the five $SU(3)$ structure torsion classes for the resolved warped deformed conifold are given as follows:
\begin{eqnarray}
\label{W_1}
& & \hskip -1.2in W_1 = \frac{1}{6} e^{-g(\tau)-3 p-\frac{3 x(\tau)}{2}}  \big(-{\cal B}(\tau)+ {a(\tau)}^2 {\cal B}(\tau)-2 {a(\tau)} {\cal A}(\tau) e^{g(\tau)} -{\cal B}(\tau) e^{2
{g(\tau)}}- 2 {a(\tau)}{\cal A}(\tau) e^{6 {p(\tau)}+2x(\tau)}  \nonumber \\
& &\hskip -1.2in -  2{\cal B}(\tau) e^{{g(\tau)}+6{p(\tau)}+2x(\tau)} -
2 {a^\prime(\tau)}e^{6{p(\tau)}+2x(\tau)} +  2 e^{{g(\tau)}+6{p(\tau)}+2x(\tau)} ({\cal B}(\tau) {\cal A}^\prime(\tau)  -  {\cal A}(\tau) {\cal B}^\prime(\tau) )\big)\nonumber\\
& & \hskip -1.2in \sim \frac{e^{-3\tau}}{\sqrt{4\pi g_s N}}<<1\ ({\rm in\ the\ UV});\nonumber\\
& &\hskip -1.2in W_2 = -\frac{2}{3}  e^{-{g(\tau)}-3  {p(\tau)}-\frac{3  x(\tau)}{2} }  G_5 \wedge G_6    \
\big(- {\cal B}(\tau)+ {a(\tau)}^2  {\cal B}(\tau) - 2 {a(\tau)}  {\cal A}(\tau)  e^{g(\tau)}
-  {\cal B}(\tau) e^{2{g(\tau)}}+    {a(\tau)}  {\cal A}(\tau)  e^{6  {p(\tau)}+2  x(\tau)} \nonumber \\
& &\hskip -1.2in + {\cal B}(\tau)  e^{{g(\tau)}+6  {p(\tau)}+2  x(\tau)}+  e^{6  {p(\tau)}+ 2  x(\tau)} {a^\prime(\tau)}
- e^{{g(\tau)}+6  {p(\tau)}+2  x(\tau)} ({\cal B}(\tau) {\cal A}^\prime(\tau) - {\cal A}(\tau) {\cal B}^\prime(\tau)) \big) \nonumber \\
& &\hskip -1.2in +  e^{-{g(\tau)}+3{p(\tau)}+\frac{x(\tau)}{2}}  (G_2 \wedge G_3  -  G_1 \wedge G_4 )
\big( {a(\tau)}  {\cal B}(\tau) +  {\cal A}(\tau)  {\cal B}(\tau) {a^\prime(\tau)} +   {\cal B}(\tau)^2 e^{g(\tau)} {g^\prime(\tau)} \big) \nonumber \\
& &\hskip -1.2in - \frac{1}{3}  e^{-{g(\tau)}-3  {p(\tau)}- \frac{3  x(\tau)}{2}}  G_1 \wedge G_2
\big( {\cal B}(\tau)-  {a(\tau)}^2  {\cal B}(\tau)+ 2 {a(\tau)}  {\cal A}(\tau)  e^{g(\tau)} +  {\cal B}(\tau)  e^{2  {g(\tau)}}+ 2  {a(\tau)}  {\cal A}(\tau)  e^{6  {p(\tau)}+2 x(\tau)}
\nonumber \\
& &\hskip -1.2in - {\cal B}(\tau)  e^{{g(\tau)}+6  {p(\tau)}+2  x(\tau)} + (3 {\cal A}(\tau)^2 -1)   e^{6  p + 2  x(\tau)} {a^\prime(\tau)}  +
e^{{g(\tau)}+6  {p(\tau)}+2  x(\tau)} ({\cal B}(\tau) {\cal A}^\prime(\tau) - {\cal A}(\tau) {\cal B}^\prime(\tau) + 3 {\cal A}(\tau) {\cal B}(\tau) {g^\prime(\tau)}) \big) \nonumber \\
& &\hskip -1.2in +  \frac{1}{3}  e^{-{g(\tau)}-3  {p(\tau)}- \frac{3  x(\tau)}{2}}  G_3 \wedge G_4
\big(- {\cal B}(\tau) + {a(\tau)}^2  {\cal B}(\tau) - 2 {a(\tau)}  {\cal A}(\tau)  e^{g(\tau)} -  {\cal B}(\tau)  e^{2  {g(\tau)}}+ 4  {a(\tau)}  {\cal A}(\tau)  e^{6  {p(\tau)}+2 x(\tau)}
\nonumber \\
& &\hskip -1.2in + {\cal B}(\tau)  e^{{g(\tau)}+6  {p(\tau)}+2  x(\tau)} + (3 {\cal A}(\tau)^2 + 1)   e^{6  p + 2  x(\tau)} {a^\prime(\tau)}  +
e^{{g(\tau)}+6  {p(\tau)}+2  x(\tau)} (- {\cal B}(\tau) {\cal A}^\prime(\tau) + {\cal A}(\tau) {\cal B}^\prime(\tau) + 3 {\cal A}(\tau) {\cal B}(\tau) {g^\prime(\tau)}) \big)\nonumber\\
& & \hskip -1.2in \left(4\pi g_sN\right)^{\frac{1}{4}}e^{-3\tau}\left(d\tau\wedge e_\psi + e_1\wedge e_2 + \epsilon_1\wedge e_2\right)<<1\ ({\rm in\ the\ UV});\nonumber\\
& &\hskip -1.2in W_3 =   -\frac{1}{4}  e^{-{g(\tau)}- 3  {p(\tau)}- \frac{3  x(\tau)}{2}} ( G_1 \wedge G_3
-  G_2 \wedge G_4 ) \wedge  G_6
\big(-{\cal B}(\tau)+ {a(\tau)}^2  {\cal B}(\tau) - 2  {a(\tau)}  {\cal A}(\tau)  e^{g(\tau)} - {\cal B}(\tau)  e^{2  {g(\tau)}} \nonumber \\
& &\hskip -1.2in +
 2  {a(\tau)}  {\cal A}(\tau)  e^{6  {p(\tau)}+2  x(\tau)}  +
 2  {\cal B}(\tau)  e^{{g(\tau)}+6  {p(\tau)}+2  x } - 2  e^{6  {p(\tau)}+2 x(\tau)}{a^\prime(\tau)}  +   2  e^{{g(\tau)}+6  {p(\tau)}+2  x(\tau)} ({\cal B}(\tau) {\cal A}^\prime(\tau) - {\cal A}(\tau)
 {\cal B}^\prime(\tau))  \big)  \nonumber \\
& &\hskip -1.2in +
\frac{{\cal A}(\tau)}{2}e^{-{g(\tau)}-3  {p(\tau)}-\frac{3  x(\tau)}{2}} ( G_1 \wedge G_2  -  G_3 \wedge G_4 )
\wedge G_5  \nonumber\\
   & & \hskip -1.2in \times \big(-1+{{a(\tau)}^2}+e^{2  {g(\tau)}}-2  {\cal B}(\tau)  e^{6  {p(\tau)}+2  x(\tau)}{a^\prime(\tau)} +  2  {\cal A}(\tau)
 e^{{g(\tau)}+6  {p(\tau)}+2  x(\tau)} {g^\prime(\tau)} \big) \nonumber \\
& &\hskip -1.2in +
\frac{1}{4}  e^{-{g(\tau)}-3  {p(\tau)}-\frac{3  x(\tau)}{2}}  G_2 \wedge G_3 \wedge G_5
 \big({\cal B}(\tau)- {{a(\tau)}^2}  {\cal B}(\tau)-2  {a(\tau)}  {\cal A}(\tau)  e^{g(\tau)}-3  {\cal B}(\tau)  e^{2  {g(\tau)}}+ 2  {a(\tau)}
{\cal A}(\tau)  e^{6  {p(\tau)}+2  x(\tau)}  \nonumber \\
 & &\hskip -1.2in +   2  {\cal B}(\tau)  e^{{g(\tau)}+6  {p(\tau)}+2  x(\tau)}+ 2 ({\cal B}(\tau)^2 -{\cal A}(\tau)^2)  e^{6  {p(\tau)}+2 x(\tau)} {a^\prime(\tau)}
 + 2    e^{{g(\tau)}+6  {p(\tau)}+2  x(\tau)} ({\cal B}(\tau) {\cal A}^\prime(\tau) - {\cal A}(\tau) {\cal B}^\prime(\tau) - 2 {\cal A}(\tau) {\cal B}(\tau) {g^\prime(\tau)})\big) \nonumber \\
& & \hskip -1.2in+
\frac{1}{4}  e^{-{g(\tau)}-3  {p(\tau)}-\frac{3  x(\tau)}{2}}  G_1 \wedge G_4 \wedge G_5
 \big(-3{\cal B}(\tau) +3 {{a(\tau)}^2}  {\cal B}(\tau) - 2  {a(\tau)}  {\cal A}(\tau)  e^{g(\tau)} +   {\cal B}(\tau)  e^{2  {g(\tau)}}+ 2  {a(\tau)}
{\cal A}(\tau)  e^{6  {p(\tau)}+2  x(\tau)}  \nonumber \\
& &\hskip -1.2in + 2  {\cal B}(\tau)  e^{{g(\tau)}+6  {p(\tau)}+2  x(\tau)} - 2 ( 1 + 2{\cal B}(\tau)^2)  e^{6  {p(\tau)}+2 x(\tau)} {a^\prime(\tau)}
 + 2    e^{{g(\tau)}+6  {p(\tau)}+2  x(\tau)} ({\cal B}(\tau) {\cal A}^\prime(\tau) - {\cal A}(\tau) {\cal B}^\prime(\tau) + 2 {\cal A}(\tau) {\cal B}(\tau) {g^\prime(\tau)}) \big);
\nonumber\\
& & \hskip -1.2in \left.\sqrt{4\pi g_sN}\left(32\sqrt{\frac{2}{3}}e^{-3\tau}\left(e_1\wedge \epsilon_1 + e_2\wedge \epsilon_2\right)\wedge e_\psi + 2\sqrt{\frac{2}{3}}e_2\wedge \epsilon_1\wedge d\tau e^{-\tau} + 32e_1\wedge \epsilon_2\wedge d\tau e^{-3\tau} \right)\right|_{\theta_1\sim0;\ {\rm UV}}<<1;\nonumber\\
& & \hskip -1.2in W_4  =  \frac{1}{2}  e^{-{g(\tau)}-3  {p(\tau)}-\frac{3  x(\tau)}{2}}  G_5
 \big( - {\cal A}(\tau)+ {{a(\tau)}^2}  {\cal A}(\tau)+2  {a(\tau)}  {\cal B}(\tau)  e^{g(\tau)} - {\cal A}(\tau)  e^{2 {g(\tau)}}+
2 e^{{g(\tau)}+6  {p(\tau)}+2  x(\tau)}  x^\prime(\tau) \big) \nonumber\\
& & \hskip -1.2in \sim - \frac{2}{3}e^{-g(\tau)}d\tau = 2 W^3_4 = 2 W^{\bar{3}}_4;\nonumber\\
& & \hskip -1.2in W_5^{(\bar 3)} = \frac{1}{4}  e^{-{g(\tau)}+3{p(\tau)}+\frac{x(\tau)}{2}} (G_5-i G_6)
\big(2  {a(\tau)}  {\cal B}(\tau)  - 2 {\cal A}(\tau)  e^{{g(\tau)}} - 6  e^{{g(\tau)}} {p^\prime(\tau)} + e^{{g(\tau)}} x^\prime(\tau) \big)\nonumber\\
& & \hskip -1.2in \sim -\frac{1}{2}\left(d\tau - i e_\psi\right).
\end{eqnarray}
From (\ref{W_1}), one sees that:
\begin{equation}
\label{SUSY}
W^{\bar{3}}_4 = \frac{2}{3}\Re e W^{\bar{3}}_5,
\end{equation}
matching with column B of Table 2 of \cite{Butti-et-al}.

\section{Embedding of $T^2$-Invariant sLag in $T^* S^3$}
\setcounter{equation}{0}\seceqcc

Based on \cite{T2slag}, the explicit embedding of the maximal $T^2$-invariant special Lagrangian three-cycle in a deformed conifold is given by the following:
\begin{eqnarray}
\label{theta1-embedding_i}
& &  r = \left(\frac{c_1}{cos(\phi_1+\phi_2)} + \frac{c_2}{cos(\phi_1-\phi_2)}\right)^{\frac{7}{3}};  {\rm In\ large}\ r\ {\rm limit}:\ c_1=c_2\sim (r^{<}_\Lambda)^{\frac{7}{3}},c_3\sim (r^{<}_\Lambda)^6\ \{{\rm e.g.}r^{<}_{\Lambda}\sim r_\Lambda^{\frac{1}{4}}\}:-\nonumber\\
& &\cos\theta_1=\nonumber\\
& & \frac{1}{8} \Biggl\{\sec^2 (\psi ) \sec ({\phi_1}-{\phi_2}) \Biggl[32 \cos ({\phi_1}-{\phi_2}-\psi )-8 \cos ({\phi_1}+{\phi_2}-\psi )+32 \cos ({\phi_1}-{\phi_2}+\psi
   )\nonumber\\
   & &-8 \cos ({\phi_1}+{\phi_2}+\psi ) +\Biggl\{\left(-32 \cos ({\phi_1}-{\phi_2}-\psi )+8 \cos ({\phi_1}+{\phi_2}-\psi )-32 \cos ({\phi_1}-{\phi_2}+\psi )+ \right.\nonumber\\
   & & \left. 8 \cos
   ({\phi_1}+{\phi_2}+\psi )+3\ 2^{3/7} \cos (2 {\phi_1}) \sec ^{\frac{3}{7}}({\phi_1}+{\phi_2})+3\ 2^{3/7} \cos (2 {\phi_2}) \sec
   ^{\frac{3}{7}}({\phi_1}+{\phi_2})\right.\nonumber\\
   & &  \left.+2^{3/7} \cos (2 (2 {\phi_1}+{\phi_2})) \sec ^{\frac{3}{7}}({\phi_1}+{\phi_2})+2^{3/7} \cos (2 ({\phi_1}+2 {\phi_2})) \sec
   ^{\frac{3}{7}}({\phi_1}+{\phi_2})\right)^2\nonumber\\
   & &  +64 \cos (\psi ) \cos ({\phi_1}-{\phi_2}) \left(-32 \cos ({\phi_1}-{\phi_2}-\psi )+16 \cos ({\phi_1}+{\phi_2}-\psi
   )-32 \cos ({\phi_1}-{\phi_2}+\psi )\right.\nonumber\\
   & & \left.+16 \cos ({\phi_1}+{\phi_2}+\psi )+3\ 2^{3/7} \cos (2 {\phi_1}) \sec ^{\frac{3}{7}}({\phi_1}+{\phi_2})+3\ 2^{3/7} \cos (2
   {\phi_2}) \sec ^{\frac{3}{7}}({\phi_1}+{\phi_2})\right.\nonumber\\
   & & \left.+2^{3/7} \cos (2 (2 {\phi_1}+{\phi_2})) \sec ^{\frac{3}{7}}({\phi_1}+{\phi_2})+2^{3/7} \cos (2 ({\phi_1}+2
   {\phi_2})) \sec ^{\frac{3}{7}}({\phi_1}+{\phi_2})\right)\Biggr\}^{\frac{1}{2}}\nonumber\\
   & & -3\ 2^{3/7} \cos (2 {\phi_1}) \sec ^{\frac{3}{7}}({\phi_1}+{\phi_2})-3\ 2^{3/7} \cos (2 {\phi_2}) \sec
   ^{\frac{3}{7}}({\phi_1}+{\phi_2})\nonumber\\
   & &-2^{3/7} \cos (2 (2 {\phi_1}+{\phi_2})) \sec ^{\frac{3}{7}}({\phi_1}+{\phi_2}) -2^{3/7} \cos (2 ({\phi_1}+2 {\phi_2})) \sec
   ^{\frac{3}{7}}({\phi_1}+{\phi_2})\Biggr]\Biggr\}^{\frac{1}{2}};
   \end{eqnarray}
\begin{eqnarray}
\label{theta1-embedding_ii}
   & &  \cos\theta_2=\nonumber\\
   & &  \frac{1}{8} \Biggl\{\sec^2 (\psi ) (-\sec ({\phi_1}-{\phi_2})) \Biggl[-32 \cos ({\phi_1}-{\phi_2}-\psi )+8 \cos ({\phi_1}+{\phi_2}-\psi )-32 \cos
   ({\phi_1}-{\phi_2}+\psi )\nonumber\\
   & & +8 \cos ({\phi_1}+{\phi_2}+\psi )+\Biggl\{\left(-32 \cos ({\phi_1}-{\phi_2}-\psi )+8 \cos ({\phi_1}+{\phi_2}-\psi )-32 \cos
   ({\phi_1}-{\phi_2}+\psi )\right.\nonumber\\
   & &   \left.+8 \cos ({\phi_1}+{\phi_2}+\psi )+3\ 2^{3/7} \cos (2 {\phi_1}) \sec ^{\frac{3}{7}}({\phi_1}+{\phi_2})
    +3\ 2^{3/7} \cos (2 {\phi_2}) \sec^{\frac{3}{7}}({\phi_1}+{\phi_2})\right.\nonumber\\
    & &   \left.+2^{3/7} \cos (2 (2 {\phi_1}+{\phi_2})) \sec ^{\frac{3}{7}}({\phi_1}+{\phi_2})+2^{3/7} \cos (2 ({\phi_1}+2 {\phi_2})) \sec
   ^{\frac{3}{7}}({\phi_1}+{\phi_2})\right)^2\nonumber\\
   & &  \left.+64 \cos (\psi ) \cos ({\phi_1}-{\phi_2}) \left(-32 \cos ({\phi_1}-{\phi_2}-\psi )+16 \cos ({\phi_1}+{\phi_2}-\psi)\right.-32 \cos ({\phi_1}-{\phi_2}+\psi )\right.\nonumber\\
   & &  \left.+16 \cos ({\phi_1}+{\phi_2}+\psi )+3\ 2^{3/7} \cos (2 {\phi_1}) \sec ^{\frac{3}{7}}({\phi_1}+{\phi_2})+3\ 2^{3/7} \cos (2
   {\phi_2}) \sec ^{\frac{3}{7}}({\phi_1}+{\phi_2})\right.\nonumber\\
   & &  \left.+2^{3/7} \cos (2 (2 {\phi_1}+{\phi_2})) \sec ^{\frac{3}{7}}({\phi_1}+{\phi_2})+2^{3/7} \cos (2 ({\phi_1}+2
   {\phi_2})) \sec ^{\frac{3}{7}}({\phi_1}+{\phi_2})\right)\Biggr\}^{\frac{1}{2}}\nonumber\\
   & &   +3\ 2^{3/7} \cos (2 {\phi_1}) \sec ^{\frac{3}{7}}({\phi_1}+{\phi_2})+3\ 2^{3/7} \cos (2 {\phi_2}) \sec
   ^{\frac{3}{7}}({\phi_1}+{\phi_2})\nonumber\\
   & &+2^{3/7} \cos (2 (2 {\phi_1}+{\phi_2})) \sec ^{\frac{3}{7}}({\phi_1}+{\phi_2}) +2^{3/7} \cos (2 ({\phi_1}+2 {\phi_2})) \sec
   ^{\frac{3}{7}}({\phi_1}+{\phi_2})\Biggr]\Biggr\}^{\frac{1}{2}}.
   \end{eqnarray}
From \cite{Gwyn+Knauf}, the K\"{a}hler form $J$ and the nowhere-vanishing holomorphic three form $\Omega$ for a deformed conifold are given by:
\begin{eqnarray}
   \label{J}
& & \hskip -1.3in J = -\frac{r^6\hat\gamma'+\mu^4\hat\gamma-r^2\mu^4\hat\gamma'}{2r^5\sqrt{1-\mu^4/r^4}}\, dr\wedge (d\psi+\cos\theta_1\, d\phi_1+\cos\theta_2\ d\phi_2)\nonumber\\
& & \hskip -1.3in
   +\,\frac{\hat\gamma}{4}\,\sqrt{1-\frac{\mu^4}{r^4}}\,(\sin\theta_1\,d\theta_1\wedge d\phi_1 +
\sin\theta_2\,d\theta_2\wedge d\phi_2)\,,
\end{eqnarray}
and
\begin{eqnarray}
\label{Omega}
& &  \Omega = 2{\cal T}\,\frac{{\cal S}\cos\psi-\imath\sin\psi}{r{\cal S}}\,dr\wedge(\sin\theta_1\,d\theta_2\wedge d\phi_1
  - \sin\theta_2\,d\theta_1\wedge d\phi_2) \nonumber\\
& &  + \frac{2\imath{\cal T}}{r{\cal S}}\,(\cos\psi-\imath{\cal S}\sin\psi)\,dr\wedge(d\theta_1\wedge d\theta_2
  -\sin\theta_1\sin\theta_2\,d\phi_1\wedge d\phi_2)
  \nonumber\\
& & -\frac{2\mu^2{\cal T}}{r^3{\cal S}}\,dr\wedge (\sin\theta_1\,d\theta_1\wedge d\phi_1-\sin\theta_2\,
  d\theta_2\wedge d\phi_2)\nonumber\\
& &  +  \frac{\imath\mu^2{\cal T}}{r^2}\big[\sin\theta_1\,d\theta_1\wedge d\phi_1\wedge(d\psi+\cos\theta_2\,d\phi_2)
  - \sin\theta_2\,d\theta_2\wedge d\phi_2\wedge(d\psi+\cos\theta_1\,d\phi_1)\Big] \nonumber\\
& & {\hskip -0.2in} +  {\cal T}(\imath\cos\psi+{\cal S}\sin\psi)\Big[\sin\theta_2\,d\theta_1\wedge d\phi_2\wedge
  (d\psi+\cos\theta_1\,d\phi_1)-\sin\theta_1\,d\theta_2\wedge d\phi_1\wedge(d\psi+\cos\theta_2\,d\phi_2)\Big] \nonumber\\
& &  {\hskip -0.2in} +  {\cal T}({\cal S}\cos\psi-\imath\sin\psi)\Big[d\theta_1\wedge d\theta_2\wedge
  (d\psi+\cos\theta_1\,d\phi_1+\cos\theta_2\,d\phi_2)-\sin\theta_1\sin\theta_2\,d\phi_1\wedge d\phi_2\wedge d\psi\Big],\nonumber\\
\end{eqnarray}
where ${\cal S}=\sqrt{1-\mu^4/r^4}\stackrel{r>>1}{\longrightarrow}1$ and ${\cal T}=\hat\gamma\sqrt{\hat\gamma+(r^2\hat\gamma'-\hat\gamma)(1-\mu^4/r^4)}/8\stackrel{r>>1}{\longrightarrow}r^2$ as
$\hat{\gamma}\stackrel{r>>1}{\longleftarrow}r^{\frac{4}{3}}$, and the deformation parameter $\mu$ is defined via $z_1^2+z_2^2+z_3^2+z_4^2=\mu^2$.

\section{Some Intermediate Steps Pertaining to $R$-Charge Gauge Field EOM's Solution}
\setcounter{equation}{0}\seceqdd

Plugging (\ref{cal-A_alpha_soln-i}) into (\ref{cal-A_alpha-eom-i}) yields for $\tilde{\cal A}(\tau)_\alpha^{(0,0),(9,1),(2,0)}(u)$ relevant to the $\tilde{\cal A}(\tau)_\alpha(u)$'s EOM, near $u=0$ up to ${\cal O}(u)$:
\begin{eqnarray}
\label{cal-A_alpha_soln_iii}
& & (1 - u)^{-\frac{i w_3}{4} - 1}\tilde{\cal A}(\tau)_\alpha^{(0,0)}(u) = c_2 + c_2\left(1+ \frac{i }{4}w_3\right) u +
 {\cal O}(u^2);\nonumber\\
& & (1 - u)^{-\frac{i w_3}{4} - 1}\tilde{\cal A}(\tau)_\alpha^{(0,1)}(u) = \left(-\frac{1}{64} i \pi ^2 c_1-\frac{3 \pi  c_2}{16}+\frac{1}{4} i \pi  c_3+c_4\right)\nonumber\\
& & {\hskip -0.2in} +\frac{1}{256} \Biggl[\pi ^2 (w_3{w_3}-4 i) c_1-4 i \pi  (w_3{w_3}-4 i)
   \left(3 c_2-4 i c_3\right)+64 i \biggl((w_3{w_3}-4 i) c_4-c_2\biggr)\Biggr] u+O\left(u^2\right);\nonumber\\
   & & (1 - u)^{-\frac{i w_3}{4}}\tilde{\cal A}(\tau)_\alpha^{(2,0)}(u) = \frac{1}{24} \Biggl[3 \biggl(2 {Li}_3(-i)+2 {Li}_3(i)+3 \zeta (3)\biggr) c_1-i \pi ^3 c_1+\pi ^2 c_2+6 i \pi  c_3+24 c_4\Biggr]\nonumber\\
    & & +   \frac{1}{96} {w_3} \Biggl[\pi ^3 c_1+i \pi ^2 c_2-6 \pi  c_3+3 i \biggl(\left(2 {Li}_3(-i)+2 {Li}_3(i)+3 \zeta (3)\right) c_1+8 c_4\biggr)\Biggr]u + {\cal O}(u^2)
\end{eqnarray}

One can show that $\tilde{\cal A}(\tau)_{x_1}^{(0,1)\ \prime}(u)$, up to ${\cal O}(u)$, will be given by the expansion of the following up to ${\cal O}(u)$:
\begin{eqnarray}
\label{Ax'01_soln-i}
& & 2 \sqrt{2} e^{-\frac{4 u-2}{4 \sqrt{2}}} (2 u-1)^{3/2}+e^{-\frac{4 u-2}{4 \sqrt{2}}} (4 u-2)^{3/2} c_1 U\left(\frac{1}{8} \left(10+3
   \sqrt{2}\right),\frac{5}{2},\sqrt{2} u-\frac{1}{\sqrt{2}}\right)\nonumber\\
   & & +e^{-\frac{4 u-2}{4 \sqrt{2}}} (4 u-2)^{3/2} c_2 L_{\frac{1}{8} \left(-10-3
   \sqrt{2}\right)}^{\frac{3}{2}}\left(\sqrt{2} u-\frac{1}{\sqrt{2}}\right)\Biggl[ U\left(\frac{1}{8} \left(10+3 \sqrt{2}\right),\frac{5}{2},\sqrt{2} u-\frac{1}{\sqrt{2}}\right){\cal I}_1 \nonumber\\
    & & + {\cal I}_2 L_{\frac{1}{8} \left(-10-3 \sqrt{2}\right)}^{\frac{3}{2}}\left(\sqrt{2} u-\frac{1}{\sqrt{2}}\right)\Biggr],
\end{eqnarray}
where $U(a,b,z)$ are Tricomi confluent hypergeometric functions defined via \\  $U(a,b;z)\equiv \frac{\Gamma(b-1)}{\Gamma(a)}z^{1-B}(\tau) \ _1F_1
(a-b+1;2;2-b;z) + \frac{\Gamma(1-b)}{\Gamma(a-b+1)} \ _1F_1(a;b;z), b$ not being an integer; $L^a_b(z)$ are associated Laguerre polynomials, and ${\cal I}_{1,2}$ are defined in (\ref{Ax'01-I_1}).
In the context of the equation of motion (\ref{Ax-decoupled-eom_i}) for the R-charge gauge,
the exact solutions being intractable, the same corresponding to $\tilde{\cal A}(\tau)_{x_1}^{(0,0)\ \prime}$ in (\ref{sol_calA_x-i}), near $u=0$, is given as under:
\begin{eqnarray}
\label{calAx_solutions_near_u_0}
& & \hskip -0.3in \tilde{\cal A}(\tau)_{x_1}^{(0,0)\ \prime}= -i e^{\frac{1}{2} i \left(i+\sqrt{11}+i \sqrt{15} \pi \right)}\left\{U\left(\frac{1}{2}+\frac{7 i}{\sqrt{11}}+\frac{i \sqrt{15}}{2},1+i \sqrt{15},-i \sqrt{11}\right)
   c_1+c_2 L_{-\frac{1}{2}-\frac{7 i}{\sqrt{11}}-\frac{i \sqrt{15}}{2}}^{i \sqrt{15}}\left(-i \sqrt{11}\right)\right\}\nonumber\\
   & & \hskip -0.3in +\frac{1}{2} e^{\frac{1}{2} i \left(i+\sqrt{11}+i
   \sqrt{15} \pi \right)} \left[-\left\{\left(-2 i+\sqrt{11}+\sqrt{15}\right) U\left(\frac{1}{2}+\frac{7 i}{\sqrt{11}}+\frac{i \sqrt{15}}{2},1+i \sqrt{15},-i\sqrt{11}\right)\right.\right.\nonumber\\
   & & \hskip -0.3in \left.\left.+\left(14 i+\sqrt{11}+i \sqrt{165}\right) U\left(\frac{3}{2}+\frac{7 i}{\sqrt{11}}+\frac{i \sqrt{15}}{2},2+i \sqrt{15},-i \sqrt{11}\right)\right\}
   c_1\right.\nonumber\\
      & & \hskip -0.3in \left.-c_2 \left\{2 \sqrt{11} L_{-\frac{3}{2}-\frac{7 i}{\sqrt{11}}-\frac{i \sqrt{15}}{2}}^{1+i \sqrt{15}}\left(-i \sqrt{11}\right)+\left(-2
   i+\sqrt{11}+\sqrt{15}\right) L_{-\frac{1}{2}-\frac{7 i}{\sqrt{11}}-\frac{i \sqrt{15}}{2}}^{i \sqrt{15}}\left(-i \sqrt{11}\right)\right\}\right] u\nonumber\\
      & & \hskip -0.3in +\frac{1}{4}
   e^{\frac{1}{2} i \left(i+\sqrt{11}+i \sqrt{15} \pi \right)} \left[\left\{\left(14 i+2 \sqrt{11}+\sqrt{15}+i \sqrt{165}\right) U\left(\frac{1}{2}+\frac{7i}{\sqrt{11}}+\frac{i \sqrt{15}}{2},1+i \sqrt{15},-i \sqrt{11}\right)\right.\right.\nonumber\\
   & & \hskip -0.3in \left.\left.+i \left(\left(39+27 i \sqrt{11}+25 i \sqrt{15}+3 \sqrt{165}\right) U\left(\frac{3}{2}+\frac{7i}{\sqrt{11}}+\frac{i \sqrt{15}}{2},2+i \sqrt{15},-i \sqrt{11}\right)\right.\right.\right.\nonumber\\
   & & \hskip -0.3in \left.\left.\left.+2 \left(-82+14 i \sqrt{11}+11 i \sqrt{15}-7 \sqrt{165}\right) U\left(\frac{5}{2}+\frac{7
   i}{\sqrt{11}}+\frac{i \sqrt{15}}{2},3+i \sqrt{15},-i \sqrt{11}\right)\right)\right\} c_1\right.\nonumber\\
   & & \hskip -0.3in \left.+c_2 \left\{22 i L_{-\frac{5}{2}-\frac{7 i}{\sqrt{11}}-\frac{i
   \sqrt{15}}{2}}^{2+i \sqrt{15}}\left(-i \sqrt{11}\right)+\left(22 i+4 \sqrt{11}+2 i \sqrt{165}\right) L_{-\frac{3}{2}-\frac{7 i}{\sqrt{11}}-\frac{i
   \sqrt{15}}{2}}^{1+i \sqrt{15}}\left(-i \sqrt{11}\right)\right.\right.\nonumber\\
   & & \hskip -0.3in \left.\left.+\left(14 i+2 \sqrt{11}+\sqrt{15}+i \sqrt{165}\right) L_{-\frac{1}{2}-\frac{7 i}{\sqrt{11}}-\frac{i
   \sqrt{15}}{2}}^{i \sqrt{15}}\left(-i \sqrt{11}\right)\right\}\right] u^2+O\left(u^3\right)\nonumber\\
   & & {\hskip -0.3in} \approx -(80.5 - 11.6 i)c_2 + (136.7 + 32.7 i) c_2 u - (28.1 + 22.2 i) c_2 u^2 + {\cal O}(u^3)\nonumber\\
   & & {\hskip -0.3in}  \equiv a + b u + c u^2 + {\cal O}(u^2).
\end{eqnarray}

One can similarly show that $\tilde{\cal A}(\tau)_{x_1}^{(2,0)\ \prime}(u)$, up to ${\cal O}(u)$, will be given by the expansion of the following up to ${\cal O}(u)$:
\begin{eqnarray}
\label{Ax'20_soln-i}
& & \tilde{\cal A}(\tau)_{x_1}^{(2,0)\ \prime}(u) = 2 \sqrt{2} e^{-\frac{4 u-2}{4 \sqrt{2}}} (2 u-1)^{3/2}+e^{-\frac{4 u-2}{4 \sqrt{2}}} (4 u-2)^{3/2} c_1 U\left(\frac{1}{8} \left(10+3
   \sqrt{2}\right),\frac{5}{2},\sqrt{2} u-\frac{1}{\sqrt{2}}\right)\nonumber\\
   & & +e^{-\frac{4 u-2}{4 \sqrt{2}}} (4 u-2)^{3/2} c_2 L_{\frac{1}{8} \left(-10-3
   \sqrt{2}\right)}^{\frac{3}{2}}\left(\sqrt{2} u-\frac{1}{\sqrt{2}}\right)\left(U\left(\frac{1}{8} \left(10+3 \sqrt{2}\right),\frac{5}{2},\sqrt{2} u-\frac{1}{\sqrt{2}}\right){\cal I}_1^\prime \right.\nonumber\\
    & & \left.+ L_{\frac{1}{8} \left(-10-3 \sqrt{2}\right)}^{\frac{3}{2}}\left(\sqrt{2} u-\frac{1}{\sqrt{2}}\right){\cal I}_2^\prime\right),
\end{eqnarray}
where ${\cal I}_{1,2}^\prime$ are defined in (\ref{integrals1-2}).

The integrals ${\cal I}_{1,2}$ appearing in (\ref{Ax'01_soln-i}) and (\ref{Ax'20_soln-i}) are defined as:
\begin{eqnarray}
\label{Ax'01-I_1}
& & {\cal I}_1\equiv\int_1^u d\lambda_1\Biggl[ -\frac{i a e^{-\frac{1-2 \lambda_1}{2 \sqrt{2}}} L_{\frac{1}{8} \left(-10-3 \sqrt{2}\right)}^{\frac{3}{2}}\left(\sqrt{2} \lambda_1-\frac{1}{\sqrt{2}}\right)}{4 \sqrt{2} (2
   \lambda_1-1)^{3/2}{\cal D}_1 } -\frac{i c e^{-\frac{1-2 \lambda_1}{2 \sqrt{2}}} L_{\frac{1}{8} \left(-10-3 \sqrt{2}\right)}^{\frac{3}{2}}\left(\sqrt{2}
   \lambda_1-\frac{1}{\sqrt{2}}\right)}{\sqrt{2} (2 \lambda_1-1)^{3/2} {\cal D}_1}\nonumber\\
   & & -\frac{5 i a e^{-\frac{1-2 \lambda_1}{2 \sqrt{2}}} L_{\frac{1}{8} \left(-10-3 \sqrt{2}\right)}^{\frac{3}{2}}\left(\sqrt{2}
   \lambda_1-\frac{1}{\sqrt{2}}\right)}{4 \sqrt{2} (2 \lambda_1-1)^{5/2} {\cal D}_1} -\frac{i b e^{-\frac{1-2 \lambda_1}{2 \sqrt{2}}} L_{\frac{1}{8} \left(-10-3
   \sqrt{2}\right)}^{\frac{3}{2}}\left(\sqrt{2} \lambda_1-\frac{1}{\sqrt{2}}\right)}{\sqrt{2} (2 \lambda_1-1)^{5/2} {\cal D}_1}\nonumber\\
   & & -\frac{i c e^{-\frac{1-2 \lambda_1}{2
   \sqrt{2}}} L_{\frac{1}{8} \left(-10-3 \sqrt{2}\right)}^{\frac{3}{2}}\left(\sqrt{2} \lambda_1-\frac{1}{\sqrt{2}}\right)}{\sqrt{2} (2 \lambda_1-1)^{5/2} {\cal D}_1}\Biggr];\nonumber\\
   & & {\cal I}_2\equiv \Biggl[\frac{i a e^{-\frac{1-2 \lambda_2}{2 \sqrt{2}}} U\left(\frac{1}{8} \left(10+3 \sqrt{2}\right),\frac{5}{2},\sqrt{2} \lambda_2-\frac{1}{\sqrt{2}}\right)}{4 \sqrt{2} (2
   \lambda_2-1)^{3/2} {\cal D}_2}+\frac{i c e^{-\frac{1-2 \lambda_2}{2 \sqrt{2}}} U\left(\frac{1}{8} \left(10+3 \sqrt{2}\right),\frac{5}{2},\sqrt{2}
   \lambda_2-\frac{1}{\sqrt{2}}\right)}{\sqrt{2} (2 \lambda_2-1)^{3/2} {\cal D}_2}\nonumber\\
     & & +\frac{5 i a e^{-\frac{1-2 \lambda_2}{2 \sqrt{2}}} U\left(\frac{1}{8} \left(10+3 \sqrt{2}\right),\frac{5}{2},\sqrt{2}
   \lambda_2-\frac{1}{\sqrt{2}}\right)}{4 \sqrt{2} (2 \lambda_2-1)^{5/2} {\cal D}_2}+\frac{i b e^{-\frac{1-2 \lambda_2}{2 \sqrt{2}}} U\left(\frac{1}{8} \left(10+3
   \sqrt{2}\right),\frac{5}{2},\sqrt{2} \lambda_2-\frac{1}{\sqrt{2}}\right)}{\sqrt{2} (2 \lambda_2-1)^{5/2} {\cal D}_2}\nonumber\\
   & & +\frac{i c e^{-\frac{1-2 \lambda_2}{2
   \sqrt{2}}} U\left(\frac{1}{8} \left(10+3 \sqrt{2}\right),\frac{5}{2},\sqrt{2} \lambda_2-\frac{1}{\sqrt{2}}\right)}{\sqrt{2} (2 \lambda_2-1)^{5/2} {\cal D}_2}\Biggr];
\end{eqnarray}
\begin{eqnarray}
\label{integrals1-2}
& & \hskip -0.4in  {\cal I}_1^\prime\equiv \int_1^u d \lambda_1\Biggl[ \frac{3 a e^{-\frac{1-2 \lambda_1}{2 \sqrt{2}}} L_{\frac{1}{8} \left(-10-3 \sqrt{2}\right)}^{\frac{3}{2}}\left(\sqrt{2} \lambda_1-\frac{1}{\sqrt{2}}\right)}{\sqrt{2} (2
   \lambda_1-1)^{3/2}{\cal D}_1 }-\frac{b e^{-\frac{1-2 \lambda_1}{2 \sqrt{2}}} L_{\frac{1}{8} \left(-10-3 \sqrt{2}\right)}^{\frac{3}{2}}\left(\sqrt{2}
   \lambda_1-\frac{1}{\sqrt{2}}\right)}{\sqrt{2} (2 \lambda_1-1)^{3/2} {\cal D}_1}\nonumber\\
   & &\hskip -0.4in  +\frac{a e^{-\frac{1-2 \lambda_1}{2 \sqrt{2}}} L_{\frac{1}{8} \left(-10-3 \sqrt{2}\right)}^{\frac{3}{2}}\left(\sqrt{2}
   \lambda_1-\frac{1}{\sqrt{2}}\right)}{\sqrt{2} (2 \lambda_1-1)^{5/2} {\cal D}_1} -\frac{b e^{-\frac{1-2 \lambda_1}{2 \sqrt{2}}} L_{\frac{1}{8} \left(-10-3 \sqrt{2}\right)}^{\frac{3}{2}}\left(\sqrt{2}
   \lambda_1-\frac{1}{\sqrt{2}}\right)}{\sqrt{2} (2 \lambda_1-1)^{5/2} {\cal D}_1}\Biggr];\nonumber
    \end{eqnarray}
   \begin{eqnarray}
 & &\hskip -0.4in  {\cal I}_2^\prime\equiv \int_1^u d \lambda_2\Biggl[ -\frac{3 a e^{-\frac{1-2 \lambda_2}{2 \sqrt{2}}} U\left(\frac{1}{8} \left(10+3 \sqrt{2}\right),\frac{5}{2},\sqrt{2} \lambda_2-\frac{1}{\sqrt{2}}\right)}{\sqrt{2} (2
   \lambda_2-1)^{3/2} {\cal D}_2} +\frac{b e^{-\frac{1-2 \lambda_2}{2 \sqrt{2}}} U\left(\frac{1}{8} \left(10+3 \sqrt{2}\right),\frac{5}{2},\sqrt{2}
   \lambda_2-\frac{1}{\sqrt{2}}\right)}{\sqrt{2} (2 \lambda_2-1)^{3/2} {\cal D}_2}\nonumber\\
   & &\hskip -0.4in  -\frac{a e^{-\frac{1-2 \lambda_2}{2 \sqrt{2}}} U\left(\frac{1}{8} \left(10+3 \sqrt{2}\right),\frac{5}{2},\sqrt{2}
   \lambda_2-\frac{1}{\sqrt{2}}\right)}{\sqrt{2} (2 \lambda_2-1)^{5/2} {\cal D}_2}+\frac{b e^{-\frac{1-2 \lambda_2}{2 \sqrt{2}}} U\left(\frac{1}{8} \left(10+3 \sqrt{2}\right),\frac{5}{2},\sqrt{2}
   \lambda_2-\frac{1}{\sqrt{2}}\right)}{\sqrt{2} (2 \lambda_2-1)^{5/2} {\cal D}_2}\Biggr].
\end{eqnarray}
wherein
\begin{eqnarray}
& &c {\cal D}_i\equiv \left(5 \sqrt{2} U\left(\frac{3}{8} \left(6+\sqrt{2}\right),\frac{7}{2},\frac{2 \lambda_i-1}{\sqrt{2}}\right) L_{\frac{1}{8} \left(-10-3
   \sqrt{2}\right)}^{\frac{3}{2}}\left(\frac{2 \lambda_i-1}{\sqrt{2}}\right)c\right.\nonumber\\
   & & \left.+3 U\left(\frac{3}{8} \left(6+\sqrt{2}\right),\frac{7}{2},\frac{2 \lambda_i-1}{\sqrt{2}}\right)
   L_{\frac{1}{8} \left(-10-3 \sqrt{2}\right)}^{\frac{3}{2}}\left(\frac{2 \lambda_i-1}{\sqrt{2}}\right)\right.\nonumber\\
   &&\left. -4 \sqrt{2} U\left(\frac{1}{8} \left(10+3
   \sqrt{2}\right),\frac{5}{2},\frac{2 \lambda_i-1}{\sqrt{2}}\right) L_{-\frac{3}{8} \left(6+\sqrt{2}\right)}^{\frac{5}{2}}\left(\frac{2
   \lambda_i-1}{\sqrt{2}}\right)\right).   \end{eqnarray}

\end{document}